\input harvmac


\input amssym
\input epsf

\lref\SavitFW{
  R.~Savit,
  ``Topological Excitations in U(1) Invariant Theories,''
Phys.\ Rev.\ Lett.\  {\bf 39}, 55 (1977).
}
\lref\OrlandKU{
  P.~Orland,
  ``Instantons and Disorder in Antisymmetric Tensor Gauge Fields,''
Nucl.\ Phys.\ B {\bf 205}, 107 (1982).
}

\lref\FrohlichCH{
  J.~Frohlich, J.~Fuchs, I.~Runkel and C.~Schweigert,
  ``Duality and defects in rational conformal field theory,''
Nucl.\ Phys.\ B {\bf 763}, 354 (2007).
[hep-th/0607247].
}
\lref\DavydovRM{
  A.~Davydov, L.~Kong and I.~Runkel,
  ``Invertible Defects and Isomorphisms of Rational CFTs,''
Adv.\ Theor.\ Math.\ Phys.\  {\bf 15} (2011).
[arXiv:1004.4725 [hep-th]].
}
\lref\BrunnerLUA{
  I.~Brunner, N.~Carqueville and D.~Plencner,
  ``Discrete torsion defects,''
[arXiv:1404.7497 [hep-th]].
}

\lref\KomargodskiPC{
  Z.~Komargodski and N.~Seiberg,
  ``Comments on the Fayet-Iliopoulos Term in Field Theory and Supergravity,''
JHEP {\bf 0906}, 007 (2009).
[arXiv:0904.1159 [hep-th]].
}

\lref\DumitrescuIU{
  T.~T.~Dumitrescu and N.~Seiberg,
  ``Supercurrents and Brane Currents in Diverse Dimensions,''
JHEP {\bf 1107}, 095 (2011).
[arXiv:1106.0031 [hep-th]].
}

\lref\RazamatOPA{
  S.~S.~Razamat and B.~Willett,
  ``Global Properties of Supersymmetric Theories and the Lens Space,''
[arXiv:1307.4381].
}

\lref\KalbYC{M.~Kalb and P.~Ramond, ``Classical direct interstring action,'' Phys.\ Rev.\ D {\bf 9}, 2273 (1974).}

\lref\KravecPUA{
  S.~M.~Kravec and J.~McGreevy,
  ``A gauge theory generalization of the fermion-doubling theorem,''
Phys.\ Rev.\ Lett.\  {\bf 111}, 161603 (2013).
[arXiv:1306.3992 [hep-th]].
}

\lref\SeibergPQ{
  N.~Seiberg,
  ``Electric - magnetic duality in supersymmetric nonAbelian gauge theories,''
Nucl.\ Phys.\ B {\bf 435}, 129 (1995).
[hep-th/9411149].
}
\lref\IntriligatorID{
  K.~A.~Intriligator and N.~Seiberg,
  ``Duality, monopoles, dyons, confinement and oblique confinement in supersymmetric SO(N(c)) gauge theories,''
Nucl.\ Phys.\ B {\bf 444}, 125 (1995).
[hep-th/9503179].
}

\lref\VillainIR{J.~Villain,``Theory of one-dimensional and two-dimensional magnets with an easy magnetization plane. 2. The Planar, classical, two-dimensional magnet,'' J.\ Phys.\ (France) {\bf 36}, 581 (1975).}

\lref\TeitelboimYA{
  C.~Teitelboim,
  ``Gauge Invariance for Extended Objects,''
Phys.\ Lett.\ B {\bf 167}, 63 (1986).
}

\lref\TeitelboimYC{
  C.~Teitelboim,
  ``Monopoles of Higher Rank,''
Phys.\ Lett.\ B {\bf 167}, 69 (1986).
}

\lref\CheegerSimons{J.~Cheeger, J.~Simons, ``Differential characters and geometric invariants,'' Geometry and
topology (College Park, Md., 1983/84), Springer, pp. 50�80.}

\lref\GaiottoBE{
  D.~Gaiotto, G.~W.~Moore and A.~Neitzke,
  ``Framed BPS States,''
Adv.\ Theor.\ Math.\ Phys.\  {\bf 17}, 241 (2013).
[arXiv:1006.0146 [hep-th]].
}

\lref\SeibergQD{
  N.~Seiberg,
  ``Modifying the Sum Over Topological Sectors and Constraints on Supergravity,''
JHEP {\bf 1007}, 070 (2010).
[arXiv:1005.0002 [hep-th]].
}

\lref\GukovZKA{
  S.~Gukov and A.~Kapustin,
  ``Topological Quantum Field Theory, Nonlocal Operators, and Gapped Phases of Gauge Theories,''
[arXiv:1307.4793 [hep-th]].
}

\lref\FreedYC{
  D.~S.~Freed, G.~W.~Moore and G.~Segal,
  ``Heisenberg Groups and Noncommutative Fluxes,''
Annals Phys.\  {\bf 322}, 236 (2007).
[hep-th/0605200].
}

\lref\FreedYA{
  D.~S.~Freed, G.~W.~Moore and G.~Segal,
  ``The Uncertainty of Fluxes,''
Commun.\ Math.\ Phys.\  {\bf 271}, 247 (2007).
[hep-th/0605198].
}

\lref\BjorkenVG{
  J.~D.~Bjorken,
  ``A Dynamical origin for the electromagnetic field,''
Annals Phys.\  {\bf 24}, 174 (1963).
}

\lref\Moorelectures{G.~Moore, ``Lecture Notes for Felix Klein Lectures,'' {\rm http://www.physics.rutgers.edu/ $\sim$gmoore/FelixKleinLectureNotes.pdf}.}

\lref\WittenAT{
  E.~Witten,
  ``Geometric Langlands From Six Dimensions,''
[arXiv:0905.2720 [hep-th]].
}

\lref\KapustinQSA{
  A.~Kapustin and R.~Thorngren,
  ``Topological Field Theory on a Lattice, Discrete Theta-Angles and Confinement,''
[arXiv:1308.2926 [hep-th]].
}

\lref\BanksZN{
  T.~Banks and N.~Seiberg,
  ``Symmetries and Strings in Field Theory and Gravity,''
Phys.\ Rev.\ D {\bf 83}, 084019 (2011).
[arXiv:1011.5120 [hep-th]].
}
\lref\MaldacenaSS{
  J.~M.~Maldacena, G.~W.~Moore and N.~Seiberg,
  ``D-brane charges in five-brane backgrounds,''
  JHEP {\bf 0110}, 005 (2001)
  [arXiv:hep-th/0108152].
}
\lref\KapustinGUA{
  A.~Kapustin and N.~Seiberg,
  ``Coupling a QFT to a TQFT and Duality,''
JHEP {\bf 1404}, 001 (2014).
[arXiv:1401.0740 [hep-th]].
}

\lref\WittenHF{
  E.~Witten,
  ``Quantum Field Theory and the Jones Polynomial,''
Commun.\ Math.\ Phys.\  {\bf 121}, 351 (1989).
}

\lref\KapustinUXA{
  A.~Kapustin and R.~Thorngren,
  ``Higher symmetry and gapped phases of gauge theories,''
[arXiv:1309.4721 [hep-th]].
}
\lref\DumitrescuIU{
  T.~T.~Dumitrescu and N.~Seiberg,
  ``Supercurrents and Brane Currents in Diverse Dimensions,''
JHEP {\bf 1107}, 095 (2011).
[arXiv:1106.0031 [hep-th]].
}

\lref\AharonyHDA{
  O.~Aharony, N.~Seiberg and Y.~Tachikawa,
  ``Reading between the lines of four-dimensional gauge theories,''
JHEP {\bf 1308}, 115 (2013).
[arXiv:1305.0318 [hep-th]].
}

\lref\CachazoZK{
  F.~Cachazo, N.~Seiberg and E.~Witten,
  ``Phases of N=1 supersymmetric gauge theories and matrices,''
JHEP {\bf 0302}, 042 (2003).
[hep-th/0301006].
}

\lref\GukovJK{
  S.~Gukov and E.~Witten,
  ``Gauge Theory, Ramification, And The Geometric Langlands Program,''
[hep-th/0612073].
}
\lref\GukovSN{
  S.~Gukov and E.~Witten,
  ``Rigid Surface Operators,''
Adv.\ Theor.\ Math.\ Phys.\  {\bf 14} (2010).
[arXiv:0804.1561 [hep-th]].
}

\lref\wittenloops{E.~Witten, Lecture II-9 in: P.~Deligne, P.~Etingof, D.~S.~Freed, L.~C.~Jeffrey, D.~Kazhdan, J.~W.~Morgan, D.~R.~Morrison and E.~Witten,
  ``Quantum fields and strings: A course for mathematicians. Vol. 1, 2,''
Providence, USA: AMS (1999) 1-1501.}

\lref\AharonyDHA{
  O.~Aharony, S.~S.~Razamat, N.~Seiberg and B.~Willett,
  ``3d dualities from 4d dualities,''
JHEP {\bf 1307}, 149 (2013).
[arXiv:1305.3924 [hep-th]].
}

\lref\IntriligatorLCA{
  K.~Intriligator and N.~Seiberg,
  ``Aspects of 3d N=2 Chern-Simons-Matter Theories,''
JHEP {\bf 1307}, 079 (2013).
[arXiv:1305.1633 [hep-th]].
}

\lref\AharonyKMA{
  O.~Aharony, S.~S.~Razamat, N.~Seiberg and B.~Willett,
  ``3$d$ dualities from 4$d$ dualities for orthogonal groups,''
JHEP {\bf 1308}, 099 (2013).
[arXiv:1307.0511, arXiv:1307.0511 [hep-th]].
}

\lref\BanksZN{
  T.~Banks and N.~Seiberg,
  ``Symmetries and Strings in Field Theory and Gravity,''
Phys.\ Rev.\ D {\bf 83}, 084019 (2011).
[arXiv:1011.5120 [hep-th]].
}

\lref\FradkinDV{
  E.~H.~Fradkin and S.~H.~Shenker,
  ``Phase Diagrams of Lattice Gauge Theories with Higgs Fields,''
Phys.\ Rev.\ D {\bf 19}, 3682 (1979).
}
\lref\BanksFI{
  T.~Banks and E.~Rabinovici,
  ``Finite Temperature Behavior of the Lattice Abelian Higgs Model,''
Nucl.\ Phys.\ B {\bf 160}, 349 (1979).
}

\lref\VafaWX{
  C.~Vafa,
  ``Modular Invariance and Discrete Torsion on Orbifolds,''
Nucl.\ Phys.\ B {\bf 273}, 592 (1986).
}

\lref\DijkgraafPZ{
  R.~Dijkgraaf and E.~Witten,
  ``Topological Gauge Theories and Group Cohomology,''
Commun.\ Math.\ Phys.\  {\bf 129}, 393 (1990).
}

\lref\KapustinDXA{
  A.~Kapustin, R.~Thorngren, A.~Turzillo and Z.~Wang,
  ``Fermionic Symmetry Protected Topological Phases and Cobordisms,''
[arXiv:1406.7329 [cond-mat.str-el]].
}

\lref\ChenPG{
  X.~Chen, Z.~-C.~Gu, Z.~-X.~Liu and X.~-G.~Wen,
  ``Symmetry protected topological orders and the cohomology class of their symmetry group,''
Phys.\ Rev.\ B {\bf 87}, 155114 (2013).
[arXiv:1106.4772 [cond-mat.str-el]].
}

\lref\KapustinTFA{
  A.~Kapustin,
  ``Symmetry Protected Topological Phases, Anomalies, and Cobordisms: Beyond Group Cohomology,''
[arXiv:1403.1467 [cond-mat.str-el]].
}

\lref\KapustinPY{
  A.~Kapustin,
  ``Wilson-'t Hooft operators in four-dimensional gauge theories and S-duality,''
Phys.\ Rev.\ D {\bf 74}, 025005 (2006).
[hep-th/0501015].
}
\lref\SeibergRS{
  N.~Seiberg and E.~Witten,
  ``Electric - magnetic duality, monopole condensation, and confinement in N=2 supersymmetric Yang-Mills theory,''
Nucl.\ Phys.\ B {\bf 426}, 19 (1994), [Erratum-ibid.\ B {\bf 430}, 485 (1994)].
[hep-th/9407087].
}

\lref\SeibergAJ{
  N.~Seiberg and E.~Witten,
  ``Monopoles, duality and chiral symmetry breaking in N=2 supersymmetric QCD,''
Nucl.\ Phys.\ B {\bf 431}, 484 (1994).
[hep-th/9408099].
}

\lref\KapustinZVA{
  A.~Kapustin and R.~Thorngren,
  ``Anomalies of discrete symmetries in various dimensions and group cohomology,''
[arXiv:1404.3230 [hep-th]].
}

\lref\GepnerKX{
  D.~Gepner,
  ``Foundations of rational quantum field theory. 1.,''
[hep-th/9211100].
}

\lref\MarinoRE{
  M.~Marino and C.~Vafa,
  ``Framed knots at large N,''
Contemp.\ Math.\  {\bf 310}, 185 (2002).
[hep-th/0108064].
}

\lref\WittenYA{
  E.~Witten,
  ``SL(2,Z) action on three-dimensional conformal field theories with Abelian symmetry,''
In *Shifman, M. (ed.) et al.: From fields to strings, vol. 2* 1173-1200.
[hep-th/0307041].
}

\lref\AcharyaDZ{
  B.~S.~Acharya and C.~Vafa,
  ``On domain walls of N=1 supersymmetric Yang-Mills in four-dimensions,''
[hep-th/0103011].
}

\lref\DieriglXTA{
  M.~Dierigl and A.~Pritzel,
  ``Topological Model for Domain Walls in (Super-)Yang-Mills Theories,''
[arXiv:1405.4291 [hep-th]].
}

\lref\SeibergDR{
  N.~Seiberg and W.~Taylor,
  ``Charge Lattices and Consistency of 6D Supergravity,''
JHEP {\bf 1106}, 001 (2011).
[arXiv:1103.0019 [hep-th]].
}

\lref\Milgram{
R.~J.~Milgram,
``Surgery with Coefficients,''
Ann. of Math. {\bf 100}, 194 (1974).
}

\lref\sltwoorder{
 {Novotn{\'y}}, P. and {Hrivn{\'a}k}, J.
``On Orbits of the Ring $\Z_m^n$ under the Action of the Group $SL(m,\Z_n)$,''
[arXiv:0710.0326]
}

\lref\Deligne{
P.~Deligne, ``Th\'{e}orie de Hodge II,'' Publ. Math. Inst. Hautes \'{E}tudes Sci. {\bf 40}, 5 (1971).
}

\lref\Beilinson{
A.~Beilinson, ``Higher regulators and values of L-functions,'' Current problems in mathematics, vol. {\bf 24}, 181, Itogi Nauki i Tekhniki, Akad. Nauk SSSR, Moscow, 1984.
}

\lref\supercohomology{
Z.~C.~Gu and X.~G.~Wen, ``Symmetry-protected topological orders for interacting fermions: Fermionic topological non-linear sigma-models and a group super-cohomology theory,'' Phys.\ Rev.\ B {\bf 90}, 115141 (2014).
[arXiv:1201.2648 [cond-mat.str-el]].
}

\lref\FreedSPT{
D.~S.~Freed, ``Short-range entanglement and invertible field theories,''
[arXiv:1406.7278 [cond-mat.str-el]].
}

\lref\Whitehead{
J.~H.~C.~Whitehead, ``On simply connected, 4-dimensional polyhedra,'' Comm. Math. Helv. {\bf 22} (1949) 48.
}

\lref\Nussinov{
Z.~Nussinov and G.~Ortiz, ``A Symmetry Principle for Topological Quantum Order,'' [arXiv:0702377 [cond-mat.str-el]].
}


\def\bb{
\font\tenmsb=msbm10
\font\sevenmsb=msbm7
\font\fivemsb=msbm5
\textfont1=\tenmsb
\scriptfont1=\sevenmsb
\scriptscriptfont1=\fivemsb
}



\def\vev#1{\left\langle #1\right\rangle}


\def\tilde{\widetilde}

\def\hat{\widehat}

\def\bar{\overline}
\def\b{\bar}
\def\bsq#1{{{\b{#1}}^{\lower 2.5pt\hbox{$\scriptstyle 2$}}}}
\def\bexp#1#2{{{\b{#1}}^{\lower 2.5pt\hbox{$\scriptstyle #2$}}}}
\def\dotexp#1#2{{{#1}^{\lower 2.5pt\hbox{$\scriptstyle #2$}}}}


\def\rt2{\sqrt{2}}
\def\half {{1 \over 2}}

\def\mod{{\rm mod}}

\def\Tr{\mathop{\rm Tr}}

\def\ra{\rightarrow}


\def\alphadot{{\dot\alpha}}


\def\CB{{\cal B}}
\def\CC{{\cal C}}
\def\CD{{\cal D}}

\def\CG{{\cal G}}
\def\CH{{\cal H}}

\def\CK{{\cal K}}
\def\CL{{\cal L}}
\def\CM{{\cal M}}
\def\CN{{\cal N}}
\def\CO{{\cal O}}
\def\CP{{\cal P}}

\def\CT{{\cal T}}

\def\CV{{\cal V}}
\def\CW{{\cal W}}


\def\1{{\ds 1}}
\def\R{\hbox{$\bb R$}}
\def\C{\hbox{$\bb C$}}

\def\Z{\hbox{$\bb Z$}}

\def\S{\hbox{$\bb S$}}


\noblackbox

\def\unit{\relax{\rm 1\kern-.26em I}}
\def\nada{\relax{\rm 0\kern-.30em l}}
\def\tilde{\widetilde}

\def\alphadot{{\dot \alpha}}

\def\mod{{\rm mod}}
\def\CP{{\cal P}}
\noblackbox
\def\IL{\relax{\rm I\kern-.18em L}}
\def\IH{\relax{\rm I\kern-.18em H}}
\def\IR{\relax{\rm I\kern-.18em R}}
\def\IC{\relax\hbox{$\inbar\kern-.3em{\rm C}$}}
\def\IZ{\relax\ifmmode\mathchoice
{\hbox{\cmss Z\kern-.4em Z}}{\hbox{\cmss Z\kern-.4em Z}} {\lower.9pt\hbox{\cmsss Z\kern-.4em Z}}
{\lower1.2pt\hbox{\cmsss Z\kern-.4em Z}}\else{\cmss Z\kern-.4em Z}\fi}
\def\CM {{\cal M}}

\def\CN {{\cal N}}

\def\CD {{\cal D}}

\def\partialslash{\not{\hbox{\kern-2pt $\partial$}}}
\def\CP {{\cal P }}
\def\CL {{\cal L}}
\def\CV {{\cal V}}
\def\CO {{\cal O}}

\def\CG {{\cal G}}
\def\CH {{\cal H}}
\def\CC {{\cal C}}
\def\CB {{\cal B}}
\def\CW{{\cal W}}

\def\CK{{\cal K}}
\def\CM {{\cal M}}
\def\CN {{\cal N}}

\def\CO {{\cal O}}

\def\CP {{\cal P }}

\def\CV{{\cal V }}

\def\frP{{\frak P}}

\def\Hom{{\rm Hom}}

\def\Tr{{\rm Tr}}

\font\manual=manfnt \def\dbend{\lower3.5pt\hbox{\manual\char127}}

\def\IZ{\relax\ifmmode\mathchoice
{\hbox{\cmss Z\kern-.4em Z}}{\hbox{\cmss Z\kern-.4em Z}} {\lower.9pt\hbox{\cmsss Z\kern-.4em Z}}
{\lower1.2pt\hbox{\cmsss Z\kern-.4em Z}}\else{\cmss Z\kern-.4em Z}\fi}
\def\half {{1\over 2}}

\def\bar{\overline}

\def\CH{{\cal H}}

\def\rt2{\sqrt{2}}
\def\irt2{{1\over\sqrt{2}}}

\def\hat{\widehat}
\def\slashchar#1{\setbox0=\hbox{$#1$}           
   \dimen0=\wd0                                 
   \setbox1=\hbox{/} \dimen1=\wd1               
   \ifdim\dimen0>\dimen1                        
      \rlap{\hbox to \dimen0{\hfil/\hfil}}      
      #1                                        
   \else                                        
      \rlap{\hbox to \dimen1{\hfil$#1$\hfil}}   
      /                                         
   \fi}

\def\gcd{{\rm gcd}}
\def\lcm{{\rm lcm}}

\def\figcaption#1#2{\DefWarn#1\xdef#1{Figure~\noexpand\hyperref{}{figure}%
{\the\figno}{\the\figno}}\writedef{#1\leftbracket Figure\noexpand~\xfig#1}%
\medskip\centerline{{\footnotefont\bf Figure~\hyperdef\hypernoname{figure}{\the\figno}{\the\figno}:}  #2 \wrlabeL{#1=#1}}%
\global\advance\figno by1}
\Title {\vbox{}}
{\vbox{\centerline{Generalized Global Symmetries}
\vskip7pt
}}

\centerline{Davide Gaiotto$^a$, Anton Kapustin$^b$\foot{On leave of absence from California Institute of Technology.}, Nathan Seiberg$^c$, and Brian Willett$^c$ }
\bigskip
\centerline{${}^a${\it Perimeter Institute for Theoretical Physics, Waterloo, Ontario, Canada N2L 2Y5}}
\centerline{${}^b${\it Simons Center for Geometry and Physics, Stony Brook, NY 11790}}
\centerline{$^c${\it School of Natural Sciences, Institute for Advanced Study, Princeton, NJ 08540, USA}}

\bigskip
\vskip.1in \vskip.1in
\noindent
A $q$-form global symmetry is a global symmetry for which the charged operators are of space-time dimension $q$; e.g.\ Wilson lines, surface defects, etc., and the charged excitations have $q$ spatial dimensions; e.g.\ strings, membranes, etc.  Many of the properties of ordinary global symmetries ($q=0$) apply here.  They lead to Ward identities and hence to selection rules on amplitudes.  Such global symmetries can be coupled to classical background fields and they can be gauged by summing over these classical fields.  These generalized global symmetries can be spontaneously broken (either completely or to a subgroup).  They can also have 't Hooft anomalies, which prevent us from gauging them, but lead to 't Hooft anomaly matching conditions.  Such anomalies can also lead to anomaly inflow on various defects and exotic Symmetry Protected Topological phases. Our analysis of these symmetries gives a new unified perspective of many known phenomena and uncovers new results.
\vfill

\Date{December 2014}

\newsec{Introduction}

The study of higher-form gauge fields is by now standard in physics and in mathematics.
In the physics literature such gauge theories were first discussed by Kalb and Ramond \KalbYC\ and were used extensively in lattice gauge theories following the work of Villain
\VillainIR.  The analysis of electric and magnetic charges of such gauge theories started in various contexts by \refs{\SavitFW\OrlandKU\TeitelboimYA-\TeitelboimYC} and found a natural setting in supergravity and string theory.  In the mathematical literature higher-form gauge fields are known as Cheeger-Simons differential characters \CheegerSimons\ or Deligne-Beilinson cocycles \refs{\Deligne,\Beilinson} and they have many applications.

The purpose of this paper is to explore the consequences of such global symmetries  \refs{\KapustinUXA,\KapustinGUA} rather than their local gauged versions.  Here the charged objects are not particles, but higher dimensional branes.  Similarly, the charged observables are not zero-dimensional local operators, but higher dimensional objects. Similar ideas have been considered in \Nussinov . 

Logically, such global symmetries should be discussed before they are being gauged.  In fact, as we will see below, often these global symmetries have 't Hooft anomalies and hence they cannot be gauged.  Furthermore, as with ordinary ($q=0$) gauge symmetries, the higher-form generalized gauge symmetries can be ambiguous.  The same physical system can have different presentations with different higher-form gauge symmetries.  On the other hand, higher-form global symmetries are unambiguous.  They lead to Ward identities, organize the spectrum into representations of the symmetry, etc.

It is common in string theory to have charged branes.  These branes differ from the objects we will discuss in two respects.  First, they are dynamical objects in the theory, while we will focus mostly on operators that represent external probe branes.  Second, in a theory of gravity just as ordinary global symmetries are absent, higher-form symmetries should also be gauged (see e.g.\ \BanksZN\ and references therein).

When the symmetry is continuous there is a Noether current, which is a high spin current. Some familiar examples of such symmetries and the corresponding brane currents are known in the study of the supersymmetry algebra and BPS states (see e.g.\ \DumitrescuIU\ and references therein), but to the best of our knowledge they are rarely discussed as global symmetries.

In order to avoid confusion we should clarify our terminology.  First, for extended observables (lines, surfaces, etc.)  we will use the phrases operator and defect interchangeably.  When the spacetime manifold is of the form $\CM_{d-1}\times \R$ or $ \CM_{d-1} \times S^1$, it is natural to refer to the factor of $\R$ or $S^1$ as (Euclidean) time.  In that case, when the extended observable is placed at a given time, it can be interpreted as an operator acting on a Hilbert space.  Alternatively, when it is stretched along the time direction, it is not an operator in the theory, but a defect -- it describes the same theory with a different Hilbert space.  Since our view will be mostly Euclidean and we will view these observables as insertions in the functional integral, we will use the terms, observable/operator/defect interchangeably.

Second, the standard terminology of observables suffers from an ambiguity.  Often, the definition of observables depends on a choice of lower codimension objects they are attached to.  A well known example occurs in the Ising model.  The spin operator is a local operator.  The disorder operator needs a line connecting it to another disorder operator.  For zero magnetic field the dependence on this line is only through its topology and therefore the disorder operator is sometimes also being referred to as a local operator.  In order to avoid confusion, following \KapustinGUA, we will refer to operators, like the spin, which are independent of the choice of line as genuine local operators.  A similar comment applies to extended observables.  Consider for example an $SU(N)$ pure gauge theories (without matter).  The Wilson line is a genuine line operator.  But the 't Hooft line depends on the topology of the surface it bounds.  On the other hand, in the $PSU(N)=SU(N)/\Z_N$ gauge theory the 't Hooft line is a genuine line operator and the basic Wilson line needs a topological surface attached to it.

Our presentation of global symmetries will not rely on an underlying Lagrangian.  Instead, we will characterize the charges and the charged objects as abstract operators. This gives an intrinsic description of the symmetry, which is valid even when there is no Lagrangian description of the theory, or when there is more than one Lagrangian (as in duality).  Also, as we will see, sometimes the charge does not have a clear action on the fundamental fields, but the corresponding charge operator still exists.

Specifically, a $q$-form symmetry in $d$ dimensions is implemented by an operator associated with a codimension $q+1$ closed manifold $M^{(d-q-1)}$,
\eqn\Uetafirst{U_g(M^{(d-q-1)})~,}
where $g \in G$ is an element of the symmetry group $G$.  The fact that this is a symmetry means that the manifold $M^{(d-q-1)}$ can be deformed slightly without affecting correlation functions -- the answers depend only on the topology of $M^{(d-q-1)}$.  Such operators can be multiplied
\eqn\Uetamfirst{U_g(M^{(d-q-1)})U_{g'}(M^{(d-q-1)})=U_{g''}(M^{(d-q-1)})}
with $g''=g g' \in G$.\foot{More generally, topological operators can have more complicated products.  For example, we can have a linear combination of operators on the right hand side of \Uetamfirst.  Here we limit ourselves to the set of topological operators with products like \Uetamfirst.  It would be nice to understand the general conditions on such operator products.}  As we will see below, for $q>0$ the symmetry group $G$ must be Abelian. The charged objects are operators supported on $q$-dimensional manifolds $V(\CC^q)$.  The dependence on $\CC^q$ is not always topological. Below we will explain it in more detail and will present several examples.

Such higher-form symmetries are important in the context of duality, where several different Lagrangians describe the same theory.  In such a situation the gauge symmetries of the dual descriptions do not have to match.  But the global symmetries must match.  We claim that the same is true for higher-form global symmetries.  The various dual descriptions should have the same such symmetries.  And the charged operators in the dual descriptions should also match.  Indeed, the analysis of \AharonyHDA\ demonstrates such mapping between charged lines in dual theories.  This nontrivial matching of operators is particularly impressive because the charged objects are not BPS.

Many of the known properties of ordinary global symmetries are true here.  First, they lead to Ward identities and selection rules of amplitudes.  Second, it is natural to couple the system to flat background gauge fields for these symmetries.  This allows us to study ``twisted sectors.''  For ordinary $q=0$ symmetries this is common in the study of orbifolds and in computing partition functions with various chemical potentials.  The analogous constructions for higher-form symmetries occur in the study of 't Hooft twisted boundary conditions and in related systems.

Concretely, we start with a gauge theory based on a simply connected gauge group $\CG$ and assume that there are no matter fields transforming under $\Gamma$, a subgroup of the center of $\CG$.  Then the system has $\Gamma$ one-form global symmetry.  We can consider the partition function $Z_a$ of a $\CG/\Gamma$ bundle that is not a $\CG$ bundles ($a$ labels such bundles).  The partition function of the original $\CG$ gauge theory is $Z(\CG)=Z_0$.  We interpret the partition functions $Z_a$ with fixed $a$ as observables in the $\CG$ gauge theory.  They describe the result of coupling the system to flat background gauge fields of $\Gamma$.

As with orbifolds, gauging these global symmetries amounts to summing over the twisted sectors with some weights $h_a$
\eqn\sumse{Z(\CG/\Gamma)=\sum_a h_aZ_a}
(and when the one-form global symmetry is continuous we might also allow nonzero field strength).  This leads to the partition function of the $\CG/\Gamma$ gauge theory.\foot{Constructing the $\CG/\Gamma$ gauge theory using gauging of higher-form symmetries is common in lattice gauge theories \VillainIR.}  Interestingly, there is some freedom in the values of the constants $h_a$. Available choices are parameterized by discrete theta-parameters. This is a higher-form generalization of the phenomenon of discrete torsion \refs{\AharonyHDA,\KapustinGUA,\KapustinQSA}.  In fact, as we will discuss, the coefficient $h_a$ are meaningful even in the untwisted $\CG$ theory.  Here they can be interpreted as contact terms, or equivalently, as nonlinear couplings of the background gauge fields of $\Gamma$.

As with ordinary global symmetries these higher-form global symmetries can be spontaneously broken.  One example of that, which we will discuss in detail, is the identification of the massless photon as a Nambu-Goldstone boson of such a spontaneously broken symmetry.\foot{This is distinct from the suggestion of \BjorkenVG\ that the photon is a Goldstone boson of broken Lorentz symmetry.}  Broken global discrete symmetries also have low energy consequences.  Spontaneously broken ordinary $q=0$ symmetries lead to domain walls.  Spontaneously broken higher-form global symmetry lead in the infrared to a topological gauge theory.

We will also phrase the classification of phases of $4d$ gauge theory using the language of spontaneous breaking of one-form global symmetries.  This will unify the Landau description of phase transitions with Wilson's and 't Hooft's description of these phases in terms of the long distance behavior of line operators.

Anomalies are important in the study of ordinary global symmetries.  We will see that higher-form global symmetries can also have anomalies.  And the low-energy theory should respect the 't Hooft anomaly matching conditions for these symmetries.  This is particularly interesting for the unbroken part of the one-form global symmetry group.  As we will show, such anomaly matching can lead to interesting boundary effects and to long range order on domain walls.

In section 2 we discuss ordinary global symmetries in a language that will make it easy for us to generalize to higher-form symmetries.  Most of this section will review known facts, but we will also have some new points.

Section 3 defines higher-form global symmetries and explores some of their general properties.  Section 4 presents examples of theories and their higher-form global symmetries.  The discussion here focuses on the kinematics; we postpone dynamical considerations to later sections.

Section 5 is devoted to the discussion of spontaneous symmetry breaking of higher-form symmetries.  We make various general comments and demonstrate them in several examples.

Section 6 presents a series of operations relating different theories.  These involve two basic operations.  First, we modify contact terms -- equivalently, we add higher order terms in background gauge fields of higher form symmetries.  Second, we gauge these symmetries by summing over these fields.  These operations are similar to Witten's operations involving ordinary global symmetries in $3d$ \WittenYA.

Section 7 discusses several applications of these global symmetries.  We point out the selection rules that they lead to, the way they help organize the phases of gauge theories, their anomalies, and the consequences of these anomalies.

In Appendices A, B, and C we review and extend various points in \KapustinGUA, which are useful for our analysis here.  They focus on the continuum description of discrete gauge theories.  In Appendix A we discuss the $\Z_n\times \Z_n$ $2d$ Dijkgraaf-Witten theory \DijkgraafPZ\ with emphasis on its broken and unbroken global symmetries and the effect on boundaries.  In Appendix B we discuss a $4d$ topological gauge theory.  Appendix C describes a construction of $SU(N)/\Z_k$ gauge theories using gauging of a one-form global symmetry.

Appendix D uses our interpretation of the confinement index to describes some of the phases found in \CachazoZK\ using our language.  Appendix E extends some of our general comments to gauge theories on non-spin manifolds.  Appendix F explains the relation of our work to the noncommuting fluxes of \refs{\FreedYA,\FreedYC}.

\newsec{Preliminaries: ordinary symmetries}

In this section we review ordinary symmetries and express them in a language that makes the generalization to higher-form symmetries easy.  In addition to phrasing many known facts in our language, we will also present a number of new results.

Symmetry transformations form a group.  If the group is continuous, for every continuous generator there is a conserved Noether current, which we write as a $d-1$-form $j$.
The conserved charge is the integral
\eqn\consc{Q(M^{(d-1)})=\oint_{M^{(d-1)}} j}
where $M^{(d-1)} $ is a $d-1$-dimensional manifold. Typically $M^{(d-1)} $ is space and it can be non-compact.  More generally, it is a closed $d-1$-dimensional space separating spacetime to two regions.

It is useful to consider the symmetry transformation as an operator associated with the manifold $U_g (M^{(d-1)})$ with $g$ a group element of the global symmetry.  In the continuous case $U_g(M^{(d-1)})$ can be obtained by exponentiating $ Q(M^{(d-1)})$. Both for discrete and continuous symmetry groups, we can define $U_g (M^{(d-1)})$ by cutting space-time along
$M^{(d-1)}$ and inserting a group transformation in the complete set of states for the Hilbert space associated to $U_g (M^{(d-1)})$.
The transformations satisfy the group law
\eqn\grouplaw{U_g (M^{(d-1)})U_{g'} (M^{(d-1)})= U_{g''} (M^{(d-1)}) }
with $g''=g g'$ in the group.

The fact that $U_g(M^{(d-1)})$ is associated with a symmetry means that the dependence on $M^{(d-1)}$ is topological -- it is unchanged when $M^{(d-1)}$ is deformed slightly.
It can change only when the deformation of $M^{(d-1)}$ crosses an operator $V$ charged under the symmetry.  Specifically, for a closed $d-1$-dimensional sphere $S^{d-1}$ surrounding a point $\CP$
\eqn\comrele{U_g(S^{d-1})\, V_i(\CP) =  R^j_i(g)\, V_j(\CP) ~, }
where $R^j_i(g)$ is the representation of the group element $g$ of $V$.
For example, when the group is $U(1)$ the parameter $g$ is a phase and $R^j_i(g) = g^{q(V)}$ with $q(V)$ the charge of $V$.  A special case of \comrele\ is in canonical quantization with $M^{(d-1)}$ being the entire space.  Then, \comrele\  implies the operator commutation relation
\eqn\comrel{U_g(M^{(d-1)})\, V_i(\CP) =  R^j_i(g)\, V_j(\CP)\, U_g(M^{(d-1)})~ \qquad {\rm at ~ equal ~ time}. }
Note that since this is an equal time relation the two operators are not space-like separated; the lack of commutativity arises from the fact that the local operator at $\CP$ is at the same spacetime point as a point on $ M^{(d-1)}$.

Equivalently, the operator $U_g(M^{(d-1)})$ introduces a discontinuity in the fields across $M^{(d-1)}$ by the symmetry transformation.

It is also common to consider operators associated with open $d-1$-dimensional spaces $\Sigma^{(d-1)}$.  Such operators are not unique.  We can add on the boundary $\gamma^{(d-2)}=\partial \Sigma^{(d-1)}$ an arbitrary operator.

One way of thinking about these operators is to couple our original system to classical background gauge fields for the global symmetry.  Then, the operator associated with the open manifold with boundary $\gamma^{(d-2)}$ can be described as a flat background gauge field with holonomy $g$ around $\gamma^{(d-2)}$.  The flatness fails at the boundary $\gamma^{(d-2)}$, reflecting the fact that one needs to specify additional information there to get a well-defined observable.

More generally, we can always think about a background flat connection for the global symmetry as a network of $U_g (M^{(d-1)})$ defects.\foot{Similar constructions have appeared in the study of $2d$ RCFT (see e.g.\ \refs{\FrohlichCH,\DavydovRM}) and $2d$ TQFT (see e.g.\ \BrunnerLUA).}
The flat connection is defined by partitioning the manifold into open sets ${\cal U}_\alpha$ and
prescribing transition functions $g_{\alpha \beta}$ in the intersection of two open sets ${\cal U}_\alpha \cap {\cal U}_\beta$, constrained to agree on triple intersections ${\cal U}_\alpha \cap {\cal U}_\beta\cap {\cal U}_\gamma$. We can implement the $g_{\alpha \beta}$ transition functions by $U_{g_{\alpha \beta}}$ operators lying in the appropriate intersection, joined at junctions of various codimensions.
This description allows one to encode the coupling of the theory to a background flat connection in terms of the $U_g(M^{(d-1)})$ and their junctions. As we will see in examples momentarily,
global 't Hooft anomalies may also be encoded in the absence of appropriate topological junctions for the $U_g(M^{(d-1)})$.\foot{For continuous symmetries, one can of course consider more general background connections which are not flat, and thus cannot be encoded as network of $U_g$ defects. }

We can also make this gauge symmetry dynamical. As the $U_g (M^{(d-1)})$ operators now implement a gauge transformation, and all surviving operators are gauge-invariant, the
$U_g (M^{(d-1)})$ become trivial in the gauged theory. And the operators with $\gamma^{(d-2)}=\partial \Sigma^{(d-1)}$ are twisted sector operators associated with $\gamma^{(d-2)}$.
Gauging discrete symmetries means to sum over all possible flat connections on the manifold, i.e.\ summing over all possible
insertions of the operators $U_g (M^{(d-1)})$.

Often, there are inequivalent ways to gauge a symmetry. A classical example is the choice of discrete torsion in discrete orbifolds of $2d$ CFTs \VafaWX: the torus partition function for the orbifolded theory is a sum over torus partition functions of the original theory twisted by a flat connection on the torus with general commuting holonomies $(g_1,g_2)$:
\eqn\orbit{
Z_{G} = \sum_{g_1,g_2} \epsilon(g_1,g_2) Z_{g_1,g_2}~.}
The choice of extra phases $\epsilon(g_1,g_2)$ is the discrete torsion. The possible choices of discrete torsion and the phases required in higher genus orbifold partition functions
can be understood in the language of the $U_g$ defects: the flat connections can be realized by a network of $U_g$ defects and the extra phases
associated to each network can be written as a product of individual contributions from the junctions in the network.
The contributions from each junction must be such that the final answer does not depend on the actual topology of the network,
but only on the choice of flat connection, and is covariant under modular transformations
of the Riemann surface.\foot{This can be expressed as well in the language of the open sets $\;{\cal U}_\alpha$ and transition functions $g_{\alpha \beta}$,
as a topological action written as a sum of contributions from the triple intersections ${\cal U}_\alpha \cap {\cal U}_\beta\cap {\cal U}_\gamma$.}

All intersections in the network can be resolved as trivalent vertices, with defects $U_{g_i}$ coming together at a point and $g_1 g_2 g_3=1$, weighed by some phase
$\alpha(g_1, g_2, g_3)$. These factors are invariant with respect to cyclic permutations of $g_1,g_2,g_3$. There is also a gauge symmetry: if $b(g):G\to U(1)$ is an arbitrary function satisfying $b(g^{-1})=b(g)^{-1}$, the transformation
\eqn\coboundary{
\alpha(g_1,g_2,g_3)\mapsto \alpha(g_1,g_2,g_3) b(g_1)b(g_2)b(g_3)
}
leaves the overall phase factor attached to the network unchanged.

For any network of defects representing a flat $G$-connection, the phase factor is a product of $\alpha$'s. For example, for a flat connection on a torus one gets
\eqn\discrete{
\epsilon(g_1,g_2) = \alpha(g_1,g_2,g_2^{-1} g_1^{-1}) \alpha(g_1^{-1},g_2^{-1},g_2 g_1).
}
Invariance under change of topology of the network and modular invariance are guaranteed, if the two distinct resolutions of a quadruple intersection,
with defects $U_{g_i}$ coming together and $g_1 g_2 g_3 g_4=1$, give the same answer:
\eqn\cohomology{
\alpha(g_1,g_2,g_3 g_4) \alpha(g_1 g_2, g_3, g_4) = \alpha(g_1,g_2 g_3,g_4)\alpha(g_4 g_1,g_2,g_3).
}
The inequivalent choices of phases $\alpha$ for a group $G$ define the second group cohomology  $H^2(G,U(1))$ \VafaWX.\foot{This is slightly nonobvious, since usually group cohomology is computed using cocycles that are not cyclically symmetric. The key observation is that for such cocycles the gauge symmetry is also larger: the functions $b(g)$ are not required to satisfy $b(g^{-1})=b(g)^{-1}$. Using this additional gauge freedom one can always find a cyclically-symmetric $U(1)$-valued 2-cocycle in a given cohomology class.}

An alternative, useful point of view is that the phases $\epsilon(g_1,g_2)$ by themselves are the partition function of a very simple gapped
``theory'' $T^\alpha_G$ with global symmetry group $G$, defined by some topological action $S_\alpha$ depending on a flat $G$ connection, the Dijkgraaf-Witten action
\DijkgraafPZ. An orbifold with discrete torsion of some $\rm{CFT}_2$ is the same as a vanilla orbifold of a product theory $T^\alpha_G \times \rm{CFT}_2$.
The theories $T^\alpha_G$ play a useful role in physics on their own: they describe two-dimensional Symmetry Protected Topological phases of matter.

A Symmetry Protected Topological (SPT) phase of matter is a gapped phase, which is indistinguishable from the trivial phase, if one disregards symmetry (in particular, the partition function of the low-energy TQFT is the identity on any closed manifold), but cannot be deformed to the trivial phase without undergoing a quantum phase transition and while preserving the symmetry \ChenPG.  To every SPT phase with symmetry $G$ one can attach a classical topological action describing the coupling of the system to a background
geometry as well as a flat background $G$-connection \KapustinTFA.

For ordinary discrete global symmetries and in sufficiently low dimension $d$, Symmetry Protected Topological (SPT) phases without fermions are classified by Dijkgraaf-Witten (DW) actions built from group cohomology elements in
$H^d(G,U(1))$. The precise classification in generic dimension is not yet settled: very general topological actions can be described by elements in
certain cobordism groups \KapustinTFA, but it is not clear at this point whether these actions can all be given a local formulation.
This approach can be generalized to SPT phases with time-reversing symmetries and fermionic degrees of freedom \refs{\KapustinTFA, \KapustinDXA} (see \supercohomology\ and \FreedSPT\ for alternative proposals).

SPT phases of matter are of particular interest because of their boundary behavior: every boundary condition must either break the global symmetry or support gapless or topologically ordered
degrees of freedom. In high energy physics language, the topological actions for SPT phases always have a boundary 't Hooft anomaly,
which must be compensated by an anomaly inflow mechanism involving anomalous degrees of freedom at the boundary.

We can give a simple $2d$ example. A $2d$ boundary condition preserves the symmetry $G$ if the $U_g$ line defects can end topologically on the boundary.
This allows one to draw the networks of $U_g$ defects which, say, are required to define partition functions of an orbifold theory on a Riemann surface with boundaries.
In order to have topological and modular invariance, there must be a relation between any phases $\beta(g)$ associated to the boundary endpoints of $U_g$
defects and $\alpha(g_1,g_2,g_3)$:
\eqn\boundarytop{\beta(g_1) \beta(g_2) =\alpha(g_1,g_2,g_2^{-1} g_1^{-1}) \beta(g_1 g_2)}
This is precisely the condition for $\alpha(g_1,g_2,g_3)$ to be cohomologically trivial.
Graphically, this condition means that the boundary endpoints of two lines can be merged and split into a single boundary endpoint joined to a triple intersection.

In order to have a non-trivial $\alpha(g_1,g_2,g_3)$ we must allow for some boundary degeneracy: we tensor the $\rm{CFT}_2$ boundary condition
with some vector space $B$, the ground states of the $1d$ boundary topological quantum field theory (TQFT), and promote $\beta$ to linear operators on $B$. Then \boundarytop\ is the condition for
$\beta$ to define a projective $G$ representation on $B$ of type $\alpha$. Notice that $H^2(G,U(1))$ indeed classifies projective representations of $G$.

In two dimensions, the choice of discrete torsion affects how the global symmetry group acts on twisted sector operators
(or at least the subgroup of $G$ that commutes with the holonomy around the twist operator) and thus determines the twisted sector operators
in the orbifolded theory. As an example, consider two copies of the $2d$ Ising model. This theory has a $\Z_2\times\Z_2$ symmetry, whose generators multiply the spin operators $\sigma_1$ and $\sigma_2$ by $-1$. In the language of line defects, a spin operator changes sign when it crosses the corresponding line defect. Since $H^2(\Z_2\times\Z_2,U(1))=\Z_2$, there are two inequivalent ways to gauge the symmetry. A concrete form of a nontrivial 2-cocycle is not unique; one possible choice is
\eqn\alphacocycle{
\alpha(g,g',(g g')^{-1})=i^{a_1 a_2'-a_2 a_1'},
}
where $g=(a_1,a_2)$ and $g'=(a_1',a_2')$ are arbitrary elements of $\Z_2\times\Z_2$. Then it is easy to see that an intersection of two line defects $U_{g}$ and $U_{g'}$ is assigned a phase
\eqn\epsiloncocycle{
\epsilon(g,g')=\alpha(g,g',(gg')^{-1})\alpha({g}^{-1},{g'}^{-1}, g' g)=(-1)^{a_1a_2'-a_2 a_1'}.
}
Consider now possible terminations of a line defect $U_{g}$. In a gauged theory, $U_{g}$ becomes invisible, so the termination point must be a genuine local operator. In particular, one should be able to move it across some other line defect $U_{g'}$ without changing the partition function. However, moving a termination point across $U_{g'}$ either creates or destroys an intersection of line defects, and, if $\epsilon(g,g')$ is nontrivial, changes the sign of the partition function. To compensate for this, we need to insert a nontrivial spin operator at the termination point. For example, choosing $g=(1,0)$ and $g'=(0,1)$, we see that we need to place $\sigma_2$ at the termination point of $U_{g}$ and $\sigma_1$ at the termination point of $U_{g'}$.

\newsec{Higher form symmetries: kinematics}

This discussion is easily extended to higher-form symmetries. Intuitively, a $q$-form symmetry is a symmetry that acts on
operators $V(\CC^{(q)})$ supported on $q$-dimensional manifolds $\CC^{(q)}$. The symmetry parameter is a closed $q$ form $\lambda$
and the transformation of an operator is controlled by the pairing $\int_{\CC^{(q)}} \lambda$. A theory equipped with a non-anomalous
higher-form symmetry can be coupled to a background $(q+1)$-form connection.

A continuous $q$-form symmetry is simply associated to a conserved $q+1$-form current or, equivalently, a closed $d-q-1$-form current $j$.
Given the closed $q$-form $\lambda$ we can define a standard charge operator $U_\lambda$ by integrating $\lambda \wedge j$
on space. It is often more useful, though, to think in terms of charge operators $U_g(M^{(d-q-1)})$ associated with co-dimension $q+1$ manifolds $M^{(d-q-1)}$,
obtained by integrating the conserved current on $M^{(d-q-1)}$. If we embed $M^{(d-q-1)}$ in some constant time slice, $U_g(M^{(d-q-1)})$ will coincide with
$U_\lambda$ for some form $\lambda$ dual to $M^{(d-q-1)}$ in the constant time slice.

In this paper, we will essentially define a general higher form symmetry as the existence of topological operators $U_g(M^{(d-q-1)})$ associated with co-dimension $q+1$ manifolds $M^{(d-q-1)}$
that fuse according to a group law
\eqn\grouplawq{U_g (M^{(d-q-1)})\, U_{g'} (M^{(d-q-1)})= U_{g''} (M^{(d-q-1)}) }
(compare with \grouplaw).

The expression \grouplawq\ can be interpreted as the multiplication rule of two operators acting at a given time $t$ along the manifold $M^{(d-q-1)}$.  The ordering of the operators in \grouplawq\ can be studied by inserting the two operators at slightly different times, say $t$ and $t+\epsilon$, and then the standard time-ordering determines the operator ordering.  For ordinary symmetries, the manifold $M^{(d-q-1)}$ is of co-dimension $1$ and the operators $U_g (M^{(d-q-1)})$ at the different times might not commute.  Hence, $G$ can be non-Abelian.  On the other hand, for $q>0$ the manifold $M^{(d-q-1)}$ at time $t+\epsilon$ can be continuously deformed to $M^{(d-q-1)}$ at time $t-\epsilon$.  Therefore, the two operators in \grouplawq\ must commute and hence $G$ must be Abelian. (We will see in Appendix F how the symmetry can become non-Abelian when the theory is placed on a manifold with non-trivial topology.)

The existence of such symmetries does not contradict the Coleman-Mandula theorem.  The charged objects are $q$-branes rather than particles.  Therefore, the ordinary charge obtained by integrating the current over all of space is infinite; only the charge per unit $q$-volume (along the brane) is finite.

The charged operators are supported on $q$-dimensional manifolds $\CC^{(q)}$.  The analog of \comrele\ is obtained for a small sphere linking $\CC^{(q)}$
\eqn\comrelqe{U_g(S^{d-q-1}) \, V(\CC^{(q)}) =  g(V) \, V(\CC^q)  }
where $g(V)$ is the representation of $g$ (typically a phase).  The analog of the equal time commutator \comrel\ is
\eqn\comrelq{U_g(M^{(d-q-1)}) \, V(\CC^{(q)}) =  g(V)^{( \CC^{(q)}, M^{(d-q-1)})}\, V(\CC^q) \, U_g(M^{(d-q-1)}) \qquad {\rm at ~ equal ~ time,} }
where $( \CC^{(q)}, M^{(d-q-1)})$ is the intersection number.

The interpretation of \comrelq\ is clear.  The operator $U_g(M^{(d-q-1)})$ implements the $q$-form symmetry transformation as we cross $M^{(d-q-1)}$.  The factor of $ g(V)^{( \CC^{(q)}, M^{(d-q-1)})}$ arises from the action of this operator on $  V(\CC^{(q)})$.

An equivalent way to phrase \comrelqe\ and \comrelq\ is obtained by coupling the system to flat background gauge fields for the $q$-form symmetry. These are $q+1$-form gauge fields, whose holonomies are measured on $q+1$-dimensional manifolds.  Then, the operator along $M^{(d-q-1)}$ can be described as generating holonomy $g$ in these gauge fields for every $q+1$ dimensional manifold that crosses $M^{(d-q-1)}$. In order to couple the theory to a general background $(q+1)$-form connection, we will generally need to stretch a topological network of $U_g$ defects over the
manifold. In particular, we need the theory to be equipped with topological junctions between $U_g$ defects that can be reorganized freely to change the topology of the
network.

The set of topological classes of such networks of $U_g$ defects in a spatial slice $\CM_{d-1}$, which define charge operators, is in one-to-one correspondence with $H^q(\CM_{d-1},G)$.  Namely, given an element in  $H^q(\CM_{d-1},G)$, represented as \v{C}ech cocycle, we can place appropriate $U_g$ defects inside $(q+1)$-fold intersections of sets in the open cover.  The condition that the cocycle is coclosed ensures that we can consistently join these defects at junctions contained in the $(q+2)$-fold intersections.  One can check conversely that all $U_g$ networks arise in this way.  Therefore let us denote the operator corresponding to such a network as $U(\alpha)$, for $\alpha \in H^q(\CM_{d-1},G)$.  For non-torsion cocycles, one can construct the corresponding operator by simply placing a $U_g$ defect on the Poincare dual cycle, while for torsion cocycles one must use a network of defects.  Some aspects of this construction in the torsion case, and a connection to the work of \refs{\FreedYA,\FreedYC}, are discussed in Appendix F.  The charged operators are similarly associated to elements $\beta$, in the homology, $H_q(\CM_{d-1},\hat{G})$ in an obvious way, and one finds the following generalization of \comrelq:

\eqn\comrelqgen{
U(\alpha) \, V(\beta) =  \langle\alpha,\beta\rangle\,V(\beta) \,U(\alpha)  \qquad {\rm at ~ equal ~ time,}
}
where the intersection number is replaced by the pairing $\langle.,.\rangle$ between $H^q(\CM_{d-1},G)$ and $H_q(\CM_{d-1},\hat{G})$.

The non-existence of consistent junctions signals an 't Hooft anomaly for the higher form symmetry -- the symmetry exists, but it cannot be gauged. A notable example is a $3d$ theory with a one-form symmetry
such that the $U_g$ operators themselves are charged under the symmetry: the $U_g$ operators cannot be deformed with impunity
across each other. We will encounter an example of this phenomenon in a later section.
This can be generalized to a theory with a $q$-form and an $(d-q-1)$-form symmetry: if the $U_g$ operators for one symmetry
are charged under the other symmetry, the theory has a mixed 't Hooft anomaly.

One can classify possible 't Hooft anomalies under the assumption that the anomaly can be canceled by an anomaly inflow from a classical topological field theory in one dimension higher.\foot{This assumption may not hold in sufficiently complicated theories. Some examples for the case of $0$-form global symmetry have been given in \KapustinZVA .}  If this is the case, an anomaly can be characterized by the action of this theory. For example, consider a theory in $d$ dimensions that has a $U(1)$ global $q$-form symmetry. Then, possible 't Hooft anomalies correspond to topological actions in $d+1$ dimensions built out of a $q+1$-form gauge field $C$. These actions should be viewed as classical.  They might not lead to consistent quantum theories.  For example, 't Hooft anomalies for an ordinary $U(1)$ global symmetry in $4d$ are described by a $5d$ Chern-Simons action $C(dC)^2$ for a background $U(1)$ gauge field $C$.  Since it is cubic, it is hard to make sense of quantum mechanically. Another fairly simple case if when $G$ is discrete. Then, as explained in \refs{\KapustinQSA, \KapustinUXA}, possible topological actions depending only on the $q+1$-form  gauge field correspond to elements of $H^{d+1}(B^{q+1} G,U(1)),$ where $B^{q+1} G$ is the iterated classifying space of $G$ (i.e. an Eilenberg-MacLane space of type $K(G,q+1)$). If one allows dependence on the topology of the space-time manifold (which corresponds to taking into account discrete gravitational and mixed gauge-gravitational anomalies), then 't Hooft anomalies are classified by the degree $d+1$ cobordism group of $B^{q+1} G$ with $U(1)$ coefficients \KapustinTFA. When symmetries with more than one value of $q$ are present, one can classify anomalies in a similar way, but it is difficult to write down a general answer valid in all dimensions. In the case when only $q=0$ and $q=1$ discrete symmetries are present, topological actions in low dimensions have been classified in \KapustinUXA .

We can also consider the operators on open $d-q-1$-dimensional manifolds $\Sigma^{(d-q-1)} $ with $\gamma^{(d-q-2)} =\partial \Sigma^{(d-q-1)} $.
Then $g$ is the holonomy of the flat background gauge field around $\gamma^{(d-q-2)}$. When we gauge a $q$-form global symmetry, (some of) these twisted sector operators associated to $\gamma^{(d-q-2)} $ will become true operators in the gauged theory. As for Abelian orbifolds in $2d$, every time we gauge a discrete $q$-form symmetry we gain a new quantum $(d-q-2)$-form symmetry, whose charged objects are
the original twisted sector operators. The ${}^\vee U_{g'}(M^{(q+1)})$ operators for the new symmetry are simply the $q$-form version of Wilson line operators for the
gauged symmetry. They can end on the charged operators for the gauged symmetry, which play the role of twisted sectors for the dual symmetry.
The analogy with $2d$ Abelian orbifolds extends to the observation that gauging the dual symmetry gives back the original theory. We will discuss this example further in the next section.

As for standard global symmetries, there may be several inequivalent ways to gauge $q$-form symmetries, which can be thought as tensoring the theory by some
non-trivial SPT phase before gauging the global symmetry. Such $q$-form SPT phases should be associated to boundary 't Hooft anomalies for the
$q$-form symmetry. We refer to \KapustinUXA\  for a general discussion. We will encounter some neat examples of $4d$ SPT phases for one-form global symmetries in
the next section.

\newsec{Examples of higher form global symmetries}

\subsec{$U(1)$ gauge theory in $4d$}

The first example is a $U(1)$ gauge theory without matter.  Here we have two $U(1)$ one-form symmetries.  One of them, the electric symmetry, is associated with $j_e={2\over  g^2} {}^*F$, which is closed by the equation of motion. Here $g$ is the gauge coupling constant.  It is generated by
\eqn\Uele{U^E_{g =e^{i\alpha}}(M^{(2)}) = e^{i{2\alpha\over g^2 }\int_{M^{(2)}} {}^*F}.}
Clearly, $\int_{M^{(2)}} {}^*F$ is the electric flux through $M^{(2)}$. In terms of action on the dynamical fields, this symmetry shifts the electric gauge field by a flat connection.

The other one-form symmetry is referred to as magnetic. It is associated with $j_m={1\over 2\pi} F$, which is closed because of the Bianchi identity. The symmetry is generated by
\eqn\Umag{U^M_{g =e^{i\eta}}(M^{(2)}) = e^{i{\eta\over 2\pi}\int_{M^{(2)}} F}.}
Clearly, $\int_{M^{(2)}} F$ is the magnetic flux through $M^{(2)}$.  The symmetry action on the fields is simple only after duality and then it shifts the magnetic photon by a flat gauge field.\foot{In $d$-dimensions the electric symmetry is still a one-form symmetry, while the magnetic symmetry is a $d-3$-form symmetry.}

These two operators were considered in \refs{\GukovJK,\GukovSN} and will be referred to as Gukov-Witten operators.  (Our normalizations of $\alpha$ and $\eta$ differ from \refs{\GukovJK,\GukovSN} by a factor of $2 \pi$; they are $2\pi $ periodic.)

The charged objects under these two symmetries are Wilson loops and 't Hooft loops respectively.

An open magnetic surface $U^M$ is bounded by an improperly quantized Wilson loop.  To see that, write $U_{g=e^{i\eta}}^M(\Sigma^2)=e^{i {\eta\over 2\pi} \int_{\Sigma^2}F} = e^{i{\eta\over 2\pi} \oint_{\gamma^1} A}$.  Similarly, an open electric surface $U^E$ is bounded by an improperly quantized 't Hooft loop.  Using the terminology of \KapustinGUA\ such lines that bound a topological surface are not genuine line operators.

It is nice to track these symmetries as we compactify the system on $S^1$ to $3d$.  Denoting the component of the gauge field around the $S^1$ as $A_4$, the electric one-form symmetry becomes an electric zero-form symmetry with charge $\int {}^*d A_4$, and an electric one-form symmetry associated with the electric flux $\int {}^*F$. The magnetic one-form symmetry becomes a magnetic zero-form symmetry, whose charge is the magnetic flux $\int F$, and a one-form symmetry, which is the winding of $A_4$ with charge $ \int d A_4$.

The two one-form symmetries have a mixed 't Hooft anomaly.\foot{This discussion is identical to a free compact scalar in $2d$, which has two global symmetries -- winding and momentum.  Either one of them can be gauged, but an anomaly prevents us from gauging both of them. } For example, if we couple the theory to a generic background connection $B_m$ for the
magnetic one-form symmetry by a ${i \over 2\pi}\int B_m \wedge F$ coupling, we make the gauge current $J = d B_m$ non-zero and thus $j_e$ is not conserved anymore.
We could try to modify $j_e$ to ${2\over  g^2} {}^*F- B_m$, but it would not be a well-defined 2-form (the modification would be akin to shifting the axial current
of QCD by the Chern-Simons form of the gauge fields to hide the axial anomaly). The corresponding anomaly polynomial is the 6-form $dB_e \wedge dB_m$.

We can also gauge only a $\Z_n$ subgroup of the $U(1)$ one-form magnetic symmetry.  (Alternatively, we can do it with the electric global symmetry.)  This can be done by adding to the $F^2 $ Lagrangian
\eqn\addedterms{{i \over 2\pi} B_m \wedge F + {in \over 2\pi} B_m \wedge F_m ~,}
where $F_m=dA_m$ is an ordinary $U(1)$ gauge field, whose equation of motion makes $B_m$ a $\Z_n$ two-form gauge field \refs{\MaldacenaSS,\BanksZN}.  Now, the equation of motion of $B_m$ sets $F+n F_m =0$ and therefore the sum over topological sectors of the original $U(1)$ gauge theory is constrained to include only sectors, whose first Chern-class is divisible by $n$.  This is an explicit example demonstrating that the sum over topological sector does not have to include all of them \refs{\SeibergQD,\BanksZN}.  (In this particular case we can simply eliminate $A=nA_m$ and use the gauge field $A_m$ as the basic degree of freedom.)

It is amusing to consider the supersymmetric version of these couplings.  As we said above, a $U(1)$ gauge theory has a magnetic $U(1)$ one-form global symmetry with current $F$, which can be coupled to a background two-form gauge field $B_m$.  The supersymmetrization of the $F\wedge B_m$ coupling was discussed in \BanksZN.
The gauge field $A$ belongs to a vector superfield $\CV$ and the background
two-form $B_m$ belongs to a chiral spinor superfield $\CB_\alpha$ (satisfying $\bar D_\alphadot \CB_\alpha= 0$) with the gauge symmetry\foot{The gauge invariant field strength $H_m=dB_m$ is embedded in the real linear superfield $\CH=  D^ \alpha \CB_\alpha + h.c.$.}
\eqn\Psigauge{\CB_\alpha \to \CB_\alpha + \bar D^2 D_\alpha L}
with arbitrary real $L$.  Then $ B_m\wedge F$ is included in
\eqn\BFSus{ \int d^4 \theta \CH \CV=  \int d^2 \theta \CB_\alpha W^\alpha + h.c.~,}
which is fully gauge invariant.  Denoting the bottom component of $\CH$ (which is also the Lorentz scalar $\theta$-component of $\CB$) by $\zeta$, we recognize in \BFSus\ the FI-term $\zeta \int d^4\theta \CV$.  In other words, an FI-term can be interpreted as coupling the gauge theory to a background supersymmetric two-form gauge field and as such, there is an independent FI-term for each global $U(1)$ one-form symmetry.  This interpretation of the FI-term is tied with the fact that it is related to a one-form central charge in the supersymmetry algebra \refs{\KomargodskiPC,\DumitrescuIU} (and hence to BPS vortices) -- the central charge equals a sum of products of an FI-term and the corresponding one-form symmetry charge.

Next, we add to our theory charge $n$ matter fields.  The magnetic symmetry and $U^M$ are unchanged.  But the electric symmetry is explicitly broken to $\Z_n$.  In this case the dependence of $U^E(M^{(2)})$ on the surface $M^{(2)}$ is not topological unless $g$ is an $n$'th root of unity.  The Gukov-Witten operators \Uele\Umag\ are still meaningful, but they are not always topological.

If we introduce both electrically and magnetically charged matter, we break both electric and magnetic one-form symmetries to a discrete subgroup.
In general, if $\Gamma_g$ is the full lattice of charges of some $U(1)^r$ gauge theory, and $\Gamma'_g$ the sublattice of charges generated by the matter fields,
the theory has an unbroken one-form symmetry valued in $\Gamma_g/\Gamma'_g$.

A simple example of this setup is $4d$ $\CN=2$ theory with $SU(2)$ pure gauge group without matter \refs{\SeibergRS}.
Its low energy Abelian description has a $\Z_2$
one-form symmetry, which matches the UV global one-form symmetry we will discuss momentarily.  (The theory with fundamental matter \SeibergAJ\ does not have such a symmetry.)
Similar considerations apply to general $\CN=2$ theories.

\subsec{Non-Abelian gauge theory in $4d$}

Before selecting the global form of the gauge group, a non-Abelian gauge theory based on a Lie algebra $\frak{g}$ potentially admits
't Hooft-Wilson line defects labeled by generic charges $(m,e) \in \Lambda^w_m \times \Lambda^w_e$, defined modulo the action of the Weyl group \KapustinPY.
Here $\Lambda^w_e$ is the weight lattice of $\frak{g}$ and $\Lambda^w_m$ the magnetic weight lattice, i.e.\ the dual of the
root lattice. General 't Hooft-Wilson line defects are not pairwise local. Mutual locality is controlled by the integrality of the Dirac pairing
$m \cdot e' - m'\cdot e$.

In the absence of matter fields, for simply-connected gauge group $\CG$, one allows Wilson loops of any possible representation.
Thus $e$ is unconstrained and hence $m$ must be in the magnetic root lattice, i.e.\ the dual of the weight lattice. The theory has a one-form ``electric'' symmetry
valued in the center $Z(\CG)$ of $\CG$. Its generators are Gukov-Witten surface operators \refs{\GukovJK,\GukovSN} associated to elements of $Z(\CG)$.\foot{Generic Gukov-Witten operators are labeled by conjugacy classes in $\CG$, but such operators are not topological, in general, and thus do not generate one-form symmetries.}
This symmetry acts on the gauge fields by shifting them by a flat $Z(\CG)$-valued gauge field \KapustinGUA.  If we regard such a gauge field as a \v{C}ech 1-cocycle with values in $Z(\CG)$, the symmetry transformation acts by multiplying each transition function defining a $\CG$-bundle by the corresponding value of the \v{C}ech 1-cocycle.  The Lagrangian, which depends only on the curvature of the gauge field, is not affected by this transformation, hence it is a symmetry.
The one-form charge of an 't Hooft-Wilson line defect is simply given by $e$ modulo the root lattice $\Lambda^r_e$.

Open Gukov-Witten surface operators associated to some element $g$ in $Z(\CG)$ are bounded by 't Hooft-Wilson line defects with magnetic charge $m=g$ modulo
the magnetic root lattice $\Lambda^r_m$. Notice that $Z(\CG) = \Lambda^w_m/\Lambda^r_m$ and the character group $\hat Z(\CG) = \Lambda^w_e/\Lambda^r_e$.
For topological considerations such as in this section, only the charges $(m,e)$ modulo $\Lambda^r_m \times \Lambda^r_e$ matter.
They belong to
\eqn\latt{Z^\sharp = Z(\CG) \times \hat Z(\CG) ~.}
In the rest of this section we will label the charges by \latt.

Next, if we gauge a subgroup $\Gamma\subset Z(\CG)$ of the global one-form symmetry, we obtain a gauge theory with gauge group $\CG/\Gamma$.  It turns out that there are several ways to do this labeled by a discrete theta parameter \AharonyHDA.  One way to describe it is in terms of the choice of lines.  The gauging restricts the allowed Wilson lines to be in $\widehat {Z(\CG)/\Gamma}\subset \hat Z(\CG)$, but this means that additional lines can now be introduced.  They have to be local relative to these Wilson lines and also local relative to each other. Furthermore, we want this choice of lines to be maximal.  Using the Dirac pairing, we can describe the choice of lines as choosing a maximal Lagrangian subgroup $L \subset Z^\sharp$ of \latt.

The gauging clearly reduces the electric one-form symmetry from $Z(\CG)$ to $Z(\CG)/\Gamma$.  For trivial discrete theta parameter the additional lines have magnetic charges $m \in \Gamma$, while their electric charges can be screened by the allowed Wilson lines.  Hence, we gain a new one-form magnetic symmetry valued in
$\hat \Gamma$, the character group of $\Gamma$.  And the full one-form global symmetry is $Z(\CG)/\Gamma\times \hat \Gamma$.
Open magnetic surfaces labeled by an element $\hat \gamma \in \hat \Gamma$ end on line defects with Wilson loop charge $e$
which maps to $\hat \gamma$ under the obvious map $\hat Z(\CG) \to \hat \Gamma$.

For other values of the discrete theta parameter there is another one-form global symmetry with group $ L$ associated to appropriate Gukov-Witten surfaces.
Some of these choices are related to others by shifts of the ordinary theta angle by multiples of $2 \pi$.
Other choices are simply inequivalent \AharonyHDA.  Appendix C includes a nontrivial example.

A convenient way to describe $L$ is to note that the projection of $L$ to $Z(\CG)$ is $\Gamma$. The electric charge of a line with a magnetic charge
$m\in\Gamma$ is defined modulo charges of Wilson lines and thus can be thought of as an element of $\Hom(\Gamma,U(1))$. It must also be a linear function of $m$. Thus we can encode $L$ into a choice of a bilinear form $\eta:\Gamma\times\Gamma\to U(1)$.

For example, for $\CG=SU(4)$ and $\Gamma=\Z_4$ we have four different choices of $\eta$. If we identify $U(1)$ with $\R/\Z$, we can write these four choices as follows:
\eqn\etasufour{
\eta(x,y)=p xy/4,\quad x,y,p\in \Z_4  ~.
}
They are related by shifting the ordinary theta angle in $SU(4)/\Z_4$ theory by an integer multiple of $2\pi$.  The one-form global symmetry group in all these theories is $\Z_4$.  On the other hand, for $\Gamma=\Z_2$ there are two values of the discrete theta parameter.  When it vanishes the global symmetry group is $\Z_2\times \Z_2$ and when it is non-trivial the global symmetry group is $\Z_4$.  These two theories are not related by a shift of the ordinary theta angle \AharonyHDA.

This discussion hides a subtlety which occurs on 4-manifolds without spin structure (which are allowed if one is dealing with pure Yang-Mills theory without fermions). Namely, in general the resulting theory depends not just on the bilinear form $\eta$, but on its quadratic refinement. This is discussed in more detail below. In general, there is more than one choice of a quadratic refinement, so the choice of lines, or equivalently the choice of the one-form symmetry group $\hat L$, does not completely determine the theory. We will return to this issue below.

When we add matter fields that transform under the center the electric one-form symmetry is explicitly broken to the subgroup $F\subset Z(\CG)$ that does not act on the matter fields. Locality with the matter fields restricts $m$ to live in $F$, while the possibility of screening suggests restricting $e$ to the group $\hat F$ of electric charges modulo the charges of the matter fields. This case can be analyzed as before, with the replacement $Z(\CG)$ by $F$.

\subsec{Discrete Abelian gauge theories}

\centerline{\it Theories with a trivial action}
\bigskip

A theory with a $q$-form discrete gauge symmetry $\CG$ has a gauge field, which is a $q+1$-form (more precisely, if $\CG$ is finite, it is a $(q+1)$-cochain). We will call such theories $(q+1)$-form gauge theories. For $q=0$, $\CG$ can be non-Abelian, but since we are mostly interested in $q>0$, we will only consider Abelian $\CG$. Topological $(q+1)$-form gauge theories are interesting TQFTs. In this section we discuss their global symmetries. We distinguish between theories with a trivial action, which we discuss first, and theories with non-trivial action, which we will discuss below.

All these theories have magnetic symmetry.  It is a $(d-q-2)$-form global symmetry generated by $\hat U_{\hat g}(M^{(q+1)})$, $\hat g \in \hat \CG$.  It is generated by Wilson-like operators, which measure the holonomy of the $\CG$ flat connection along a $(q+1)$-cycle $M^{(q+1)}$. For a discrete gauge theory these operators are indeed topological. This global symmetry group is $\hat \CG$.

The second ``electric'' symmetry is special to pure discrete gauge theories with a trivial action.  (In the presence of matter, these operators might not be topological.)  It is a $(q+1)$-form symmetry generated by the Gukov-Witten
twist operators $ U_g(M^{(d-q-2)})$ with $g \in \CG$.   Note that they are distinct from the generators of the $(d-q-2)$-form magnetic symmetry $\hat U_{\hat g}$.  This electric global symmetry is isomorphic to $\CG$.\foot{One should not confuse the gauge symmetry $\CG$ with the electric symmetry which is also $\CG$. Apart from the fact that the former is a gauge symmetry and the latter is a global symmetry, they are also distinguished by the fact that the former is a $q$-form symmetry, while the latter is a $(q+1)$-form symmetry.}

The two types of global symmetries are dual to each other -- the $ U_g$ are charged under the $\hat \CG$ global symmetry
and the $\hat U_{\hat g}$ are charged under the $\CG$ global symmetry. As we will discuss below, this means that both global symmetries are spontaneously broken --
charged operators acting on the vacuum do not create charged excitations. We will discuss this interpretation further in the next section.

The non-trivial correlation functions of linked $ U_g$ and $\hat U_{\hat g}$ defects are the basic observables of this TQFT.
In the language of condensed matter physics, a TQFT endowed with extra global symmetries can be used to describe the IR physics of a Symmetry
Enriched Topological (SET) phase. Hence, discrete Abelian gauge theory can describe a somewhat exotic SET, enriched by higher form symmetries.
Notice that there is a perfect symmetry between $\CG$ $q$-form gauge theories and $\hat \CG$ $(d-q-3)$-form gauge theories (with a trivial action) --
they describe the same $d$-dimensional SET phase.

Two useful examples of such theories are discussed in two Appendices.  In Appendix A we discuss a two-dimensional $q=0$ $\Z_{n_1}\times \Z_{n_2}$ gauge theory and in Appendix B we discuss a four-dimensional $q=1$ $\Z_n$ gauge theory.  The trivial action is obtained by setting the parameter $p$ in these actions to zero.

\bigskip
\centerline{\it Twisted discrete Abelian gauge theories}
\bigskip

As in the work of Dijkgraaf and Witten \DijkgraafPZ, we can also consider topological discrete Abelian gauge theories with a nontrivial action. For $q=0$, such actions are classified by $H^d(B\CG,U(1))$ \DijkgraafPZ, where $B\CG$ is a classifying space of $\CG$, i.e.\ a  topological space, whose fundamental group is $\CG$ and all other homotopy groups are trivial.\foot{More precisely, if we require the action to be defined on orientable $d$-manifolds, rather than general topological spaces, and allow for topological couplings to geometry, actions are classified by the cobordism group of $B\CG$ with $U(1)$ coefficients \refs{\DijkgraafPZ,\KapustinTFA}.} For general $q$, similar arguments \KapustinUXA\ show that topological actions are classified by $H^d(B^{q+1}\CG,U(1))$, where $B^{q+1}\CG$ is an iterated classifying space (its only nontrivial homotopy group occurs in dimension $q+1$ and is isomorphic to $\CG$). We will call such theories twisted discrete gauge theories.

Although the magnetic global symmetry $\hat \CG$ of the twisted discrete gauge theory is unaffected by the twisting, only a certain quotient of it $\hat \CG/\CK$ acts in the topological theory.  Correspondingly, some of the electric global symmetry generators are not present.  This can be interpreted as $\CK \subset \hat \CG$ being gauged in the twisted theory.  Alternatively, as we explain in Section 5 and Appendices A and B, this can be interpreted to mean that the symmetry $\hat \CG$ is broken to $\CK$ and hence only $\hat \CG/\CK$ acts nontrivially in the topological theory.  The twisting also affects the electric global symmetry and breaks it to a subgroup.  As a result, the twisted topological theory has fewer observables than the untwisted theory.

Although the subgroup $\CK\subset \hat \CG$ seems absent in the TQFT, it turns out to have interesting effects when the theory is placed on a manifold with a boundary.  If we pick boundary conditions that preserve $\hat \CG$, there is a nontrivial TQFT on the boundary, which depends on $\CK$.  This can be interpreted as if the global symmetry $\CK$ is unbroken, but it has nontrivial 't Hooft anomalies.  The boundary theory is needed in order to satisfy the 't Hooft anomaly matching conditions.  It is natural to describe this phenomenon as a SPT phase for the higher form global symmetry.

We will not discuss the most general case, but simply refer the reader to the two examples: $2d$ $\Z_{n_1}\times \Z_{n_2}$ gauge theory in Appendix A, and $4d$ $\Z_n$ gauge theory in Appendix B.

\subsec{$3d$ Chern-Simons theory and 't Hooft anomaly}

 In this section we discuss discrete one-form symmetry of $3d$ Chern-Simons theory. It is perhaps the simplest example of a higher form global symmetry afflicted with an 't Hooft anomaly.

We start with $U(1)$ Chern-Simons theory at level $k$.  This theory has a global one-form $\Z_k$ symmetry.  Its generators are the Wilson lines
\eqn\CSUU{U_{e^{2\pi i n/k}}(M^{(1)})=\exp(in \oint_{M^{(1)}} A) ~,}
with $n=0,1,...,k-1$.  One way to see that is to note that the action is invariant under shifting the gauge field $A \to A+ {1\over k} \epsilon$,
with $\epsilon$ a properly normalized flat $U(1)$ gauge field.
Alternatively, compactify to $2d$ to find the Lagrangian ${k \over 2 \pi } A_3 F$.  It
describes a system with an ordinary $\Z_k$ global symmetry \KapustinGUA.

The charged objects $V$ are the same Wilson lines \CSUU\ with the analog of \comrelqe
\eqn\comrelqeCS{\eqalign{
&U_{e^{2\pi i n/k}}(M^{(1)}) \, U_{e^{2\pi i m/k}}(\CC^{(1)}) =  e^{2\pi i mn\over k} \, U_{e^{2\pi i m/k}}(\CC^{(1)}) \cr
&U_{e^{2\pi i n/k}}(M^{(1)}) = V_{n}(M^{(1)})}}
with $ M^{(1)}$ a small circle around $\CC^{(1)}$.

Equation \comrelqeCS\ means that the generators of the one-form $\Z_k$ symmetry are charged under it.  We interpret it to mean that the global $\Z_k$ symmetry has an 't Hooft anomaly.
One way to think about it is as follows.  Gauging the $\Z_k$ symmetry amounts to summing over all possible insertions of $U_{e^{2\pi i/k}}$.  But because of \comrelqeCS\ this has the effect of making the functional integral vanish.  Alternatively, gauging is the same as passing from $U(1)$ to its quotient $U(1)/\Z_k$ (which is again isomorphic to $U(1)$). This means that $A$ is no longer a well-defined gauge field, but $A'=kA$ is. Expressing the action in terms of $A'$, we get
\eqn\CSaction{
S_{CS}^{gauged}={1\over 4\pi k} \int A' dA'.
}
Since the Chern-Simons level for $A'$ is fractional, the action is not gauge-invariant, implying 't Hooft anomaly.

As a second example consider the $ SU(2)$ Chern-Simons theory with level $k$.  This theory has a $\Z_2$ global symmetry generated by the Wilson line associated with the $SU(2)$ representation with $j=k/2$. The charged objects are the various Wilson lines and the analog of \comrelqe\ is
\eqn\comrelqeCSS{U_{g=-1}(M^{(1)}) V_j(\CC^{(1)}) =  (-1)^{2j} V(\CC^{(1)})  }
with $ M^{(1)}$ a small circle around $\CC^{(1)}$.\foot{As a check, consider the expectation value of two linked loops in the representations $j$ and $j'$ in $S^3$.  It is $S_{jj'}=\sqrt{2\over k+2}\sin \left({\pi (2j+1)(2j'+1)\over k+2}\right)$ \WittenHF.  For $j'={k\over 2}$ we have $S_{j{k\over 2}}=(-1)^{2j} S_{j0}$, which is consistent with \comrelqeCSS.}  In other words, the integer $j$ lines are $\Z_2$ even and the half integer $j$ lines are $\Z_2$ odd. For odd $k$, this one-form symmetry has an 't Hooft anomaly.

Notice that this $\Z_2$ one-form symmetry coincides with the $\Z_2$ one-form symmetry associated to the center of the $SU(2)$ gauge theory
and that gauging the $\Z_2$ one-form symmetry will produce an $SO(3)$ Chern-Simons theory. The anomaly for odd $k$ is
again related to the fact that the level quantization is different for $SO(3)=SU(2)/\Z_2$ Chern-Simons theory.

This analysis generalizes readily to an $SU(N)$ Chern-Simons theory. This theory has a $\Z_N$ one-form symmetry associated to its center,
such that Wilson loops in representations with $\ell$ boxes have charge $\ell$. The topological lines for this symmetry can be interpreted as
Wilson lines as well: the generator $U_1$ is a Wilson loop with $k$ boxes in a symmetric representation, the other group elements $U_n$ are labeled by rectangular
Young Tableaux with $n k$ boxes.

Notice that the fusion of the $U_1$ Wilson loop with Wilson loops $V_R$ labeled by a generic representation $R$ gives a single Wilson loop labeled
by a representation $\sigma(R)$ obtained from $R$ by adding a row of $k$ boxes and removing any columns of height $N$ \GepnerKX.

The $\Z_N$ symmetry is again afflicted with 't Hooft anomaly for general $k$, \eqn\comrelqeCS{U_{1}(M^{(1)}) \, U_1(\CC^{(1)}) =  e^{-2\pi i k\over N} \, U_1(\CC^{(1)})}
We can compute the phase factor readily by comparing with the WZW dimensions $\Delta_n = {n (N -n)k \over 2 N}$ of the simple currents associated to $U_n$.

If $k$ is a multiple of $N$, we can gauge the $\Z_N$ one-form symmetry to recover a $PSU(N)$ Chern-Simons theory. Similarly, for appropriate values of $k$ we can
also gauge other subgroups $\Gamma$ of the $\Z_N$ one-form symmetry to recover an $SU(N)/\Gamma$ Chern-Simons theory.\foot{As we gauge a one-form symmetry in three dimensions, we expect to find a dual ``magnetic'' zero-form symmetry carried by monopole operators
and measured by appropriate GW operators. Because of the CS coupling, the monopole operators will also carry gauge charge, and there will be no
gauge-invariant local operators charged under the dual magnetic global symmetry. }

Finally, we can consider $U(N)$ Chern-Simons theories at level $k$. These theories are usually described by realizing $U(N) = {SU(N) \times U(1) \over \Z_N}$
and combining appropriately an $SU(N)$ and $U(1)$ Chern-Simons theories. In our language, the combination is realized as follows:
we start from an $SU(N)$ CS theory at level $k$ and an $U(1)$ theory at level $(k+N)N$ and gauge the diagonal combination of the
$\Z_N$ one-form global symmetry of the $SU(N)$ CS theory and the $\Z_N$ subgroup of the $\Z_{N(k+N)}$ one-form global symmetry
of the $U(1)$ CS theory.

The required level for the $U(1)$ theory is well-known \refs{\GepnerKX, \MarinoRE},
but can also be understood as due to an 't Hooft anomaly cancellation: we are gauging composite $U_1$ operators which have anomaly
$e^{-2\pi i k\over N}$ from the $SU(N)_k$ factor and $e^{-2\pi i (N+k)^2 \over N(k+N)}$ from the $U(1)_{N(k+N)}$ factor.

As a result, the $U(N)$ Chern-Simons theories at level $k$ has a residual $\Z_{k+N}$ one-form symmetry generated by the
$U(1)$ Wilson loops of charge $N$, i.e. $U(N)$ Wilson loops in the $N$-th antisymmetric power of a fundamental representation of $U(N)$.
This symmetry has anomaly $e^{-2\pi i N \over k+N}$.

Thus a $U(N)$ Chern-Simons theory at level $k$ (or better, renormalized level $k+N$) has the same global 't Hooft anomaly as $N$ copies of a $U(1)_{k+N}$ theory,
or a product of $U(n_a)$ CS theories at level $k-n_a+N$ (or better, renormalized level $k+N$) with $\sum_a n_a = N$.
It should be possible to explain this observation in terms of an 't Hooft anomaly matching for an RG flow starting from an
$U(N)$ Chern-Simons theory coupled to adjoint matter with a vev for the adjoint field which Higgses it to $\prod_a U(n_a)$.

\subsec{$6d$ $(2,0)$ theories}

As another example, consider the $6d$ $(2,0)$ theory. For a recent review see \refs{\GaiottoBE,\Moorelectures} and references therein.  Before we start discussing this theory we should clarify that when referring to the $(2,0)$ theory one can mean two different things.  First, we can mean a superconformal field theory satisfying all the standard axioms of quantum field theory.  Alternatively, we can express such a field theory in terms of conformal blocks\foot{We caution the reader that the term ``conformal block'' here is used in a somewhat non-standard fashion, following the terminology introduced in \WittenAT\ in analogy to conformal blocks in two-dimensional RCFTs, which have a specific relation to $3d$ TQFTs. It should not be confused with the conformal blocks which are used in dimension higher than two to decompose CFT correlation functions into the contributions of individual (super)conformal primaries, and have no known relation to higher-dimensional TQFTs.} and refer to the theory of these conformal blocks.

This is analogous to a theory of chiral bosons in $2d$.  In this theory the bosons take values in a torus associated with a lattice.  Locality of the vertex operators forces the lattice to be integral.  When spacetime is a torus, invariance under the $S$ transformation of the torus forces the lattice to be self-dual.  This is enough to have a standard CFT on a space with a spin structure.  If one also wants the theory to be fully modular invariant, then the lattice has to be even but this is not essential.  Consider for example a theory of a single left-moving scalar.  The lattice is self-dual when it is at the ``free fermion radius.'' This is a standard CFT (even though the lattice is not even).  When the radius squared is rational (but not equal to one) the theory of the chiral boson does not make sense as a CFT.  In this case one has two options.  First, we can tensor it with another CFT (e.g. another chiral boson) and make a standard CFT.  Alternatively, we can view the theory as a generalization of a CFT where one studies the space of conformal blocks.

The situation in the $6d$ $(2,0)$ theory is similar.  In the rest of this section we will limit ourselves to standard field theory.  The more general situation, where one considers more generalized theories that do not satisfy the standard axioms of quantum field theory was studied in \WittenAT.  In any such theory we can always add free tensor multiplets, whose periodicities are correlated with that of the generalized theory and thus form a standard field theory.  It was emphasized in \SeibergDR\ that in the context of string theory the necessary free tensors are always present such that we have a ``standard theory.''

The $6d$ theory is characterized by a Lie group $\cal G$ of
rank $n$, such that after compactification to $5d$, this group is the gauge
group.\foot{We cannot rule out the possibility that there are other $(2,0)$
theories, which are not characterized by such a group.}

The theory has surface operators $V(M^{(2)})$ and volume operators $U(M^{(3)})$.
One way to think about them is by moving along the Coulomb branch of the theory,
where the low-energy dynamics is that of $n$ approximately free tensor multiplets,
which include $n$ two-form gauge fields $B_i$, whose field strengths $H_i=dB_i$ are
self-dual.  Then, the surface and volume operators are given by
\eqn\tzsurf{\eqalign{
&V(M^{(2)} ) =e^{i a_i \oint_{M^(2)} B_i} \cr
&U(M^{(3)} ) =e^{i b_i \oint_{M^(3)} H_i} ~.}}

Because of the self-duality of $H_i$, the parameters $a_i$ must belong to a self-dual
lattice. Also, locality of $V(M^{(2)})$ forces that lattice to be integral, but not
necessarily even. (See the discussion in \SeibergDR\ and references therein.) This
lattice is the weight lattice of $\cal G$ and this restricts the allowed choices of
$\cal G$.

Similarly, unlike the $4d$ $U(1)$ gauge theory, where we can use both $F$ and
${}^*F$ \Uele\Umag, here because of the self-duality we have only one kind of
operator $U$.  As in all our examples, the topological $U(M^{(3)})$ generate a
two-form global symmetry $G$ and the surface operators $V(M^{(2)} )$ are charged
under it.

The simplest example of this theory is the $A_{n-1}$ theory.  Since we want the
weight lattice to be self-dual, we take ${\cal G}=U(n)$.  (Note that its weight
lattice is integer but not even.)  A low energy observer along the flat directions
has volume operators $U(M^{(3)} )$ labeled by $n$ parameters $b_i$.  Most of them
are not topological.  The topological $U(M^{(3)} )$ depend on a single compact
parameter $b=b_i$ associated with $\sum_i H_i$.  It generates the two-form global
symmetry $G=U(1)$.

Upon compactification on a circle to $5d$, the two-form $G=U(1)$ global symmetry
leads to a two-form $U(1)$ global symmetry and a one-form $U(1)$ global symmetry.
These act as a magnetic symmetry and an electric symmetry of the $5d$ $U(n)$ gauge
theory.  More explicitly, they are associated with the $U(1)\subset U(n)$ gauge
theory.

Upon further compactification on a circle to $4d$ we have four different $U(1)$
global symmetries.  They are a two-form symmetry, two different one-form symmetries
and a zero-form (ordinary) symmetry.  The $U(1)\times U(1)$ one-form global
symmetries are the ordinary electric and magnetic symmetries associated with
$U(1)\subset U(n)$.  The ordinary global $U(1) $ symmetry shifts the compact scalar
in the vector multiplet of $U(1)\subset U(n)$.  The two-form symmetry measures
the winding number of that compact scalar.  As the torus we compactified on goes to
zero size, the ordinary $U(1)$ symmetry becomes noncompact -- it becomes $\R$.  The winding symmetry (the two-form global $U(1)$) acts trivially in this limit.

This description of the symmetries is easily generalized.  The two-form $G$ global
symmetry in $6d$ leads to a two-form and a one-form $G$ global symmetries in $5d$
and to several symmetries in $4d$: an ordinary $G$ global symmetry, a two-form $G$
global symmetry and a $G\times G$ one-form global symmetry.  The latter is the
electric and magnetic global symmetries of the low energy ${\cal N}=4$ theory.

As other examples based on $SU$ groups we can take ${\cal G}=SU(k^2)/\Z_k$ with
$G=\Z_k$ or ${\cal G}=SU(k)\times SU(k)/\Z_k$ with $G=\Z_k$.  Their lattices are
integer and self-dual and they lead to the same picture as above.  In fact, given the conformal blocks of the $SU$ $6d$ $(2,0)$ theories, we can construct the quantum field theories associated with these $\CG$s and we conjecture that there exist string constructions realizing them.

We can also consider the $D_n$ theories.  Here ${\cal G}=SO(2n)$ and $G=\Z_2$.
Again, in $4d$ we find a $\Z_2\times \Z_2$ one-form global symmetry.  There are two
different $4d$ theories with gauge group $SO(2n)$, which differ by a discrete
$\theta$-parameter \AharonyHDA.  This construction leads to one of them.

All these examples lead to fully $SL(2,\Z)$ invariant theories in $4d$.  Other $4d$
${\cal N}=4$ theories are invariant only under a subgroup $\Gamma \subset SL(2,\Z)$
and they reside in an $SL(2,\Z)/\Gamma$ orbit of theories.  All the
theories in the same orbit must have the same global symmetries.  For example, the
$4d$ $SU(4)$ theory is in the same orbit as 4 different $SU(4)/\Z_4$ (with different
discrete $\theta$-parameter) and another $SU(4)/\Z_2 = SO(6)$ theory (again with a
different discrete $\theta$-parameter \AharonyHDA. All of them have a $\Z_4$
one-form global symmetry.

One can also consider a compactification of a $(2,0)$ theory on a 3-manifold $M$. The 2-form global symmetry in 6d may give rise to $q$-form symmetries with $q=0,1,2$ in the effective 3d theory. Consider for simplicity the case when $M=S^3/\Z_k$, i.e. a lens space, and a 6d theory with $U(1)$  2-form global symmetry. Then the effective 3d theory has $U(1)$ 2-form symmetry and $\Z_k$ one-form symmetry. Indeed, if $X$ is a 3d manifold, by K\"unneth formula we have
\eqn\homologyreduction{
H_2(X\times M)=H_1(X)\otimes H_1(M)+H_2(X)=H_1(X)\otimes\Z_k+H_2(X),
}
and therefore
\eqn\cohomologyreduction{
H^2(X\times M, U(1))=H^1(X,\Z_k)+H^2(X,U(1)).
}

\subsec{Higher-form global symmetries without gauge fields}

In order to demonstrate that higher-form global symmetries are not specific to gauge theories we present two examples of $d$-dimensional theories based purely on scalars with such symmetries.

Consider a theory based on a compact scalar field $\phi \sim \phi+2\pi$ with the Lagrangian $\CL=\half f (\partial \phi)^2 + V(\phi)$ with a dimensionful constant $f$.  If $V=0$ the system has an ordinary ($0$-form) symmetry with $Q =f \int {}^* d \phi$.  In addition, even for nonzero $V$ there is a $d-2$-form $U(1)$ (winding) symmetry with $Q = {1\over 2\pi} \int d\phi$.  The charged excitations are domain walls interpolating between $\phi$ and $\phi + 2\pi$. Charged objects are vortex operators.

Our second example is based on the $d$ dimensional ${\bf CP}^n$ sigma-model.  Let $\omega$ be the pull-back to spacetime of the ${\bf CP}^n$ K\"ahler form.  $\omega$ is the current of a global $d-3$-form $U(1)$ symmetry.  In $3d$ this is an ordinary symmetry, whose charge $Q=\int \omega$ counts the number of times space wraps the non-trivial 2-cycle.  In $4d$ we have a one-form global symmetry under which strings are charged.

An alternate presentation of this non-linear model uses $n+1$ complex scalars $z^i$ constrained to satisfy $\sum_i |z^i|^2 =1$ and coupled to a $U(1)$ gauge field $A$.  In this presentation $\omega \sim dA$ and the symmetry above is the magnetic global $U(1)$ symmetry of the gauge theory.  Below we will refer to this symmetry as magnetic also in the presentation without a gauge field.

\newsec{Spontaneous breaking of higher-form global symmetries}

Just as an ordinary global symmetry can be spontaneously broken, so can higher-form symmetries. To be concrete, we will focus mostly on $4d$ gauge theories and their one-form symmetries.  We will use the behavior of large loops as the diagnostic of such breaking.  And in order to avoid confusion, we will focus here only on the genuine line operators, whose definition does not depend on a choice of surface.  Such lines can exhibit area law, perimeter law or Coulomb law.

We interpret an area law for a charged loop operator as reflecting the fact that the corresponding one-form symmetry is unbroken.  Indeed, the expectation value of the loop vanishes as its size is taken to infinity.  Perimeter law can be set to zero by redefining the operator by a local geometric counterterm.  Then, perimeter law and Coulomb behavior mean that the loop has a nonzero expectation value when it is large and correspondingly the symmetry is spontaneously broken.

Equivalently, when a one-form global symmetry is unbroken, the charged states are strings.  They lead to an area law for some loop operators.  If the symmetry is spontaneously broken, there are no such strings and hence there is no area law.

More precisely, a one-form global symmetry $G$ can break to a subgroup $\CK$.  In that case the loops charged under $\CK$ exhibit area law.  And loops charged under $G$, but un-charged under $\CK$ exhibit a perimeter or Coulomb law -- renormalizing the perimeter law to zero a large loop has an expectation value breaking $G$ to $\CK$.

Although we will mostly be interested in one-form symmetries in $4d$, we point out that the generalization of this discussion is straightforward.  The key point is whether the charged objects have a non-zero vev when they are large.  Using this terminology Coulomb behavior in $2d$ and $3d$ means that the corresponding global symmetry is unbroken.

A standard argument shows that when a continuous one-form global symmetry is spontaneously broken the system should have a Goldstone boson.  This Goldstone boson is a massless photon, as can be seen from the matrix element of the two-form Noether current $j_{\mu\nu}$ between a photon with polarization $\zeta$ and momentum $p$ and the vacuum
\eqn\matele{\langle 0| j_{\mu\nu}(x) |\zeta,p \rangle = (\zeta_\mu p_\nu -\zeta_\nu p_\mu) e^{ip\cdot x}~.}

It is straightforward to imitate the analysis of Coleman, Mermin and Wagner about the possibility of spontaneous symmetry breaking.  The generalization of the $q=0$ statement to higher $q$ is that a continuous symmetry is always unbroken for $d-q < 3$ and a discrete symmetry is always unbroken for $d-q <2$.  For example, continuous one-form symmetries are always unbroken in $2d$ and $3d$.  And discrete one-form symmetries are always unbroken in $2d$.  (They do not exist in $1d$.)  One way to see that is to compactify the $d$ dimensional system on tori to $d-q$ dimensions and then use the Coleman-Mermin-Wagner analysis.  But even without using compactification, we note that Coulomb behavior in $d$ dimensions leads to the Wilson loop $\langle W\rangle \sim e^{-V(r)}$ with $V(r) \sim  {1\over r^{d-3}}$ (for $d = 3 $ it is $\log r$).  For $d=2,3$ the potential $V(r)$ diverges as $r \to \infty$ and hence $\langle W\rangle \to 0$ and the symmetry is unbroken. Note that the arithmetics of this computation is identical to the Coleman-Mermin-Wagner analysis of the IR behavior of expectation values of charged order parameters.

Let us demonstrate this general discussion in several examples.

\subsec{$U(1)$ gauge theory}

As our first example we consider the pure gauge $U(1)$ gauge theory (see section 4.1).  This system has an electric and a magnetic $U(1)$ one-form global symmetries.  As we said, we interpret the fact that the system is in a Coulomb phase to mean that its global symmetries are spontaneously broken and the massless photon is a Goldstone boson. This can be seen by noticing that the dynamical gauge field is shifted by the action of the symmetry -- it transforms inhomogeneously.  More precisely, we can use \matele\ both for the electric and the magnetic currents to show that the photon is the Goldstone boson of these two symmetries.

Further insight into this breaking is obtained by compactifying the system on a circle to $3d$. As we discussed above, the $4d$ free $U(1)$ gauge theory has a global $U(1)$ electric one-form symmetry and a global $U(1)$ magnetic one-form symmetry.  The electric symmetry leads to two symmetries in $3d$.  These are a spontaneously broken ordinary global symmetry with charge $\int {}^*d A_4$, with $A_4$ being the Goldstone boson and an unbroken electric one-form symmetry, whose charge is the electric flux $\int {}^*F \sim \int da$, which is winding for the dual photon $a$.  Similarly, the magnetic one-form symmetry leads to two symmetries in $3d$. These are a spontaneously broken magnetic zero-form symmetry, whose charge is the magnetic flux $\int F \sim \int{}^*da $ with $a$ being the Goldstone boson and an unbroken one-form symmetry, winding of $A_4$ with charge $ \int d A_4$.  Note that a $4d$ symmetry can be split to two different symmetries in $3d$; one of them is spontaneously broken and the other is not.  Clearly, this conclusion is consistent with the general discussion above about the the behavior of the loops in $4d$ and in $3d$.

Next, we discuss a $U(1)$ gauge theory with charge $n$ matter.  Because of the presence of the matter fields, the electric one-form global symmetry is explicitly broken to $\Z_n$.  But the magnetic one-form symmetry remains $U(1)$ with the charge being the magnetic flux.  If the matter is massive the low energy spectrum includes only the photon and the situation is similar to the pure $U(1)$ gauge theory.  The IR theory has an accidental electric $U(1)$ one-form global symmetry.  Both the electric and the magnetic global symmetries are spontaneously broken in this Coulomb phase.

When the matter fields condense the $U(1)$ gauge symmetry is Higgsed to $\Z_n$.  In this case the spectrum is gapped, but the low energy theory includes a $\Z_n$ gauge theory.  As discussed in section 4.3 and in the next subsection, this gauge theory represents the spontaneously broken one-form $\Z_n$ global symmetry.  The magnetic $U(1)$ one-form symmetry must be unbroken.  If it had been broken, there would have been a massless photon.  Indeed, the spectrum includes strings, which are charged under this symmetry.  Furthermore, the 't Hooft loop, which is charged under this symmetry, exhibits an area law in this phase.

The ${\bf CP}^n$ sigma-model we discussed in section 4.6 fits this description.  In its weakly coupled phase the ordinary global $SU(n+1)$ symmetry is spontaneously broken to $SU(n)$.  However, the magnetic symmetry generated by $\int \omega$ is unbroken.  This is clear because the spectrum includes states charged under it.  In $3d$ these are particles and in $4d$ these are strings.  Now, it is known that when this model is placed on the lattice it has a strong coupling phase in which the ordinary $SU(n+1)$ symmetry is unbroken and the magnetic $U(1)$ can be broken.\foot{This is familiar in the continuum $2d$ version of this system, but here there is no magnetic symmetry and the emergence of the photon is not as dramatic as in higher dimensions.}  In that phase there must be a massless photon.  In $3d$ it is a scalar and in $4d$ it is a massless emergent gauge field.  Normally, this point is argued by ``integrating in'' the $U(1)$ gauge field (as we mentioned in section 4.6) and then finding it as a massless excitation.  From our perspective, the existence of the massless photon follows even without integrating it in.  It follows from the spontaneously broken global symmetry.

\subsec{$\Z_N$  gauge theory}

The discussion of $4d$ discrete ordinary gauge theories in section 4.3 and in Appendix B can be interpreted to mean the following.  The untwisted theory with $p=0$ has a spontaneously broken one-form $\Z_N$ global symmetry and a two-form $\Z_N$ global symmetry.

Quite generally, a low energy observer can detect spontaneously broken ordinary discrete symmetries by finding their domain walls.  A spontaneously broken higher-form discrete symmetry leads in the IR to a TQFT.  This theory has several dual presentations (see e.g.\ \KapustinGUA) including a discrete gauge theory.  So a UV theory with a one-form discrete global symmetry, which is spontaneously broken must lead in the IR to a TQFT and hence to long range topological order.

For nonzero $p$ the bulk of the system exhibits only $\Z_L$ one-form and two-form global symmetries with $L=\gcd(p,N)$.  Both of these symmetries are spontaneously broken.  As we discuss in Appendix $B$, we can interpret this to mean that we start with a UV theory with a global $\Z_N$ one-form symmetry, which is spontaneously broken to $\Z_K$ with $K=N/L$.  The unbroken $\Z_K$ symmetry does not act in the IR.  However, as we will discuss below, this symmetry can have anomalies, which lead to some degrees of freedom on boundaries or domain walls.

\subsec{$SU(N)$ gauge theory}

Non-Abelian gauge theories can have both electric and magnetic one-form global symmetries.  Any of them could be spontaneously broken to a subgroup.

An $SU(N)$ gauge theory with matter fields in the adjoint representation has a global $\Z_N$ one-form electric symmetry and no magnetic symmetry.  The order parameter for its breaking is the Wilson loop $W$.  In the standard confining phase $W^l$ with $l=1,2,..., N-1$ exhibits an area law.  We take it to mean that the electric one-form symmetry is unbroken.  In this case the spectrum includes $\Z_N$ strings.

We can also have a phase where $W^l$ with $l\not= 0\, \mod\, t$ with some $t$ a factor of $N$ has area law, but $W^t$ has a perimeter law \CachazoZK.  $t$ was referred to in \CachazoZK\ as the confinement index.  From our perspective, the global $\Z_N$ one-form symmetry is broken to $\Z_t$.  $W^t$ is the order parameter for this breaking, and the $t$ different strings, whose tension is probed by $W^l$ with $l=1,2,...,t-1$ are charged under this unbroken symmetry. We will demonstrate this discussion in the model of \CachazoZK\ in Appendix D.

Our discussion leads to a new perspective on Polyakov's confinement/de-confinement transition. Compactifying the $SU(N)$ gauge theory on a circle, the one-form $\Z_N$ global symmetry leads to an ordinary $\Z_N$ symmetry as well as a $\Z_N$ one-form symmetry.  The Polyakov loop is an order parameter of the former, and its expectation value serves as a diagnostics of the de-confinement transition.  In the confining phase the expectation value of the Polyakov loop vanishes, thus signaling that the ordinary $\Z_N$ symmetry is unbroken and in the de-confined phase the vacuum expectation value of the Polyakov loop is nonzero, thus breaking the ordinary $\Z_N$ symmetry in $3d$.  Our $4d$ discussion above identifies the $4d$ precursor of Polyakov's ordinary $\Z_N$ symmetry as a one-form symmetry.  And it identifies the fact that it is broken or unbroken as a statement about the $4d$ one-form symmetry.

Next, we can add fundamental matter to the $SU(N)$ gauge theory.  Now, there is neither electric nor magnetic global symmetry (see section 4.2) and therefore there is no symmetry that might or might not be broken. This allows us to state neatly the statement of \refs{\FradkinDV,\BanksFI} in our language.  These authors showed that one can continuously interpolate from Higgs to confinement in an $SU(N)$ gauge theory with matter fields in the fundamental representation.  In our language, when there are matter fields in the fundamental representation there is no electric one-form global symmetry and therefore there is no order parameter for a broken symmetry and we cannot unambiguously define Higgs and confinement. Furthermore, since the gauge group is simply connected, there is no magnetic symmetry either. So the massive phase in this case is ``featureless'': it cannot be characterized in terms of the action of global symmetries on the vacuum, because there are no global symmetries beyond the Lorentz symmetry.

\subsec{$PSU(N)$ gauge theory}

As we discussed in section 4.2, the $PSU(N)=SU(N)/\Z_N$ theory has a global one-form magnetic $\Z_N$ symmetry.  This symmetry could be spontaneously broken to $\Z_K$ for any $K$ a divisor of $N$.  To see how this happens recall that the $PSU(N)$ theory is characterized by a discrete theta-parameter $p=0,...,N-1$ (for simplicity, we limit ourselves to spin manifolds).  Denoting the fundamental Wilson line by $W$ and the electrically neutral 't Hooft line by $H$, the genuine line operators in the $PSU(N)$ theory labeled by $p$ are powers of $HW^p$ \AharonyHDA.  Of course, $H^N$ and $W^N$ are also genuine line operators, which we will view as trivial because they are invariant under the global symmetry.

In the standard confining vacuum of the theory we have condensation of magnetic monopoles with no electric charge.  Their charges are the same as those of $H^N$. Let us examine the unbroken subgroup of the global one-form $\Z_N$ symmetry. Using $L=\gcd(p,N)$ and $K=N/L$ we have $(HW^p)^K = H^K (W^N)^{p /\gcd(p,N)}$.  The second factor is a power of $W^N$, which is invariant under the global symmetry and hence it has a perimeter law.  The first factor is not invariant under $\Z_N$, but since its electric-magnetic charge is aligned with that of the condensing monopole, it also has a perimeter law.  We conclude that $(HW^p)^K $ has a perimeter law, while $(HW^p)^l$ with $l\not=0\, \mod\, K$ have area law.  This means that the one-form $\Z_N$ symmetry is spontaneously broken to $\Z_K$.

As in our previous examples, this means that the low energy theory realizes a spontaneously broken $\Z_N/\Z_K=\Z_L$ one-form symmetry.  We will discuss this system in more detail below and we will show that the unbroken $\Z_K$ symmetry is also important in the IR.  This symmetry has anomalies that are saturated by excitations on boundaries or domain walls.  Therefore, the low-energy theory is not merely a $\Z_L$ gauge theory but is the twisted $\Z_N$ theory of Appendix B.

We can also consider other phases in which dyons condense.  In the phase labeled by $k$ the condensed dyons have the charges of $(HW^k)^N$.  By shifting $\theta \to \theta - 2\pi k/N$ we shift $p\to p-k$ and the condensed dyons charges are those of $H^N$.  Therefore, in this case the unbroken symmetry is $\Z_{N/\gcd(p-k,N)}$.

\newsec{$S$ and $T$ transformations}

In this section we generalize Witten's operations on $3d$ theories with global symmetries \WittenYA\ to $4d$ theories with one-form global symmetries.

When we consider the partition function on a manifold ${\cal M}$ of a theory with $G$-valued $q$-form symmetry, it is natural to couple it to a flat background $(q+1)$-form gauge field $\lambda$.
Such a gauge field  is a $(q+1)$-dimensional cocycle with values in $G$. Gauge transformations change $\lambda$ by an exact cocycle. Since the partition function $Z_{\CM}(\lambda)$ is gauge-invariant, it can be regarded as a function on $H^{q+1}(\CM, G)$:
\eqn\partfunc{
{Z}_{\CM}:  H^{q+1}(\CM,G)\ra \C,\quad [\lambda]\mapsto {Z}_{\CM}(\lambda).
}
If $G$ is finite, gauging the symmetry amounts to summing over flat $(q+1)$-form gauge fields with some weight.  The simplest possibility is to assign equal weight to all gauge fields. The resulting theory has a
$\hat{G}$-valued $(d-q-2)$-form global symmetry, where $\hat G=\Hom(G,\R/\Z)$ is the Pontryagin-dual of $G$. The partition function of the new theory twisted by a $(d-q-1)$-form gauge field $\hat\lambda$ is
\eqn\partfuncgauge{
\hat{Z}_{\CM}(\hat\lambda) \sim \sum_{[\lambda]} Z_{\CM}(\lambda) \exp(2\pi i \langle \hat\lambda, \cup\lambda\rangle).
}
We leave the normalization factor arbitrary. Thus $\hat Z_{\CM}$ is a discrete Fourier transform of $Z_{\CM}$. Note that this makes sense because by Poincare duality $H^{d-q-1}(\CM,\hat G)$ is the Pontryagin-dual of $H^{q+1}(\CM,G)$. We may call this operation $S$.

If $d$ is even and $q=d/2-1$, $\lambda$ and $\hat\lambda$ have the same degree, and both the original theory and its $S$-transform have a $q$-form symmetry. It is easy to check that
\eqn\Ssquare{
(S^2Z_{\CM})(\lambda)\sim Z_{\CM}(-\lambda),
}
i.e.\ up to a numerical factor we have $S^2=C$, where $C$ is charge-conjugation.

Let us further assume that $d$ is divisible by $4$, and we are given a quadratic function $\sigma: G\ra \R/\Z$ such that the corresponding symmetric bilinear form $\eta$ is non-degenerate. We can use $\eta$ to identify $G$ and $\hat G$.  Then the operation $S$ acts as follows:
 \eqn\partfunS{
{SZ}_{\CM}(\mu)\sim \sum_{[\lambda]} Z_{\CM}(\lambda) \exp\left(2\pi i \int_\CM \eta(\mu, \cup\lambda)\right).
}
Further, the Poincare-Pontryagin bilinear form
\eqn\PP{
H^q(\CM,G)\times H^q(\CM,G)\ra \R/\Z, \quad (\mu,\lambda)\mapsto \int_\CM \eta(\mu,\cup\lambda)
}
has a quadratic refinement \Whitehead
\eqn\Psq{
\lambda\mapsto \int_\CM\frP_\sigma(\lambda),
}
where $\frP_\sigma: H^{d/2}(\CM,G)\to H^d(\CM,\R/\Z)$ is the Pontryagin square operation associated with $\sigma$. A detailed discussion of the
Pontryagin square can be found in \KapustinQSA .
We can use this quadratic refinement to define an operation $T$ as follows:
\eqn\partfunT{
T: Z_{\CM}(\lambda)\mapsto Z_{\CM}(\lambda) \exp\left(2\pi i\int_\CM \frP_\sigma(\lambda)\right).
}

In the special case $G=\Z_N$, we can take the quadratic function $\sigma$ to be $(1-N)x^2/2N$ if $N$ is odd and $x^2/2N$ if $N$ is even, so that in both cases $\eta(x,y)=xy/N$. We claim that in both cases $S$ and $T$ satisfy the $SL(2,\Z_N)$ relations.

The easiest way to see this is to consider a $5d$ topological 2-form gauge theory with gauge group $\Z_N$. Such a gauge theory can be described in the continuum by an action
\eqn\CStwoform{
S={iN\over 2\pi}\int_X B_1 dB_2,
}
where $B_1$ and $B_2$ are $U(1)$ 2-form gauge fields. (A somewhat related discussion appeared in \KravecPUA.)  One can regard either $B_2$ or $B_1$ as a Lagrange multiplier field which constrains $B_1$ or $B_2$, respectively, to be flat and have holonomy in $N^{\rm th}$ roots of unity.

This $5d$ theory has a $\Z_N\times \Z_N$ 2-form global symmetry generated by the surface operators $U_I=e^{i\oint B_I}$.  Note that $U_2$ is charged under the $\Z_N$ global symmetry generated by $U_1$ and viceversa. Therefore, we cannot gauge both of them.

This theory has an obvious $SL(2,\Z_N)$ symmetry, with the generators $S$ and $T$ acting by
\eqn\ST{
S: (B_1,B_2)\mapsto (B_2,-B_1),\quad T: (B_1,B_2)\mapsto (B_1+B_2, B_2).
}

For our purpose we consider the $5d$ manifold to be $X=\CM \times I$, where $\CM$ is a four manifold and $I$ is a line segment.  At one end of the segment (the left end) we place our $4d$ gauge theory with its $\Z_N$ one-form global symmetry, which is generated by the surface operator $U(M^{(2)})$ with $M^{(2)}$ a closed two-surface in $\CM$ at the left end of $I$. $U(M^{(2)})$ acts on charged lines $V$ in $\CM$ at the left end.  We couple this global symmetry to the bulk gauge field $B_1$.  This means that a line operator $V$ must be the boundary of an open $B_1$ surface.  Another consequence of this gauging is that we must impose
\eqn\Btbc{U(M^{(2)})=e^{i\oint_{M^{(2)}} B_2}~.}
In other words, the (Dirichlet) boundary conditions of $B_2$ correlate its component that is a two-form along the boundary
$B_2\big|$ with the values of the $4d$ dynamical fields there.  In order to understand why \Btbc\ should be imposed consider an open $B_1$ surface in the bulk that ends on $V$ in the left boundary.  Adding a closed $B_2$ surface in the bulk $U_2$ that links the $B_1$ surface multiplies the answer by a $\Z_N$ phase.  Then we can slide the closed surface $U_2$ and move it to the left boundary, where it links $V$.  In order to reproduce the same $\Z_N$ phase in the boundary theory, we must impose \Btbc.

Next, we have to choose boundary conditions at the other boundary (the right end of $I$).  Here we set a linear combination of $B_1\big|$ and $B_2\big|$ to zero.  These choices are related by $SL(2,\Z_N)$ of \ST.

Since the $5d$ theory is topological, we can think of $I$ as being very short and then the full $5d$ theory is effectively a four-dimensional theory on $\CM$. Therefore, the different boundary conditions lead to different effective $4d$ theories.  For example, with the right boundary condition $B_1\big|=0 $ we recover the original $4d$ theory.  To see that, note that the lines charged under the global $\Z_N$ need to be attached to surfaces constructed out of $B_1$, which can end on the right boundary and hence they behave like genuine line operators in the effective $4d$ theory.

Alternatively, we can use the boundary conditions $B_2\big|=0$ at the right boundary.  Now the $4d$ lines that need a surface constructed out of $B_1$ are not genuine lines in the effective $4d$ theory.  The needed surface cannot end on the right boundary and it is physical (although it could still be topological) in the effective $4d$ theory.

As a concrete example, let us start with a $4d$  $SU(N)$ theory on the left boundary and set $B_1\big|=0$ at the right end of $I$.  The effective theory is a $4d$ $SU(N)$ theory. The Wilson lines on the left boundary are attached to $B_1$ surfaces, which can end on the right boundary.  They are genuine lines in the effective $4d$ theory.  't Hooft lines need a surface in the $4d$ theory on the left end of $I$.  This surface is an open version of the closed surface operator $U$. Therefore, the boundary conditions \Btbc\ mean that the 't Hooft lines are attached to open $B_2$ surfaces.  Since these surfaces cannot end on the right boundary, the 't Hooft lines are not genuine line operators in the effective $4d$ theory -- they still need surfaces.  We see that the Wilson lines are attached to $B_1$ surfaces and the 't Hooft lines are attached to $B_2$ surfaces.  Note that the 't Hooft commutation relations in the effective $4d$ theory are reproduced using the nontrivial braiding of surfaces of $B_1$ and $B_2$ in the bulk.

Alternatively, with boundary conditions $B_2\big|=0$ at the right boundary Wilson lines are still attached to $B_1$ surfaces, but they cannot end on the right boundary. Then the Wilson lines need surfaces in the effective $4d$ theory.  Therefore, the effective theory is a $PSU(N)$ theory. The 't Hooft lines are the left ends of $B_2$ surfaces in the bulk.
Now, these surfaces can end on the right boundary and therefore the 't Hooft lines are genuine lines in the effective $4d$ theory.  This is consistent with the theory being a $PSU(N)$ theory.

While the $S$ operation changes the theory we are considering, $T$ acts by adding terms that depend only on the background 2-form gauge field.  This does not change the theory.  It only affects ``contact terms.''  Similarly, the group $SL(2,\Z_N)$ contains matrices ${\rm diag}(\alpha,\alpha^{-1})$ for $\alpha$ a unit in $\Z_N$.  For example, for $\alpha=-1$ this gives the charge conjugation operation.  Such operations correspond to a redefinition of the generator of $\Z_N$, and so also do not change the theory in an interesting way.  Thus the number of distinct physical theories we may obtain by acting with a general $SL(2,\Z_N)$ element is given by:
\eqn\sltzcount{
N_{th} = {|SL(2,\Z_N)| \over N \phi(N)} ~,
}
where in the denominator, $N$ comes from the action of $T$, and $\phi(N)$, the Euler totient function, counts the number of units in $\Z_N$.\foot{Euler's totient $\phi(n)$ is the number of totatives of $n$.  These are the positive integers less than or equal to $n$ that are relatively prime to $n$.}  Inserting expressions for the order of $SL(2,\Z_N)$ (see, e.g.\ \sltwoorder) and $\phi(N)$, we find:
\eqn\sltzcount{
N_{th} = { N^3\prod_{p | N}(1-p^{-2}) \over N^2 \prod_{p |N}(1-p^{-1})} = N \prod_{p | N} (1+p^{-1} )~.
}
The RHS gives precisely the index of $\Gamma_0(N)$ in $SL(2,\Z)$.  In the example of $su(N)$ ${\cal N}=4$ SYM, this index was shown in \AharonyHDA\ to give the number of distinct theories in an $SL(2,\Z)$ orbit of the theory with $SU(N)$ gauge group.  Thus by repeated operations of gauging and adding contact terms, we can obtain all the choices of gauge groups and discrete theta parameters that lie in this orbit.  For $N$ square-free, this exhausts all the theories one can define by gauging of one-form symmetries.

If $N$ is not square-free, one can also gauge subgroups of $\Z_N$ that are not factors of $\Z_N$. For example, if $N=M^2$, $\Z_M$ is a subgroup of $\Z_N$, but $\Z_N$ is not a product $\Z_M\times\Z_M$, but a nontrivial extension of $\Z_M$ by $\Z_M$. In this case gauging $\Z_M$ results in a theory with a $\Z_M\times\Z_M$ one-form symmetry with a mixed 't Hooft anomaly. This anomaly signals that one of the $\Z_M$ factors is electric, and the other one is magnetic. Such an operation takes one out of the set of theories with $\Z_N$ one-form symmetries.

It is obviously possible to give a five-dimensional description of these other theories in terms of more general choices of
boundary conditions at the right end of the interval $I$: we can simply start from the $B_1\big|=0 $ Dirichlet boundary condition,
and gauge the $\Z_M$ subgroup of the $\Z_N$ boundary one-form global symmetry. Remember that gauging
the full $\Z_N$ boundary one-form global symmetry would have the net effect of enforcing $B_2\big|=0 $,
while at the same time relaxing the constraint on $B_1\big|$. The effect of gauging the $\Z_M$ subgroup of the $\Z_N$ boundary one-form global symmetry is to partially relax the constraint on $B_1\big|$ and impose a partial constraint on $B_2\big|$.

The final result is a boundary condition which constrains both $B_1\big|$ and $B_2\big|$ to lie in the $\Z_M$ subgroup at the boundary.
This boundary condition treats $B_1\big|$ and $B_2\big|$ symmetrically, allowing
only a $\Z_M$ subset of the $B_1$ and $B_2$ surfaces to end. The boundary supports a $\Z_M\times\Z_M$ boundary
global one-form symmetry with the expected mixed 't Hooft anomaly.

With a bit of extra work it is also possible to promote this boundary condition to a duality wall between two different formulations of the same $5d$ theory \CStwoform: the standard formulation involving the two $\Z_N$ two-form connections $B_1$ and $B_2$, and a dual formulation
involving four $\Z_M$ connections $B_3$, $B_4$, $B_5$, $B_6$. Schematically, that is a theory with an action
\eqn\CSdualtwoform{
S={i\over 2\pi}\int_X \left[M B_3 dB_4 + M B_5 d B_6 + B_5 d B_4 \right],
}
which can be coupled to a $4d$ theory with $\Z_M \times \Z_M$ one-form global symmetry and a mixed 't Hooft anomaly
compensating for the extra $5d$ CS term $B_5 d B_4$.

\bigskip
\centerline {\it Example: $so(N)$ gauge theory in four dimensions}
\bigskip

As a concrete example, take a four dimensional $so(N)$ gauge theory, with vector matter, on a four manifold ${\cal M}_4$ with non-trivial $2$-cycles.
There are two choices of global form of the gauge group, $Spin(N)$ and $SO(N)$.  When the gauge group is $SO(N)$ there are four choices of the discrete theta-parameter.  On a spin manifold, there are only two inequivalent choices and the theories are denoted $SO(N)_\pm$ \AharonyHDA.  In that case, each of the three theories, $Spin(N)$ and $SO(N)_\pm$ has a $\Z_2$-valued $1$-form symmetry, and so we can compute the partition function of each theory with this symmetry coupled to a background $2$-form gauge field $B\in H^2(\CM,\Z_2)$.  Consider for example the $Spin(N)$ case.   Its partition function is
\eqn\spinNpart{ Z_{Spin(N)}(B) = Z(w_2)\bigg|_{w_2 =B}~, }
where by the RHS we mean the contribution to the path integral from $SO(N)$-valued gauge fields living in the bundles with second Stiefel-Whitney class $w_2=B$.  Note that the partition function with the background gauge field set to zero is given by the contribution from the $SO(N)$ bundles which are also $Spin(N)$ bundles, as expected.

Gauging this one-form symmetry is achieved by summing over flat 2-form gauge fields $B$ with a weight
\eqn\weighSt{
 \exp\left(\pi i \int_\CM \hat B\cup B\right),
 }
where $\hat B$ is a new background 2-form gauge field. We have called this operation $S$ above.
Applying it to the $Spin(N)$ theory gives the $SO(N)_+$ theory whose partition function is
\eqn\sonplus{
Z_{SO(N)_+}(B)\sim\sum_{w_2} Z(w_2)  \exp\left(\pi i \int_\CM B\cup w_2 \right) .
}

On the other hand, the operation $T$ acts by multiplying the partition function $Z(B)$ by a factor
\eqn\weightT{
\exp\left(2\pi i \int_\CM \frP(B)\right).
}
Here $\frP$ is the Pontryagin square operation on $H^2(\CM,\Z_2)$ taking values in $H^4(\CM,\R/\Z)$ and corresponding to the quadratic function $\sigma:\Z_2\ra\R/\Z$ given by $\sigma(x)=x^2/4$.\foot{In this special case, the Pontryagin square can be written schematically as
\eqn\pontryaginsq{
\frP(B)={1\over 4} \tilde B\cup \tilde B.
}
where $\tilde B$ is a lift of a $\Z_2$-valued 2-cocycle $B$ to an integral 2-cocycle. It is easy to check that this expression is independent of the choice of the lift. What is not clear from this schematic formula is what to do if $B$ does not lift to an integral 2-cocycle, but only to a 2-cochain closed modulo $2$. This difficulty can be circumvented, (see appendices in \refs{\KapustinQSA, \KapustinGUA} for a detailed explanation).}

The operation $T$ only affects coupling to the background 2-form gauge field, but it gives interesting results when combined with $S$. For example,
applying $T$ and then $S$ to the $Spin(N)$ theory we get the $SO(N)_-$ theory whose partition function is
\eqn\sonminus{
Z_{SO(N)_-}(B)\sim\sum_{w_2} Z(w_2)  \exp \left(2\pi i \int_\CM \frP(w_2)+\pi i \int_\CM B\cup w_2 \right)
}

When we have a duality, performing the same operation on both sides should give new dual pairs.  This gives constraints on what background terms should appear.  For example, consider the duality of \refs{\SeibergPQ,\IntriligatorID}
involving an $\CN=1$ theory with gauge Lie algebra $so(N_c)$ and $N_f$ flavors, whose dual has Lie algebra $so(\hat{N}_c)=so(N_f+4-N_c)$.  Taking into account the global form of the gauge groups, it was shown \AharonyHDA\ that the $SO_+$ theories are mapped to themselves, and the $Spin$ and $SO_-$ are exchanged.  Taking into account also the background terms, the precise mapping is:
\eqn\somap{
Spin(N_c) \leftrightarrow T(SO_-(\hat{N}_c)), \;\;\;  SO_+(N_c) \leftrightarrow T(SO_+ (\hat{N}_c)), \;\;\; SO_-(N_c) \leftrightarrow T(Spin(\hat{N}_c)).
}
One can check that applying $S$ or $T$ to a dual pair takes one to a new dual pair.

\bigskip

\newsec{Applications}

\subsec{Selection rules}

As ordinary symmetries, higher form symmetries can lead to selection rules on amplitudes. A typical example arises in the functional integral over a compact spacetime with a charged loop wrapping a homologically nontrivial cycle. The one-form global symmetry under which the loop is charged makes this amplitude vanish.  More generally, that loop might not even be connected and its amplitude must still vanish.  Similarly, the expectation values of other charged defects are also constrained by their higher-form symmetry.

A concrete example is a $U(1)$ pure gauge theory on a compact manifold with non-trivial one-cycles.  A Wilson loop wrapping a nontrivial element in $H_1(X,U(1))$ is charged under the the one-form $U(1)$ electric symmetry, and therefore its expectation value vanishes.  This fact was shown in \wittenloops\ using a change of variables in the functional integral, which amounts to using the one-form symmetry.
Similarly, an $SU(N)$ pure gauge theory has a one-form $\Z_N$ global symmetry.  We place it on a compact manifold and wrap a Wilson loop around a nontrivial element in $H_1(X,\Z_N)$.  Since it is charged under the global symmetry, its expectation value vanishes.

As another example, consider the $U(1)$ or the $PSU(N)$ theory and its 't Hooft loop.  The global one-form magnetic symmetry forces its expectation value to vanish when it wraps a homologically non-trivial one-cycle.  In terms of the fundamental gauge degrees of freedom this vanishing arises because the 't Hooft loop cannot be defined in this case.  The presentation based on the global symmetry does not assume a specific Lagrangian formulation.

Note that our discussion above about the nonzero expectation value of the Polyakov loop on $\R^3\times S^1$ and its breaking of the global symmetry does not contradict these statements.  Spontaneous symmetry breaking can take place only when the space is non-compact.  Alternatively, we can detect it with compact spacetime $\CM_3\times S^1$ by considering the expectation value of two oppositely oriented loops wrapping the $S^1$ factor.  The selection rule allows this expectation value to be nonzero.  Then we take the limit of the distance between the two loops to be very large, but still much smaller than the size of the compact three manifold $\CM_3$.  A nonzero expectation value in that limit signifies the spontaneous symmetry breaking.

\subsec{Phases of $4d$ gauge theories}

Landau's characterization of phases depends on ordinary global symmetries and whether they are spontaneously broken or not.  These phases are characterized by the expectation values of local order parameters that are charged under the symmetry.  Wilson's and 't Hooft's characterization of phases depends on the behavior of loops.  Using our discussion above, we can state it in terms similar to Landau's.  The novelty is in the fact that the relevant symmetries are higher form symmetries.

As we discussed in section 5, when a theory has a one-form global symmetry $G$, which is broken to a subgroup $\CK$ the lines charged under $\CK$ have an area law and the lines neutral under $\CK$ have a perimeter law (or Coulomb behavior).  Using different dual presentations of the same theory does not change $G$ and $\CK$.  But it typically changes their nature as electric or magnetic.  Therefore, when we describe the phase of the system using the standard terminology (confinement, Higgs, etc.) we have to assume the set of fundamental degrees of freedom as well as the long distance behavior.

For example, consider the $\CN=2$ $SU(2)$ pure gauge theory deformed by a small $\CN=1$ preserving mass.  This system as a one-form global $\Z_2$ symmetry.  It is an electric symmetry and the charged lines are Wilson lines.  After the mass deformation these lines exhibit an area law signaling confinement and the fact that the global $\Z_2$ is unbroken.  In terms of the long distance degrees of freedom, it is more natural to use different dual frames (different in the two vacua of the system) in which the fundamental $SU(2)$ Wilson loop is a $U(1)$ 't Hooft loop.  The global $\Z_2$ is still unbroken, but using these degrees of freedom the phenomenon looks like a Higgs phase.

Another lesson from our discussion is that the standard confinement-Higgs-Coulomb classification of phases is incomplete.  First, there is the known oblique confinement phase.  Second, there is the phenomenon of nontrivial confinement index, where the basic Wilson loop has an area law, but a nontrivial power of it has a perimeter law.  We also discussed examples of a similar phenomenon with 't Hooft loops.

We propose to classify the phases not in terms of these traditional terms, but instead, in terms of the one-form global symmetry $G$ and its unbroken subgroup $\CK$ (as well as their description using the UV degrees of freedom).

In fact, $G$ and $\CK$ might not be enough to characterize the phase.  This can be seen in two ways.  First, as we will discuss in the coming sections, the unbroken group $\CK$ could have 't Hooft anomalies, which should be captured by the low-energy theory.  These anomalies lead to excitations on boundaries.  Therefore, there can be distinct phases with the same $G$ and $\CK$, but different anomalies.  Specifically, for discrete $G$ and $\CK$, the low-energy bulk theory is a $G/\CK$ gauge theory and the anomaly is represented by an additional parameter in that theory.

Second, phases can also be characterized by the long distance behavior of disk operators --  line operators that bound a topological surface.  These are not genuine lines and hence do not fall into our characterization of symmetry.  Yet, since the surface is topological, they can be used to probe phases.  For example, in the $PSU(N)$ theory the Wilson line is not a genuine line and needs a topological surface.  But it can still be used to probe confinement \KapustinGUA.  It would be nice to find a simple complete characterization of phases of $4d$ gauge theories.

The $5d$ picture of the $4d$ gauge theory, which we discussed in section 6, gives a new perspective on the discussion of phases.  There we studied a $5d$ TQFT on $\CM\times I$ with $\CM$ a four-manifold and $I$ a line segment. A $4d$ theory is placed on the left end of $I$ and its global one-form symmetry is coupled to a $5d$ gauge field.  The theory is also characterized by the boundary conditions on the right end of $I$.  In the limit that $I$ is short, we find an effective $4d$ QFT.  But since the bulk theory is topological, we should find the same answers when $I$ is large.  In that limit it is clear that the local excitations of the theory are independent of the boundary conditions on the right.  This means that as we vary these boundary conditions we find different $4d$ theories, all with the same local excitations.  For example, an $SU(N)$ gauge theory and a $PSU(N)$ gauge theory with its various discrete theta parameters are all related by such a construction.  Therefore, they can exhibit the same phases.  The discussion of the long distance behavior of open disk operators above is a demonstration of this fact.  The Wilson line can be thought of as having a surface running to the right end in the $SU(N)$ theory or as having a surface in the left end or in the bulk in the $PSU(N)$ theory.  In either case, its area law behavior signals confinement.

\subsec{Topological phases protected by one-form symmetry}

Another use of higher-form symmetries is to define new classes of Symmetry Protected Topological (SPT) Phases. The notion of an SPT phase arose in condensed matter theory. These phases can be characterized by the following three properties: (1) they are gapped (i.e.\ the energy gap between the ground state and the first excited state stays nonzero in the large volume limit); (2) the ground state is non-degenerate and has no long-range entanglement; (3) the system cannot be deformed to the trivial system, i.e.\ a system whose ground state is a factorized state, without breaking a global symmetry $\Gamma$ or without closing the gap. The definition of long-range entanglement is somewhat tricky and goes beyond the scope of this paper. Suffice it to say that it implies the non-degeneracy of the ground state on a compact space of any topology.\foot{In fact, in many of our examples, the ground state has long-range entanglement and accordingly the low-energy theory is described by a nontrivial TQFT, enriched with global symmetries.  In the condensed matter literature the corresponding phases are known as Symmetry Enhanced Topological (SET) phases.}

It was proposed in \KapustinUXA\ to generalize the notion of an SPT phase by replacing ordinary global symmetries with higher-form global symmetries. Specifically, it was proposed that phases of $4d$ gauge theories with oblique confinement are examples of SPT phases protected by one-form symmetry. In this section we elaborate on this proposal.

Let $\CG$ be a compact semi-simple Lie group. Consider a $4d$ gauge theory with gauge group $\CG$. Let us assume that all matter fields transform trivially under a subgroup $\Gamma\subset Z(\CG)$ of the center of $\CG$, so that the theory has a global one-form symmetry $\Gamma$. In a confining or oblique confining phase this symmetry is unbroken and there is a mass gap. While the vacuum need not be unique, we can focus on one of them and ask whether it corresponds to a nontrivial SPT phase protected by the one-form $\Gamma$ symmetry. To determine this, we need to study the system in the presence of topological defects generating the one-form symmetry, or equivalently by coupling the system to a background 2-form gauge field $B\in H^2(\CM,\Gamma)$. The partition function of the system will be a $B$-dependent phase (because the ground state is non-degenerate even after coupling to $B$).

The computation of the phase proceeds as follows (see Appendix E for details). First of all, we twist by $B$ using a network of topological defects $U_\gamma$ implementing $\Gamma$ transformations. These defects are localized on codimension-2 submanifolds and generically intersect pairwise at points. We can eliminate the intersection by excising a disk on one of the two intersecting defects and placing a suitable loop operator on the resulting boundary. This loop must have a perimeter law, if we want the network to remain topological. The nontrivial phase of the partition function may arise because of nontrivial $\Gamma$-charge of these loop operators.

In a confining phase the partition function is trivial. Indeed, the boundaries of topological defects $U_\gamma$ implementing $\Gamma$ transformations are 't Hooft loops, which have a perimeter law and do not carry electric charge. Thus the confining phase (which is intuitively described as a phase with a monopole condensate) is a trivial SPT phase. On the other hand, in an oblique confining phase  (which is intuitively described as a phase with a dyon condensate) the usual 't Hooft loop has an area law. But for any magnetic charge $\gamma\in \Gamma$ there exists a Wilson-'t Hooft loop with electric charge $\eta(\gamma,\cdot)\in \hat \Gamma$ that has a perimeter law. Intuitively, these are loops whose charges are proportional to the charges of the condensed dyons. One can terminate $U_\gamma$ on such a Wilson-'t Hooft loop.  These loops carry $\Gamma$-charge and therefore lead to a nontrivial phase in the partition function. One can show (see Appendix E) that the corresponding partition function is given by
\eqn\partsigma{
\exp\left(2\pi i\int_\CM \frP_\sigma(B)\right),
}
where $\sigma:\Gamma\to U(1)$ is a quadratic refinement of $\eta:\Gamma\times\Gamma\to U(1)$, and $\frP_\sigma$ is the corresponding Pontryagin square. Thus oblique confining phases correspond to nontrivial SPT phases.

Note that while on spin-manifolds the partition function is determined by $\eta$ alone, on non-spin manifolds it depends also on $\sigma$.

An alternative way to detect a nontrivial SPT phase is to consider its boundary. If we assume that the symmetry is unbroken on the boundary, a nontrivial SPT phase can be characterized by the fact that the action of the symmetry on the boundary degrees of freedom has an 't Hooft anomaly. This anomaly is canceled by the anomaly inflow from the bulk.

\midinsert\bigskip{\vbox{{\epsfxsize=4.2in
        \nobreak
    \centerline{\epsfbox{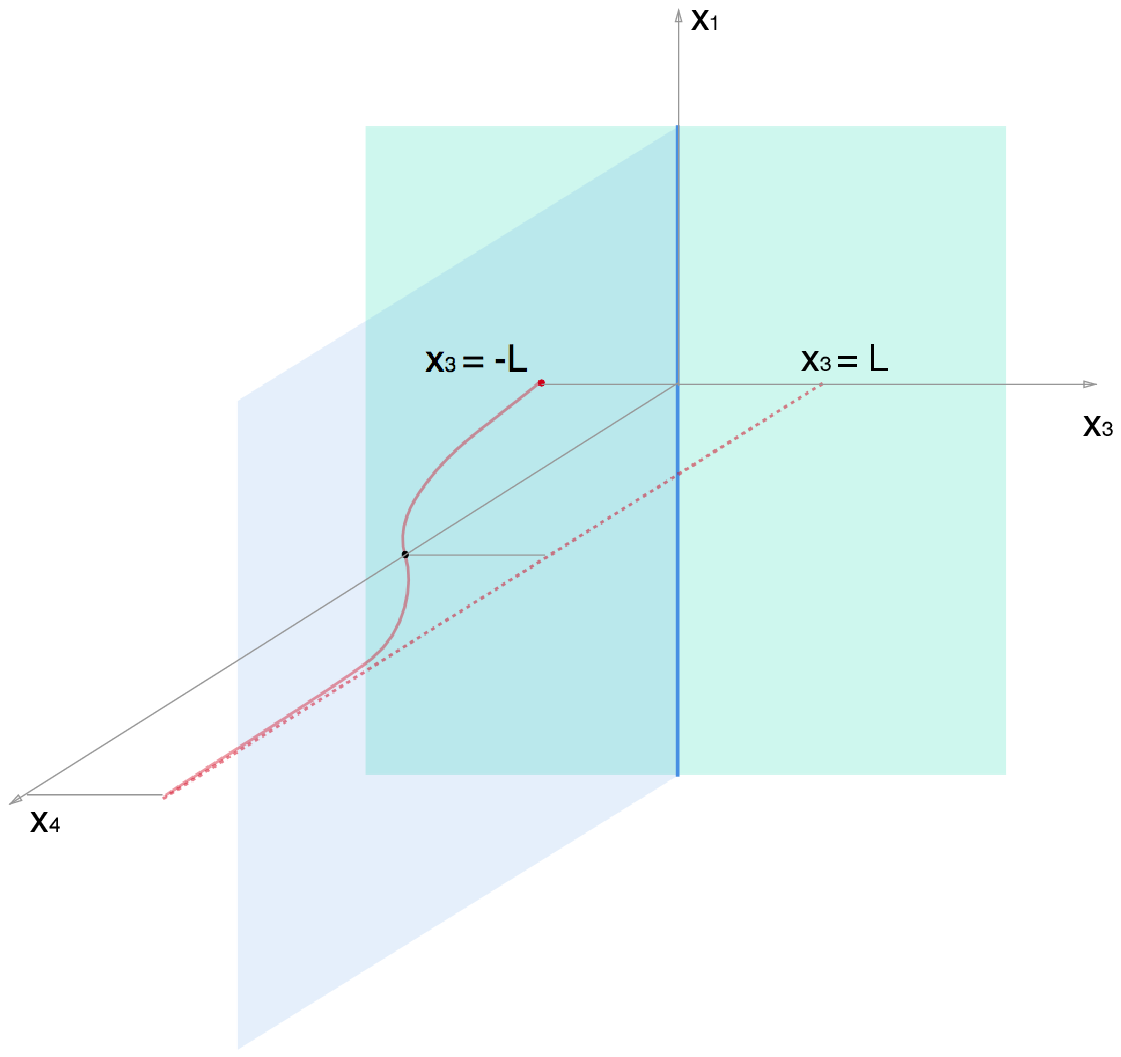}}
        \nobreak\bigskip
    {\raggedright\it \vbox{
{\bf Figure 1. Demonstrating oblique confinement as an SPT phase.} {\it The $x_2$ direction is omitted.
The green plane is the boundary.  The blue plane is a first surface defect ending on the blue line.
The red line is a second surface defect (extending along $x_2$) ending on the red dot.  The dotted red line is the original position of the second surface defect.  The black dot is the intersection point of the two surface defects.
}}}}}}
\bigskip\endinsert

Let us make this explicit in the case of oblique confining phases.
The electric surfaces should be able to end topologically on a symmetry-preserving
boundary. Concretely, suppose the theory is defined on $x_4>0$, the first surface extends
along $x_1$ and $x_4$ at $x_2=x_3=0$, ending on a topological line defect at $x_4=0$,
and the second extends along $x_2$ and $x_4$
at $x_1=0$ and $x_3=L>0$, ending on a topological line defect at $x_4=0$.  See Figure 1.

Consider a deformation of the
system that brings the second line defect from $x_3=L$ to $x_3=-L$, thus bringing it across the first line defect.  We do that only at the vicinity of the boundary without deforming the system far from the boundary.
The two surface defects now intersect at one point, and the configuration has an extra
contact term $\exp 2 \pi i \eta(\gamma, \gamma')$. Therefore, the boundaries of the $U_\gamma$ surface operators
commute only up to a phase, as in \comrelqeCS , and the symmetry has an 't Hooft anomaly on the boundary.

Alternatively, we can do a gauge transformation $B \to B + d \lambda$ for a $\Gamma$-valued
one-form $\lambda$ in the partition function. If $\CM$ has a boundary,
we are left with an anomaly
\eqn\SPTboundary{
\exp\left( 2 \pi i \int_{\partial \CM} \eta(\lambda,B)\right)
}

In order to cancel the anomaly, we need to add extra boundary degrees of freedom.
A simple possibility is a $3d$ TQFT equipped with line defects ${\cal L}_\gamma$ labeled by elements of $\Gamma$,
such that ${\cal L}_\gamma {\cal L}_{\gamma'} = {\cal L}_{\gamma+ \gamma'}$.
Such line defects can be deformed across each other, at the price of a factor controlled by
the modular matrix $S_{\gamma, \gamma'}$.

If we select the $3d$ TQFT in such a way that
\eqn\TFTboundary{e^{2 \pi i \eta(\gamma, \gamma')}S_{\gamma, 0}S_{0,\gamma'} = S_{\gamma, \gamma'}S_{0,0}}
and we dress the boundary endpoints of $U_\gamma$ operators by the ${\cal L}_\gamma$ line defects,
the bulk contact term above cancels out against the extra boundary $3d$ TQFT phase and the dressed $U_\gamma$ surfaces are fully topological.

If the $3d$ TQFT is associated to a $2d$ RCFT, such a family of line defects is associated with a collection of {\it simple currents}
of some dimensions $\Delta_\gamma$ and the phase above can be identified with $\exp 2 \pi i(\Delta_{\gamma + \gamma'} - \Delta_\gamma - \Delta_{\gamma'})$.

\subsec{BPS domain walls in $SU(N)$ SYM}

Supersymmetric Yang-Mills theory with $SU(N)$ gauge group is expected to have a rather straightforward low energy
description. The theory confines and breaks spontaneously a global $\Z_{2N}$ R-symmetry to $\Z_2$.
The spontaneous symmetry breaking is associated to the vev of a gaugino bilinear
\eqn\vev{
\langle \Tr \lambda^\alpha \lambda_\alpha \rangle = e^{{2 \pi i k \over N}} \Lambda^3~.}
The integer $k=0, \cdots, N-1$ labels the $N$ vacua of the theory and $\Lambda$ is the strong coupling scale.

As the theory contains only adjoint matter, it has stable confining strings, which can end on
Wilson loops in the (anti)fundamental representation, much as it happens for non-supersymmetric
pure $SU(N)$ Yang-Mills theory. The specific type of confinement depends on the
choice of vacuum: intuitively, confinement in the $k$-th vacuum is induced by the condensation
of a dyon with electric charge proportional to $k$.  Hence, these are oblique confinement vacua.

As in the discussion above, the one-form $\Z_N$ global symmetry is unbroken in all of these vacua, but its 't Hooft anomaly is different in the different vacua.  Roughly, the $N$ vacua of the theory correspond to distinct
SPT phases for the one-form $\Z_N$ symmetry of the theory \DieriglXTA.
More precisely, since the system has $N$ vacua this is actually a Symmetry Enhanced Topological (SET) phase. In terms of the global $\Z_{2N} \to \Z_2$ R-symmetry the TQFT is very simple: a sigma-model whose target is the set of $N$ vacua with obvious action of the global symmetry.  To that we need to add the effect of the one-form global $\Z_N$ symmetry. We will do that momentarily.

This observation agrees and partly explains
the rich topological structure of the dynamical domain walls of the theory.
Indeed, the BPS domain wall between the $k$-th and $(k+n)$-th vacua of the theory conjecturally supports
an $\CN=1$ supersymmetric $U(n)_{N}$ Chern-Simons theory \AcharyaDZ.
As we computed in section 4.4, this theory has a $\Z_N$ one-form symmetry with 't Hooft anomaly.  This anomaly matches the difference between the 't Hooft anomalies of the $\Z_N$ one-form symmetry between the $k$-th and $(k+n)$-th vacua
on the two sides of the wall.

Notice that supersymmetry is definitely instrumental in determining certain properties of the domain walls,
but supersymmetry alone does not explain why the domain wall world volume theory should include topological
degrees of freedom at low energy. The existence of topological degrees of freedom should be robust under perturbation of the theory,
irrespective of supersymmetry, as long as they don't change its universality class.
The presence of non-trivial one-form $\Z_N$ symmetry and its different 't Hooft anomalies in different vacua of the $4d$ gauge theory provides such a robust
motivation.

Topological considerations alone, of course, cannot explain why the domain walls should support $U(n)_{N}$ Chern-Simons theory
rather than, say, a $U(1)^n_{N}$ Chern-Simons theory or any other $3d$ TQFT with the same one-form symmetry and the same 't Hooft anomaly for it,
i.e.\ with a collection of line defects $L_a$ with an appropriate $S_{a a'}$ modular matrix.

It is interesting to elaborate further on the interplay between the spontaneously broken global zero-form R-symmetry
and the one-form $\Z_N$ symmetry.  As above, couple the UV action to a classical $\Z_N$ two-form gauge field $B$.  Then, since the zero-form R-symmetry rotation shifts the $\theta$-angle by $2\pi$, it performs the $T$ operation of section 6.  In other words,
a global zero-form symmetry transformation shifts the action by a quadratic function of $B$
\eqn\SPTbasic{
\delta S={i \over 2} \int {N \over 2\pi}B \wedge B ~.
}
More explicitly, the $\Z_N$ one-form symmetry acts on the UV connection as $A \to A + \lambda$.
We can write the coupling of the UV theory to the background connection $B$ by replacing the field strength $F$ in the action by the
invariant combination $F_B = F - B$.
In particular, the UV $\theta$-angle couples to $B$ as
\eqn\thetaB{
-{i N \theta \over 8\pi^2} \int B \wedge B ~.}
Because of the R-current anomaly, the $\Z_{2N}$ zero-form global symmetry holds only up to shifts of the $\theta$-angle,
and thus the UV action coupled to the $B$ two-form connection shifts as desired under the zero-form global symmetry transformations.

This shift of the action by \SPTbasic\ can be thought of as the result of a mixed cubic 't Hooft anomaly involving the global zero-form symmetry and the global one-form symmetries of the theory.

The 't Hooft anomaly matching conditions state that the same shift by \SPTbasic\ should be present also in the IR theory.  To see that, we describe the spontaneously broken $\Z_{2N}\to \Z_2$ by the effective Lagrangian ${iN\over 2\pi} \phi\ dA^{(3)}$
where $\phi \sim \phi + 2\pi$ is a scalar and $A^{(3)}$ is a properly normalized three-form gauge field.  The $N$ vacua are labeled by $\langle e^{i\phi}\rangle =e^{2\pi i k\over N}$ and $e^{i\oint A^{(3)}}$ describes a domain wall between them.  The 't Hooft anomaly can then be described by coupling this system to the classical background $\Z_N$ gauge field $B$ as
\eqn\mixedcoup{{iN\over 2\pi} \phi\ \Big( dA^{(3)} + { N \over 4\pi} B\wedge B \Big) ~.}
The one-form gauge transformation of the background $B$ shifts $B\to B+d\lambda$.  It must act on $A^{(3)}$ as
\eqn\Athrt{A^{(3)}\to A^{(3)} - {N\over 2\pi} B\wedge\lambda - {N\over 4\pi} \lambda\wedge d\lambda~.}

In \mixedcoup\ $B$ is a $\Z_N$ gauge field.  It is sometimes convenient to express it in terms of $U(1)$ gauge fields.  As in \KapustinGUA, this can be done by adding a Lagrange multiplier two-form dynamical field $f$.  Specifically, we can change \mixedcoup\ to
\eqn\mixedcoupa{{iN\over 2\pi} \phi\ dA^{(3)} + {i\over 2\pi}f \wedge(da - NB)+ { i \over 8\pi^2}\phi\ da\wedge da~.}
The first term represents the ordinary global $\Z_N$ symmetry.  $a$ is a $U(1)$ gauge field and hence the constraint $NB=da$ makes the gauge field $B$ flat and sets its periods to be in $\Z_N$.  The last term coincides with the last term in \mixedcoup\ when the constraint is used.  On a spin manifold the theory based on \mixedcoupa\ is invariant under
\eqn\mixedcoupg{\eqalign{
&a \to a + d\lambda_a +N\lambda\cr
&B \to B + d\lambda\cr
&A^{(3)}\to A^{(3)} + d\lambda^{(2)}- {1\over 2\pi} da\wedge\lambda - {N\over 4\pi} \lambda\wedge d\lambda  ~.\cr}}
Here $\lambda_a$ is an ordinary $U(1)$ gauge transformation parameter, $\lambda$ is a one-form $U(1)$ gauge transformation parameter and $\lambda^{(2)}$ is a two-form gauge transformation parameter.

The actions \Athrt\mixedcoupa\ exhibit all the anomalies in the IR theory.  They describe a SET phase with the appropriate zero-form and the one-form global symmetries.  They lead to $N$ vacua with different 't Hooft anomalies (roughly, in different SPT phases of a one-form global symmetry) corresponding to the different oblique confining vacua of the system.

This discussion makes it easy to gauge $B$.  In the UV this has the effect of changing the gauge group to $PSU(N)$ (see \KapustinGUA\ and Appendix C).  In the IR we simply need to view $B$ in \mixedcoup\ as a dynamical field.  We can also promote it to a $U(1)$ gauge field and add a one form gauge field $A$ with the term ${i N\over 2\pi} A dB$, as in Appendix B.  The coupling of $\phi$ to $B\wedge B$ can then be recognized as making the twisting parameter $p$ in Appendix B dynamical.

\bigskip
\noindent {\bf Acknowledgments}

We are grateful to S.~Razamat for collaboration at an early stage of the project and many important discussions.  We also thank O.~Aharony, N.~Arkani-Hamed, J.~Maldacena, G.~Moore, S.~Shenker, Y.~Tachikawa, and E.~Witten for helpful discussions.  We also thank Y.~Tachikawa for comments on the manuscript.
The research of DG was supported by the Perimeter Institute for Theoretical Physics.
Research at Perimeter Institute is supported by the Government of Canada through Industry Canada and by the Province of Ontario through the Ministry of Economic Development and Innovation.
The work of AK was supported in part by the DOE grant DE-FG02-92ER40701 and by Simons Foundation.
The work of NS was supported in part by DOE grant DE-SC0009988 and by the United States-Israel Binational Science Foundation (BSF) under grant number~2010/629. The research of BW was supported in part by DOE Grant DE-SC0009988 and the Roger Dashen Membership.
Opinions and conclusions expressed here are those of the authors and do not
necessarily reflect the views of funding agencies.

\appendix{A}{The $\Z_n\times \Z_m$ Dijkgraaf-Witten theory in $2d$}

\subsec{The bulk theory}

In this appendix we comment on the $2d$ Dijkgraaf-Witten theory \DijkgraafPZ\ with gauge symmetry $\Z_{n_1} \times \Z_{n_2}$.  We follow the continuum presentation of \KapustinGUA\ and extend the discussion there.

The action is
\eqn\actionDW{S ={i \over 2 \pi} \int \Big(n_1\, B_1F_1 + n_2\, B_2F_2 + p \, \lcm(n_1,n_2)A_1\wedge A_2\Big) ~,}
where $B_I$ (with $I=1,2$) are $2\pi$-periodic scalars, $A_I$ are $U(1)$ gauge fields with $F_I=dA_I$, and the parameters $n_I$ and $p$ are integers.  The theory with $p$ is the same as with $ p+\gcd(n_1,n_2)$ and hence we will take $p=0,...,\gcd(n_1,n_2)-1$. This system is invariant under the gauge symmetry
\eqn\gauget{\eqalign{
&A_I \to A_I + d f_I\cr
&B_I \to B_I - \epsilon_{IJ}{P n_J \over K}f_J ~,}}
where we use the notation
\eqn\PKdef{\eqalign{
&P={p\over \gcd(p,n_1,n_2)} \cr
&K={\gcd(n_1,n_2)\over \gcd(p,n_1,n_2)} \cr
&\gcd(P,K)=1~.}}

For $p=0$ this theory has a zero-form (ordinary) global $\Z_{n_1} \times \Z_{n_2}$ symmetry and a one-form global $\Z_{n_1} \times \Z_{n_2}$ symmetry, which are generated by
\eqn\opspz{ \eqalign{
& U_I=e^{i\oint A_I }\cr
& V_I=e^{iB_I}~,}}
respectively. They act on the fundamental fields as
\eqn\gena{\eqalign{
&B_I \to B_I + {2\pi \over n_I} \cr
&A_I \to A_I +{1\over n_I} \zeta_I}}
with $\zeta_I$ a properly normalized flat gauge field.  The charged objects are $V_I$ and $U_I$, respectively.  Using the equations of motion the generators satisfy
\eqn\Wsupng{U_I^{n_I} =V_I^{n_I}= 1~. }

For nonzero $p$ we start with the gauge invariant operators
\eqn\ops{ \eqalign{
& U_I=e^{i\oint A_I } \cr
& \hat V_I(\gamma)=e^{iB_I(\CP)}e^{-i\epsilon_{IJ}{P n_J \over K}\int_{\CP}^{\CP'} A_J }e^{-iB_I(\CP')}
~,}}
and their powers.  Here $U_I$ are functions of closed lines.  $\hat V_I$ are functions of open lines along $\gamma$ starting at $\CP$ and ending at $\CP'$.  As for $p=0$,
\eqn\Wsupn{U_I^{n_I} = 1 ~. }
Also, since our theory is topological, the correlation functions of open line operators are trivial and hence
\eqn\hatVIe{\hat V_I=1 ~.}
Equivalently, \hatVIe\ follows from the equation of motion of $A_I$
\eqn\Aeom{K dB_I + \epsilon_{IJ} P n_J A_J=0 ~.}

We should make a clarifying comment.  When we study correlation functions of operators we should exclude operators at coincident points. This means that an insertion of $\hat V_I$ or $U_I$ exclude additional operators that touch the lines along which $A_I$ are integrated.  This is the sense in which equations like \Wsupn\hatVIe\ are valid.  But we can argue more than that.  Equation \Wsupn\ follows from the equation of motion of $B_I$.  Therefore, it can be violated by contact terms only when an operator depending on $B_I$ touches the line -- not when the line is crossed by an integral of $A_I$.  Equivalently, as we cross the line of $U_I^{n_I}$, $B_I \to B_I + 2\pi$, but these two values are identified.  On the other hand, equation \hatVIe\ follows from the equation of motion of $A_I$ and therefore can be violated when the line is crossed by an integral of $A_I$.  We conclude that while \hatVIe\ can suffer from contact terms when the line is crossed, \Wsupn\ does not.

This understanding allows us to find nontrivial local operators.  The line attached to
$\hat V_I^K $ does not have contact terms and hence $\hat V_I^K $ is a product of two genuine local operators.  One of them is at $\CP$ and the other is at $\CP'$.  In what follows we will use the notation
\eqn\Vnota{V_I=e^{iKB_I}}
and will not write the line attached to it\foot{One might question this notation in situations where the total number of insertions of $V_I$ does not allow us to connect them by lines.  A global symmetry, which will be discussed below, guarantees that whenever such a question arises the correlation function vanishes.}.  The local operators $V_I$ are indeed nontrivial, for example when $U_I$ surrounds $V_I$ it leads to a phase $e^{2\pi i K/n_I}$.

We see that the theory has $n_1 n_2 \over K^2$ genuine local operators generated by $V_I$.

Let us analyze the closed lines more carefully.  We can open the closed line operator $U_I^{P n_I \over K}$ and let it end, as in \ops.  Hence, its correlation functions are trivial and $U_I^{P n_I \over K }=1$.  Note that unlike \Wsupn, this equation can suffer from contact terms.
Combining this information with \Wsupn\ we learn that the nontrivial closed lines are generated by $U_I$ and they satisfy
\eqn\closedlines{U_I^{n_I\over K}=1 ~,}
where again, the equation can be violated by contact terms.

We conclude that the theory has a global zero-form $\Z_{n_1\over K}\times \Z_{n_2\over K}$ symmetry generated by $U_I$ and a global one-form $\Z_{n_1\over K}\times \Z_{n_2\over K}$ symmetry generated by $V_I$.  The charged objects under these two symmetries are $V_I$ and $U_I$ respectively.  Comparing with the situation with $p=0$ we see that we gauged a global zero-form $\CK=\Z_{K}\times \Z_{K} \subset \Z_{n_1}\times \Z_{n_2}$ symmetry, which acts as (compare with \gena)
\eqn\newg{B_I \to B_I + {2\pi \over K} ~.}
This gauging reduces the global symmetry to the quotient
\eqn\ZNqu{\Z_{n_1}\times \Z_{n_2} \to {\Z_{n_1}\times \Z_{n_2}\over \Z_{K}\times \Z_{K}} =\Z_{n_1\over K}\times \Z_{n_2\over K}~.}
This gauging also eliminates the local operators $e^{iB_I}$ and thus
reduces the one-form $\Z_{n_1}\times \Z_{n_2}$ symmetry to a subgroup
\eqn\ZNqua{\Z_{n_1}\times \Z_{n_2} \to \Z_{n_1\over K}\times \Z_{n_2\over K} \subset \Z_{n_1}\times \Z_{n_2}~.}

\subsec{Canonical quantization on a cylinder}

The identification of the global symmetry and the operators can also be obtained using canonical quantization, with space taken to be $\S^1$ parameterized by a periodic coordinate $\sigma \in [0,2\pi)$. We choose axial gauge for $A_I$ and then impose Gauss's law.  For $p=0$ it constrains $B_I$ to be constant and the theory reduces to an ordinary quantum mechanical system of $2\pi$-periodic variables $b_I ={1\over 2\pi} \oint B_I d\sigma$ and $a_I=\oint A_I$ with the action
\eqn\canonicalDWone{
{i\over 2\pi} \int \Big(n_1\,b_1 \partial_0 a_1  + n_2\,b_2\partial_0 a_2 \Big)dt~.
}
Quantization of such a system is standard and yields a Hilbert space of dimension $n_1n_2$. The operators $U_I=e^{ia_I}$ and $V_I=e^{ib_I}$ are realized as clock and shift matrices satisfying
\eqn\clockshift{\eqalign{
&U_I^{n_I}=V_I^{n_I}=1\cr
&U_IV_J=V_J U_I  \quad {\rm for} \ I\not=J \cr
&U_I V_I=e^{2\pi i\over n_I} V_I U_I~.}}
As in the discussion of local operators above, $U_I$ generate a $\Z_{n_1}\times \Z_{n_2}$ symmetry.  The one-form global symmetry $\Z_{n_1}\times \Z_{n_2}$ becomes an ordinary symmetry in the effective quantum mechanical system.  It is generated by $V_I$.  And the central extension in the third line in \clockshift\ represents the fact that $U_I$ are charged under the symmetry generated by $V_I$ and viceversa.

For nonzero $p$ the Gauss's law constraint implies (see also \Aeom)
\eqn\gausslawconstr{
P n_I (A_I)_\sigma- \epsilon_{IJ} K \partial_\sigma B_J =0~.}
Substituting this in the action \actionDW\ we find the effective quantum mechanical system \canonicalDWone\ except that now
\eqn\qmvar{\eqalign{
&a_I=\oint A_I =  {\epsilon_{IJ} K\over P n_I}(B_J(2\pi) -B_J(0)) \cr
&b_I={1\over 2} (B_I(2\pi) + B_I(0)) ~.}}
The dependence on $p$ is through the constraint $a_I \in 2\pi {K\over P n_I}  \Z$.
Because of that, $U_I=e^{ia_I}=e^{i\oint A_I}$ satisfies
$U_I^{P n_I\over K}=  1$
and using $U_I^{n_I}=1$ we have, as in \closedlines,
\eqn\UIf{U_I^{n_I\over K}=1~.}

The constraint \UIf\ has another consequence.  Since $a_I$ is the momentum conjugate to $b_I$ it means that not only $b_I \sim b_I + 2\pi$, but also $b_I \sim b_I + {2\pi P\over K}$, or equivalently
\eqn\biden{b_I \sim b_I + {2\pi \over K}~.}
Therefore, the physical operators are not generated by $e^{ib_I}$, but by
\eqn\VIc{V_I = e^{i K b_I}.}
Again, this agrees with the analysis of the local operators above.   As in \newg, the constraint \UIf\ and the identification \biden\ can be interpreted as gauging a $\Z_K\times \Z_K$ symmetry.

For $p=0$ our system has a $\Z_{n_1}\times \Z_{n_2}$ global symmetry, which is completely spontaneously broken.  Gauging $\CK=\Z_K\times \Z_K \subset \Z_{n_1}\times \Z_{n_2}$ means that the subgroup generated by $U_I^{n_I\over K}$ does not act on the low energy fields.  This can be interpreted to mean that this symmetry is unbroken and does not act on the light modes.  In other words, the global $\Z_{n_1}\times \Z_{n_2}$ symmetry is broken to $\Z_K\times \Z_K $ and the topological theory realizes the quotient
\eqn\lowqu{{\Z_{n_1}\times \Z_{n_2} \over \Z_K\times \Z_K }= \Z_{n_1\over K}\times \Z_{n_2\over K}~.}

This discussion suggests that the theory \actionDW\ can describe the low energy dynamics of a non-topological theory with global symmetry $\Z_{n_1}\times \Z_{n_2}$, which is spontaneously broken to $\CK=\Z_K\times \Z_K$.  The UV theory has $n_1n_2$ closed line operators $U_I$, which generate the symmetry and local operators $\CO_I$, which transform under it. Because of the spontaneous breaking, no low energy mode transforms under the unbroken $\Z_K\times \Z_K$.  Hence, the correlation functions of the local UV operators that transform nontrivially under $\Z_K\times \Z_K$ decay exponentially in the distance and they vanish in the IR theory.  In our description these are the non-gauge invariant operators generated by $e^{iB_I}$.  On the other hand, the operators $V_I$, which are $\CK=\Z_K\times \Z_K$ invariant can survive in the low energy theory.  In fact, in the infinite volume limit these operators have vacuum expectation values, which are responsible for the spontaneous symmetry breaking.  Similarly, the low energy theory should not include the operators $U_I^{n_I\over K}$, which generate the unbroken $\Z_K\times \Z_K$.  Unlike the charged $\CO_I$, whose correlation functions vanish at long distance, these operators are identified as one in the IR theory and the set of nontrivial $U_I$ operators should be modded out by them.

\subsec{The system with a boundary}

\bigskip
\centerline{\it Dirichlet boundary conditions}
\bigskip

Let us discuss the system when it has a boundary.  We pick boundary conditions such that there is no surface term in the variation of the action.  The variation of \actionDW\ leads to the boundary term
\eqn\boundarysur{{i \over 2\pi } \int_{\partial M} n_I B_I \delta A_I\big| ~,}
where $\partial M$  is the boundary and $(\cdots)\big|$ means along the boundary.  We want to preserve the ordinary global symmetry $\Z_{n_1}\times \Z_{n_2}$, which shifts $B_I$.  Therefore, we pick the boundary conditions
\eqn\boundcon{A_I\big|=0 ~.}
This means that the $U(1)\times U(1)$ gauge transformation parameters $f_I$ must vanish on the boundary.

It is important that the Dirichlet boundary conditions \boundcon\ break the one-form $\Z_{n_1}\times \Z_{n_2}$ of the bulk theory.  Another possible choice is $B_I\big|=0$, which preserves the one-form global symmetry but breaks the ordinary global symmetry.  Note that there is no choice of boundary conditions that preserves the entire symmetry of the bulk.

Let us enumerate the boundary operators.  First, we can have local boundary operators
\eqn\localb{\tilde V_I(\CP) = e^{i B_I(\CP)} ~, }
where $\CP$ is a point along the boundary.  Note that it is gauge invariant.  Second, we can have line operators ending on the boundary
\eqn\lineb{\tilde U_I(\gamma) = e^{i \int_\gamma A_I} ~, }
where the line $\gamma$ starts and ends at boundary points on the same or a different component of the boundary.  Even though $\gamma$ is an open line, the fact that the gauge transformation parameters vanish at the boundary makes \lineb\ gauge invariant.  We could also try to construct line operators starting at the boundary and ending on $e^{iB_I}$ at a point in the bulk.  But these are equivalent to \localb.

Unlike the bulk operators, the boundary operators \localb\lineb\ realize the full $\Z_{n_1}\times \Z_{n_2}$ zero-form and one-form symmetries.

In order to see the symmetries and the boundary states more clearly, we analyze the system on a strip using canonical quantization.  As in the discussion about the cylinder starting around \canonicalDWone, we parameterize the strip by $\sigma\in[0,2\pi]$ and time by $t$.  We pick axial gauge and derive the effective action
\eqn\canonicalDWbb{\eqalign{
S_{eff}=&{i\over 2\pi} \int \Big(n_1\,B_1(0) \partial_0 a_1  + n_2\,B_2(2\pi)\partial_0 a_2 \Big)dt\cr
=&{i\over 2\pi} \int \Big(n_1\,B_1(2\pi) \partial_0 a_1  + n_2\,B_2(0)\partial_0 a_2 \Big)dt}}
with
\eqn\qmvart{a_I=\int A_I =  {\epsilon_{IJ} K\over P n_I}(B_J(2\pi) -B_J(0))  ~.}
(For $p=0$, $B_I(2\pi)=B_I(0)$ and $a_I$ is an independent variable.) All the variables  $B_I(0)$, $B_I(2\pi)$ and $a_I$ are $2\pi$ periodic.  These variables are related to the boundary operators\localb\lineb\ through
\eqn\bopsrel{\eqalign{&\tilde U_I(\gamma)=e^{ia_I} \cr
&\tilde V_I(0)= e^{i B_I(0) } \cr
&\tilde V_I(2\pi)= e^{i B_I(2\pi) }}}
with $\gamma$ a line from $\sigma=0$ to $\sigma=2\pi$.

The action \canonicalDWbb\ shows that we have an $n_1n_2$ dimensional Hilbert space.
$\tilde V_1(0)$, $\tilde V_2(2\pi)$ and $\tilde U_I$ (or alternatively $\tilde V_1(2\pi)$, $\tilde V_2(0)$ and $\tilde U_I$) reflect the full $\Z_{n_1}\times \Z_{n_2}$ zero-form and one-form symmetries of the bulk problem
\eqn\boundopr{\eqalign{
&\tilde V_I(0)^{n_I}=\tilde V_I(2\pi)^{n_I} =\tilde U_I^{n_I}=1 \cr
&\tilde V_1(0) \tilde U_1 = e^{2\pi i \over n_1}\tilde U_1 \tilde V_1(0) \cr
&\tilde V_2(2\pi) \tilde U_2= e^{2\pi i \over n_2} \tilde U_2 \tilde V_2(2\pi) ~.}}

We can focus on a single boundary (say, at $\sigma=0$) by writing \canonicalDWbb\ as
\eqn\boundaryfo{S_{eff}={i\over 2\pi} \int \Big(n_1\,B_1(0) \partial_0 a_1  + n_2\,B_2(0)\partial_0 a_2 - {n_1n_2P\over K} a_2 \partial_0 a_1 \Big)dt~.}
Using this action or the operator relations \boundopr\ together with
\eqn\moveB{\tilde V_2(0) = e^{iB_2(0)}= e^{iB_2(2\pi)}e^{-i {P n_1\over K} a_1}= \tilde V_2(2\pi) \tilde U_1(\gamma)^{-{P n_1\over K}}}
we find
\eqn\tildeVI{\tilde V_1(0)\tilde V_2(0) = e^{-{2\pi i P\over K}} \tilde V_2(0)\tilde V_1(0)~.}
This means that the local operators on the same boundary realize a central extension of $\Z_{n_1}\times \Z_{n_2}$.  Similarly, we find the same algebra at the other boundary of the strip, $\sigma=2\pi$.  Furthermore, operators at different boundaries commute
\eqn\dic{\tilde V_I(0)\tilde V_J(2\pi)=\tilde V_J(2\pi)\tilde V_I(0)~.}

The lack of commutativity in a single boundary can be interpreted as dependence on their order along the boundary.  Note that $\tilde V_I^K$ are central.  Indeed, the boundary operator $ \tilde V_I^K$ is the limit of the bulk operator $V_I$ \Vnota.  And as such it can smoothly move to the bulk and return to the boundary at a different point.  Therefore, the order of such an operator along the boundary is not important and it is central.

As a check of the fact that $ \tilde V_I^K$ is the limit of the bulk operator $V_I$ \Vnota\ we can verify that our algebra of operators satisfies
\eqn\VIsame{\tilde V_I(0)^K= \tilde V_I(2\pi)^K ~.}

As another check, note that there exist integers $r_I$ such that
\eqn\Vhattildes{\eqalign{
&\tilde U_1^{n_1\over K} =\tilde V_2(0)^{r_2}\tilde V_2(2\pi)^{-r_2} \cr
&\tilde U_2^{n_1\over K} =\tilde V_1(0)^{r_1}\tilde V_1(2\pi)^{-r_1} ~.
}}
This reflects that fact that the dependence on the line in $\tilde U_I^{n_I\over K}$ is topological and it factorizes to two local boundary operators.

In conclusion, $\tilde U_I=e^{ia_I}$ generate $\Z_{n_1}\times \Z_{n_2}$ (without a central extension), which can be identified as the ordinary global symmetry of the theory.  $\tilde V_I(0)=e^{iB_I(0)}$ (or alternatively $\tilde V_I(2\pi)= e^{iB_I(2\pi)}$) realize a central extension of $\Z_{n_1}\times \Z_{n_2}$.  Its $\Z_{n_1\over K}\times \Z_{n_2\over K}$ subgroup generated by $\tilde V_I(0)^K$ (or alternatively $\tilde V_I(2\pi)^K$) does not have a central extension and is identified with the one-form symmetry we saw in the bulk.  Finally, $\tilde U_I=e^{ia_I}$ transform under the symmetry generated by $\tilde V_I=e^{iB_I}$ and $\tilde V_I$ transform under the symmetry generated by $\tilde U_I$.

\bigskip
\centerline{\it Free boundary conditions}
\bigskip

Alternatively, we can study the system with free boundary conditions.  If we do that with the action \actionDW, we derive the boundary equation of motion $B\big|=0$.  Therefore, in order to have $A\big|=0$ as a result of a boundary equation of motion, we add the boundary term
\eqn\boundter{{i\over 2\pi} \int_{\partial M}( n_1B_1A_1+n_2 B_2A_2) ~.}
This added boundary term \boundter\ can be interpreted as replacing the first two term $n_IB_IdA_I$ in the action \actionDW\ with $-n_IdB_IA_I $ (integration by parts).  As with Dirichlet boundary conditions, we still need to assume that the gauge parameters $f_I$ vanish at the boundary.  The analysis of this system is identical to the discussion with Dirichlet boundary conditions.

One might want to extend the $U(1)\times U(1)$ gauge symmetry to the boundary.  To do that, we note that under a gauge transformation the action \actionDW\ plus \boundter\ is shifted by
\eqn\boundt{{i\over
2 \pi }\int _{\partial M} \Big(  p\, \lcm(n_1,n_2)f_1df_2 +n_IB_Idf_I \Big)~.}
In order to cancel it we need to add some Stueckelberg fields on the boundary.  One can regard the boundary theory as having a gauge anomaly, which is canceled by the anomaly inflow from the bulk.

Since the gauge group is $U(1)\times U(1)$, we add a pair of $2\pi$-periodic scalars $\phi_1,\phi_2$ with gauge transformations
\eqn\boundtransf{
\phi_I\to \phi_I-f_I }
and the action
\eqn\bdrya{S_{bdry}={i\over 2\pi} \int_{\partial M} \left( p\lcm(n_1,n_2)\phi_1d\phi_2 +n_IB_I(d\phi_I+A_I) \right)~,}
where we included here the term \boundter.  With the added degrees of freedom the boundary equations of motion of $B_I$ does not set $A\big|$ to zero but to a pure gauge
\eqn\boundrel{d\phi_I +A_I\big| =0 ~.}

It is important that the bulk action \actionDW\ together with the boundary action \bdrya\ are consistent with the field identifications under shifts by $2\pi$ and are also invariant under the global symmetry that shifts $B_I$ (the action depends only on $dB_I$).

We can add to the boundary theory \bdrya\ additional terms constructed out of the gauge invariant combinations $\left(p\lcm(n_1,n_2)\epsilon_{IJ}\phi_J - n_IB_I\big|\right)$ and $\left(d\phi_I + A_I\big|\right)$.  One such possibility is to replace $S_{bdry}$ \bdrya\ by
\eqn\bdryaa{{i\over 2\pi} p\lcm(n_1,n_2) \int_{\partial M} \left(-  \phi_1 d\phi_2 -\epsilon_{IJ}\phi_I A_J\right)~.}
This action was studied and analyzed in \KapustinGUA.  Unlike our system, it is equivalent to the Dirichlet boundary conditions $B\big|=0$ rather than $A\big|=0$ and therefore it breaks the global discrete symmetry.

The analysis of the operators in the theory with the boundary terms \bdrya\ is virtually identical to that with Dirichlet boundary conditions \localb\lineb.  These operators are made gauge invariant using $\phi_I$:
\eqn\UIt{\eqalign{
&\tilde U_I(\gamma) = e^{-i\phi_I(\CP)} e^{i\int_\CP^{\CP'} A_I}e^{i\phi_I(\CP')} \cr
&\tilde V_I(\CP)=e^{iB_I(\CP)}\ e^{ - i\epsilon_{IJ}{n_J P\over K }\phi_J(\CP)}~.}}

We can also use canonical quantization of the theory on a strip.  As in the discussion around \canonicalDWbb, we pick axial gauge in the bulk (leaving the gauge fields along the boundaries $A_I(0)$ and $A_I(2\pi)$ unfixed) and find the effective quantum mechanical action
\eqn\stripwp{\eqalign{
S_{eff}={i\over 2\pi} \int dt \Bigg(&n_1\Big(B_1(0) -{n_2P\over K}\phi_2(0)\Big) \partial_0 \Big(a_1 -\phi_1(0)+\phi_1(2\pi)\Big)\cr
+ &n_2\Big(B_2(2\pi) +{n_1 P\over K}\phi_1(2\pi)\Big) \partial_0 \Big(a_2 - \phi_2(0) +\phi_2(2\pi)\Big) \Bigg) ~. }}
Even though this Lagrangian is independent of $A_I(0)$ and $A_I(2\pi)$, it is
is invariant under gauge symmetries with $f_I(0)$ and $f_I(2\pi)$.  Therefore, we can easily fix the gauge $\phi_I(0)=\phi_I(2\pi)$ to find the same problem we had with Dirichlet boundary conditions \canonicalDWbb.

As expected, the canonical quantization on the strip with free boundary conditions is identical to that with Dirichlet boundary conditions.

\appendix{B}{$\Z_n$ Topological 2-Form Gauge Theory in $4d$}

In this appendix we review and extend the discussion of \KapustinGUA\ of a $\Z_n$ topological 2-form gauge theory in $4d$.  Since many of the points are discussed in \KapustinGUA\ and others are very similar to the $2d$ example in Appendix A, we will be rather brief.

Following \refs{\GukovZKA,\KapustinUXA,\KapustinGUA} we take the action
\eqn\fourda{
S={i n\over 2\pi}\int B\wedge dA+{i pn\over 4\pi} \int B\wedge B~,}
where $A$ is a one-form gauge field, $B$ is a 2-form gauge field, and $n$ and $p$ are integers. The system is invariant under zero-form gauge transformations of $A$ as well as the one-form gauge symmetry
\eqn\Btra{
B\to B+d\lambda,\quad A\to A-p \lambda~.}
The action \fourda\ is gauge-invariant on an arbitrary 4-manifold provided
\eqn\pqu{{n p\over 2}\in\Z }.
If we only allow spin manifolds, $p$ can be an arbitrary integer.
There is also a periodic identification of the parameter $p$.
\eqn\piden{p\sim p+2n~}
on a general 4-manifold and
\eqn\pidenspin{p\sim p+n~}
on a spin 4-manifold.

We will find it convenient to define
\eqn\LKdef{\eqalign{
&L=\gcd(n,p)\cr
&K={n\over \gcd(n,p)}~.}}

For $p=0$ this system is equivalent to a $\Z_n$ gauge theory \refs{\MaldacenaSS,\BanksZN}.  It has both a one-form and a two-form $\Z_n$ global symmetries, which shift $A$ and $B$ respectively.  As in Appendix A, we would like to understand the fate of these symmetries for nonzero $p$.

The theory has surface operators generated by
\eqn\surfo{U(\Sigma)=\exp(i  \oint_\Sigma B)~.}
They satisfy
\eqn\surfos{U^L=1~.}
The genuine line operators are generated by
\eqn\genline{V(\gamma)=\exp(i K \oint_\gamma A ) ~,}
where we suppressed a trivial surface bound by $\gamma$.  It satisfies
\eqn\linesa{V^L=1~.}
These line and surfaces and their correlation functions are identical to a $\Z_L$ gauge theory.  They realize a $\Z_n/\Z_K = \Z_L$ one-form global symmetry (the quotient is done by the relation \surfos) and a $\Z_L\subset \Z_n$ two-form global symmetry (the restriction to the subgroup is a result of the fact that the basic Wilson line $\exp(i\oint A)$ is not gauge invariant).

This restriction of the global symmetries can be interpreted as gauging a $\CK=\Z_K \subset \Z_n$ of the one-form global symmetry.  As in the $2d$ example of Appendix A, this system can be the low energy approximation of a UV system with a global $\Z_n$ one-form symmetry, which is spontaneously broken to $\CK=\Z_K$.  The low energy theory describes only the quotient $\Z_n/\Z_K = \Z_L$.

On a manifold $\CM$ with a boundary $\partial \CM$ we need to set boundary conditions.  With Dirichlet boundary conditions we have the option of setting $A\big|=0$, which preserves the two-form global symmetry or $B\big|=0$, which preserves the one-form global symmetry.  We choose the second option $B\big|=0$.  This restricts the one-form gauge symmetry \Btra\ to be trivial along the boundary, but the zero-form gauge symmetry of $A$ is not affected.

With these boundary conditions we have new boundary operators.  We can have surface operators in the bulk, which end on a line $\gamma $ on the boundary $ \partial \CM$
\eqn\surfb{\tilde U(\Sigma) = \exp(i \int_\Sigma B)  \qquad , \qquad \partial\Sigma = \gamma \in \partial M~.}
We can also have line operators
\eqn\Wlineb{W(\gamma) = \exp(i \oint_\gamma A) \qquad , \qquad \gamma \in \partial \CM~.}
Since the one-form gauge parameter $\lambda$ vanishes on $\partial \CM$, these operators are gauge invariant.  The operator $\tilde U(\Sigma)^L$ is interesting.  Because of \surfos, the dependence on $\Sigma $ in the bulk is trivial and hence it is a genuine line operator
\eqn\genlbo{\tilde V(\gamma) = \tilde U(\Sigma)^L ~.}
Furthermore
\eqn\genlbopL{\tilde V(\gamma)^{p\over L}= \tilde U(\Sigma)^p = W(\gamma)^{-1} ~,}
where we used the fact that in the bulk an open surface $\exp(ip\int B + i\oint A)$ is trivial.  Therefore, the line operators $\tilde V(\gamma)$ generate all genuine line operators in the boundary.  Finally the limit of the genuine bulk operators $V(\gamma) = \exp(iK\oint_\gamma A)$ as it approaches the boundary can be identified as
\eqn\bulkbouA{V(\gamma) = W(\gamma)^{K}=\tilde V(\gamma)^{-{p K\over L}} ~.}

As in Appendix A (and in \KapustinGUA), we can add Stueckelberg fields $a$ to restore the one-form gauge symmetry on the boundary.
We take the boundary action
\eqn\Sbdrya{S_{bdry}={i\over 2\pi}\int_{\partial M}\left( -{ np\over 2} a da+ n a dA\right)~,}
and then the full system is gauge invariant provided the gauge transformation \Btra\ acts also on $a$
\eqn\atoal{a\to a-\lambda~.}
This allows us to express the boundary operators \surfb\Wlineb\genlbo\ in a fully gauge invariant way
\eqn\surfba{\eqalign{
&\tilde U(\Sigma) = \exp(i \oint_\gamma a + i \int_\Sigma B) \cr
&W(\gamma) = \exp(i \oint_\gamma A - i p\oint a) \cr
&\tilde V(\gamma) = \tilde U(\Sigma)^L = \exp(i L\oint_\gamma a + i L\int_\Sigma B) ~.}}
This presentation makes it easy to explore their correlation functions.  The first term in \Sbdrya\ leads to
\eqn\sspgcd{\langle \tilde V(\gamma)^s \tilde V(\gamma')^{s'}\rangle = \exp\left( 4\pi i \ell(\gamma,\gamma') s s' L^2 \over n p\right)~,}
where $\ell(\gamma,\gamma')$ is the linking number. As a check, $\tilde V^{pK\over L} = \tilde V^{pn\over L^2}$ is central -- its correlation functions are not subject to the braiding.  Indeed, because of \bulkbouA\ it can smoothly move to the bulk and get un-braided.  Also, it is clear that we have the relation
\eqn\tildeVrel{\tilde V^{pn/L}=1~.}

\bigskip\bigskip
\centerline{\it More about the theory on compact $\CM$}
\bigskip

For general $p$, the structure of the bulk observables in this $\Z_n$ topological 2-form gauge theory is very similar to that of an ordinary one-form topological gauge theory with gauge group $\Z_L$. Nevertheless, the two theories are not equivalent, as the above discussion of the boundary conditions shows.  It is interesting to compare the partition functions of the two theories on a closed 4-manifold $\CM$.  Recall first of all that the partition function of the one-form $\Z_L$ gauge theory is
\eqn\Zoneformfourd{
{\cal Z}^{(1)}(\CM,\Z_L)={\vert H^1(\CM,\Z_L)\vert \over \vert H^0(\CM,\Z_L)\vert}.
}
The denominator is the standard normalization factor (the order of the gauge group). The partition function of the 2-form gauge theory is easy to evaluate for $p=0$. One simply counts the number of inequivalent flat 2-connections, divides by the number of global one-form gauge symmetries, and multiplies by the number of symmetries. In this way one gets
\eqn\Ztwoformpzero{
{\cal Z}^{(2)}(\CM,\Z_n,p=0)={\vert H^2(\CM,\Z_n)\vert \vert H^0(\CM,\Z_n)\vert \over \vert H^1(\CM,\Z_n)\vert }.
}
Comparing with \Zoneformfourd\ for $L=n$, we see that
\eqn\compZ{
{\cal Z}^{(2)}(\CM,\Z_n,p=0)=n^{\chi(\CM)} {\cal Z}^{(1)}(\CM,\Z_n),
}
where $\chi(\CM)$ is the Euler characteristic of $\CM$. While the two partition functions are not the same, they differ by a factor which is an exponential of a local geometric term (the integral of the Euler density).

For $p\neq 0$ the computation is more involved. The partition function of the 2-form gauge theory is
\eqn\Ztwoform{
{\cal Z}^{(2)}(\CM,\Z_n,p)={\vert H^0(\CM,\Z_n)\vert \over \vert H^1(\CM,\Z_n)\vert } \sum_{B\in H^2(\CM,\Z_n)} \exp\left(2\pi i \int_\CM \frP_\sigma(B)\right),
}
where $\frP_\sigma: H^2(\CM,\Z_n)\ra H^4(\CM,\R/\Z)$ is the Pontryagin square operation corresponding to the quadratic function
\eqn\sigmanp{
\sigma:\Z_n\ra \R/\Z,\quad \sigma: x\mapsto { p x^2\over 2n}.
}
Since the  Pontryagin square is a quadratic operation, the partition function is proportional to a Gauss sum over $H^2(\CM,\Z_n)$. The evaluation of such Gauss sums is a standard problem in number theory and topology. For simplicity, let us consider the case when the integral cohomology has no torsion. The quadratic function in the exponential takes the form
\eqn\quadratic{
\sum_a {p\over 2n} (B^a)^2 I_{aa}+\sum_{a<b} {p\over n} B^a B^b I_{ab},
}
where $a$ labels a basis of integral 2-cycles, $B^a$ are integers modulo $n$, and $I_{ab}$ is the intersection form of $\CM$.  It turns out that if both the integers $K$ and $p/L=p/\gcd(n,p)$ and the form $I_{ab}$ are odd, the Gauss sum vanishes. Indeed, suppose $I_{aa}$ is odd for some $a$ and $K$ and $p/L$ are also odd. Then shifting $B^a\mapsto B^a+K$ shifts the quadratic function by $1/2$, which means that the terms in the Gauss sum related by such a shift cancel pairwise. Suppose now that either the form $I_{ab}$ is even (which happens if $\CM$ is a spin manifold) or $K$ is even, or $p/L$ is even. Then the shift $B^a\mapsto B^a+K$ for any $a$ leaves the quadratic function unchanged (modulo integers). Therefore, the Gauss sum becomes $\vert H^2(\CM,\Z_L) \vert$ times a similar Gauss sum over $H^2(\CM,\Z_K)$. The quadratic function on $H^2(\CM,\Z_K)$ has a non-degenerate associated bilinear form, which implies \Milgram\ that the Gauss sum is equal to
\eqn\gaussnondeg{
{\sqrt {\vert H^2(\CM,\Z_K)\vert}} \exp(2\pi i\sigma(\CM)/8).
}
Note that the signature of a spin 4-manifold is always divisible by 16, so the phase of the partition function is nontrivial only for a non-spin manifolds (see Appendix E).
Putting all of this together, we find (in the torsionless case):
\eqn\Ztwoformtwo{
{\cal Z}^{(2)}(\CM,\Z_n,p)=K^{\chi(\CM)/2} L^{\chi(\CM)}e^{i\sigma(\CM)/8} {\vert H^1(\CM,\Z_L)\vert \over \vert H^0(\CM,\Z_L)\vert}.
}
Thus, unless $K$, $p/L$ and $I^{ab}$ are all odd, the partition function agrees with the partition function of the $\Z_L$ topological one-form gauge theory up to local geometric counterterms.

It might appear peculiar that for odd $K$ and $p/L$ the partition function of the $\Z_n$ 2-form gauge theory agrees with that of the $\Z_L$ one-form gauge theory on spin manifolds, but not in general. This disagreement is ultimately due to the fact that the $\Z_L$ one-form gauge theory has $\Z_L$ one-form global symmetry, but for odd $K$ and $p/L$ the 2-form gauge theory has only $\Z_{L/2}$ one-form global symmetry.  (Using that fact that $np$ must be even, it is easy to see that if $K$ and $p/L$ are odd, $L$ is even.) We exploited the non-invariance of the action \quadratic\ under $\Z_L$ one-form global symmetry to argue that the partition function vanishes on a non-spin manifold. One can bring the two theories in agreement by modifying the action of the 2-form gauge theory by a term
\eqn\modZLS{
i\pi \int_\CM w_2\cup B_2,
}
where $B_2\in H^2(\CM,\Z_2)$ is a reduction of $B$ modulo $2$, and $w_2\in H^2(\CM,\Z_2)$ is the 2nd Stiefel-Whitney class of $\CM$. By Wu's formula, this is the same as $i\pi\int_\CM B_2\cup B_2$, and it is easy to see that the additional term restores the $\Z_L$ one-form global symmetry for all manifolds.

\appendix{C}{$SU(N)/\Z_k$}

Our goal is to construct the $SU(N)/\Z_k$ in terms of continuum fields and to identify its operators and its global symmetry.  We will construct the theory in two ways.  First, we will start with an $SU(N)$ theory with its $\Z_N$ one-form electric global symmetry and gauge a $\Z_k$ subgroup of it.  The second construction will start with an $SU(N)/\Z_N$ theory with its $\Z_N$ one-form magnetic global symmetry and will gauge a $\Z_{N/k}$ subgroup of it.

We follow \KapustinGUA\ and construct the theory by starting with a
\eqn\Gdef{\CG={SU(N)\times U(1)\over \Z_k}}
gauge theory with gauge field
\eqn\UNg{a +{1\over k} \tilde A \unit ~,}
where $\unit$ a unit matrix and $a$ is traceless.  Next, we remove the $U(1)$ degrees of freedom by imposing a one-form gauge symmetry
\eqn\Ahatg{\tilde A \to \tilde A - k \lambda}
with $\lambda$ a properly normalized $U(1)$ gauge field and $k$ is a divisor of $N$.  (In \KapustinGUA\ only the special case $k=N$ was discussed.)

In this theory the magnetic flux through the closed two-surface $M^{(2)}$ is $\oint_{M^{(2)}} d \tilde A$ and because of the gauge symmetry \Ahatg\ only
\eqn\UmNk{ e^{{im\over k} \oint_{M^{(2)}} d \tilde A} \qquad m \in \Z}
are meaningful.  They lead to a global one-form magnetic $\Z_k$ symmetry.

We can also add to the action the Pontryagin square term
\eqn\Potns{{i p \over 4\pi k} \,d\tilde A \wedge d\tilde A ~,}
where gauge invariance under \Ahatg\ demands that $p$ is an integer and $p \sim p+k$.  (For simplicity we assume that our spacetime is spin.  Otherwise the conditions are slightly more complicated.)

Because of the one-form gauge symmetry \Ahatg\ the fundamental Wilson loop needs a surface $\Sigma$ with $\partial\Sigma =\gamma$
\eqn\fundW{\CW_f(\gamma,\Sigma) = \left( \Tr_f \CP e^{i\oint_\gamma a}\right) e^{{i \over k}\oint_\gamma \tilde A} e^{-{i \over k} \int_\Sigma d\tilde A}= \CW_f^\CG(\gamma)\, e^{-{i \over k} \int_\Sigma d\tilde A}}

It is convenient to introduce a gauge field $B$ for the one-form gauge symmetry \Ahatg\
\eqn\Bgauges{B \to B + d\lambda}
and to add to the Lagrangian
\eqn\addLag{ {i\over 2\pi} F\wedge (d\tilde A + k B) +{ i p k \over 4\pi} B \wedge B ~,}
where $F$ is a two-form Lagrange multiplier that transforms as
\eqn\Fgauges{F \to F - p d\lambda ~.}
We can also integrate out $\tilde A$ and express the answer in terms of the dual gauge field $A$ such that $F=dA$ with
\eqn\Agauges{A \to A - p \lambda ~.}
This turns \addLag\ to
\eqn\addLagB{ {i k\over 2\pi} B \wedge d A+{ i p k \over 4\pi} B \wedge B ~,}
which is identified as a $\Z_k$ gauge theory.

In this presentation the basic surface operator in \UmNk\ can be written as
\eqn\UmNka{ U^M( M^{(2)})= e^{{i\over k} \oint_{M^{(2)}} d \tilde A} = e^{i\oint_{M^{(2)}} B} }
and it is clear from \addLagB\ that
\eqn\Usupk{ U^M( M^{(2)})^k=1  ~.}
Also, the Wilson operator \fundW\ can be written as
\eqn\fundWa{\CW_f(\gamma,\Sigma) = \CW_f^\CG(\gamma)\, e^{i \int_\Sigma B} }
and it is clear that
\eqn\electricop{V(\gamma) =\CW_f(\gamma,\Sigma)^k}
is a genuine line operator; i.e.\ it is independent of $\Sigma$.

The 't Hooft line is expressed in terms of $A$ as
\eqn\tHooftl{\CT(\gamma,\Sigma)= e^{i\oint_\gamma A} e^{ip\int_\Sigma B }~,}
where again, we need the surface $\Sigma$.  Using $d\tilde A =-kB$ in \fundW\ we see that the dyonic line
\eqn\dyonl{\CD(\gamma)=\CT(\gamma,\Sigma) \CW_f(\gamma,\Sigma)^{-p}}
is a genuine line operator, which is independent of $\Sigma$.

Another perspective on this construction can be obtained by starting with an $SU(N)/\Z_N$ theory.  This theory can be described by starting with a $U(N)$ gauge theory ($k=N$ above) with the $U(1)\subset U(N)$ gauge field $\hat A$ with the one-form gauge symmetry $\hat A \to \hat A - N\lambda$ \KapustinGUA.  This theory has a global one-form $\Z_N$ symmetry generated by $e^{{i\over N} \oint d\hat A}$.  The $SU(N)/\Z_k$ theory is obtained by gauging a $\Z_{N/k}$ subgroup it. As in (7.17) of \KapustinGUA, we do it by adding to the Lagrangian
\eqn\addNK{ {i N\over 2\pi k } \tilde B\wedge d\tilde A  +{1 \over 2\pi} \tilde B\wedge d \hat A~.}
Here $\tilde A$ and $\hat A$ are properly normalized $U(1)$ gauge fields and $\tilde B$ is a properly normalized two-form gauge field.  The first term in \addNK\ is the $\Z_{N/k}$ gauge theory and the second term couples it to the $SU(N)/\Z_N$ theory. The field $\tilde B$ is a Lagrange multiplier implementing the constraint ${N\over k} d\tilde A= d\hat A$, which has the effect of restricting the periods $\oint d\hat A$ to be multiples of $N/k$ and directly leads to \UNg\ and to our construction above.  Note that the one-form gauge symmetry \Ahatg\ also follows from the invariance of \addNK.

Let us discuss the global one-form symmetries of these theories.  Electric surface operators can be constructed out of the $SU(N)$ degrees of freedom.  When the gauge group is $SU(N)$, the fundamental Wilson loop \fundW\fundWa\ is a genuine line and the basic electric surface operator $U^E$ acts on it with a phase $e^{2\pi i/N}$, and so $U^E$ generates a one-form global $\Z_N$ symmetry.  The magnetic operator, $U^M$ in \UmNka, is trivial in this theory.  In the $SU(N)/\Z_N$ theory, $U^M$ generates a $\Z_N$ symmetry, while $U^E$ is trivial.  In a general theory, the $N$th powers of $U^E$ and $U^M$ are always trivial, and the one-form symmetry group is given by the quotient of the $\Z_N \times \Z_N$ group they generate by the subgroup which acts trivially on the genuine lines.

To describe this quotient for the $SU(N)/\Z_k$ theory with discrete theta angle $p$, it is convenient to write the following matrix, whose columns generate the lattice of charges in $\Z_N \times \Z_N$ of the genuine lines:
\eqn\oneformquot{
\pmatrix{ k  & p \cr 0 & N/k }
}
The one-form symmetry group of the theory will then be the quotient of $\Z_N \times \Z_N$ by the group generated by the columns of this matrix.  The matrix can be put into Smith normal form by multiplying it from the left and the right by two $SL(2,\Z)$ matrices.  This does not affect the resulting quotient group.  After doing this it becomes diagonal with entries $\gcd(k,N/k,p)$ and $N/\gcd(k,N/k,p)$.  Thus the one-form symmetry group of this theory is:
\eqn\appConeform{
\Z_{\gcd(k,N/k,p)} \times \Z_{N/\gcd(k,N/k,p)} ~.
}

\appendix{D}{Super Yang Mills with adjoint matter}

In this appendix we review the analysis of \CachazoZK\ in our language.
These authors studied a supersymmetric $U(N)$ gauge theory with matter fields $\Phi$ in the adjoint representation and with a superpotential $W=\Tr P(\Phi)$ with $P(\Phi)$ a polynomial.  For simplicity, we take it to be a generic cubic with critical points $a$ and $b$.  Semiclassically, the vacua are characterized by two integers $N_1$ and $N_2$, where $\langle \Phi \rangle $ has $N_1$ eigenvalues equal to $a$ and $N_2$ eigenvalues equal to $b$.  This expectation value breaks the $U(N)$ gauge symmetry to $U(N_1) \times U(N_2)$.  At lower energies, this $U(N_1) \times U(N_2)$ gauge theory becomes strongly coupled and it has $N_1 N_2$ vacua labeled by $k_1=0,2,...,N_1-1$ and $k_2=0,2,...,N_2-1$.  The low energy degrees of freedom are $U(1)\times U(1)$ gauge fields.

Our goal is to identify the higher form global symmetries and their breaking in the various vacua.

This theory has an electric $U(1)$ one-form global symmetry and a magnetic $U(1)$ one-form global symmetry.  The order parameters of this symmetry are $W^{n_e} H^{n_m}$ with $W$ and $H$ the Wilson and 't Hooft loops.  If instead, the gauge group is $SU(N)$, then the one-form electric symmetry is $\Z_N$ and the magnetic one-form symmetry is absent.  In this case we can limit ourselves to operators with $n_e=0,...,N-1$.  Operators with other values of $n_e$ transform as these and have the same long distance behavior.  Furthermore, only operators with $n_m=0$ are genuine line operators and the operators with $n_m=1,...,N-1$ need a surface.

Let us examine the behavior of the various loops in the $N_1N_2$ vacua.  First, it is clear that $W^{N_1}$ and $W^{N_2}$ cannot have an area law.  These operators include contributions from $SU(N_1) \times SU(N_2)$ singlet representations, which are not confined. Hence, these operators exhibit perimeter law or Coulomb behavior.  It follows that $W^{\gcd(N_1,N_2)}$ is also not confined.  Correspondingly, the electric symmetry is broken at least to $\Z_{\gcd(N_1,N_2)}$.  This was described in \CachazoZK\ as electric screening.

Second, in the vacuum labeled by $(k_1,k_2)$ certain dyons condense.  The condensed dyons in the $SU(N_1)$ sector have charges aligned with (a component of) $W^{k_1}H$ and the condensed dyons in the $SU(N_2)$ sector have charges aligned with (a component of) $W^{k_2}H$.  Therefore, these two operators are also not confined.  Hence, $(W^{k_1}H)(W^{k_2}H)^{-1}=W^{k_1-k_2}$ is also not confined.  This has the effect of further breaking the $\Z_{\gcd(N_1,N_2)}$ electric one-form symmetry to $\Z_{t=\gcd(N_1,N_2, k_1-k_2)}$.  In addition, the non-confining behavior of $W^{k_1}H$ completely breaks the magnetic one-form symmetry.  This was described in \CachazoZK\ as magnetic screening.

We conclude that the only loops with area law are $W^n$ with $n=1,...,t-1$ and they realize the unbroken $\Z_t$ one-form symmetry.

If the gauge group is $SU(N)$ the previous discussion still applies.  The operators $W^{k_1}H$ and $W^{k_2}H$ need a surface, but the operator $(W^{k_1}H)(W^{k_2}H)^{-1}=W^{k_1-k_2}$ is a genuine line.  In this case the electric one-form symmetry is $\Z_N$ and it is spontaneously broken to $\Z_t$.

In \CachazoZK\ the integer $t$ was described as the confinement index -- the smallest integer such that $W^t$ does not exhibit area law.  From our perspective, this is the order of the unbroken one-form symmetry.

As emphasized in \CachazoZK, a vacuum where semiclassically $U(N)$ is broken to $ U(N_1) \times U(N_2)$ can be continuously deformed to a vacuum where semiclassically $U(N)\to U(\tilde N_1) \times U(\tilde N_2)$, but the integer $t$ should be the same in these vacua.  The unbroken $\Z_t$ global symmetry cannot change under duality.

\appendix{E}{Non-Abelian gauge theory on non-spin four-manifolds}

In section 4 we discussed gauging one-form symmetry $\Gamma\subset Z(\CG)$ of a non-Abelian gauge theory with gauge group $\CG$. We argued that different choices of gauging can be encoded in a choice of a bilinear form $\eta:\Gamma\times\Gamma\to U(1)$. In fact, this is imprecise: on a non-spin manifold $X$ one needs to specify a quadratic refinement of $\eta$, i.e. a function $\sigma:\Gamma\to U(1)$ such that $\sigma(n\gamma)=n^2\sigma(\gamma)$ for any $\gamma\in\Gamma$ and
\eqn\quadrref{
\sigma(\gamma+\gamma')=\sigma(\gamma)+\sigma(\gamma')+\eta(\gamma,\gamma').
}
This was explained in \refs{\KapustinUXA,\KapustinQSA} by interpreting the gauging procedure as coupling the non-Abelian gauge theory to a $4d$ topological gauge theory with a one-form symmetry $\Gamma$.

We can also see this using the formulation in terms of topological defects $U_\gamma$, $\gamma\in\Gamma$, implementing the one-form symmetry transformations. These defects are localized on codimension $2$ submanifolds  of $X$, and in the absence of 't Hooft anomalies a background 2-form gauge field can be represented by a network of such defects. Generically two defects $U_\gamma$ and $U_{\gamma'}$ intersect at points, and when computing the partition function one needs to choose a phase for each such point. This phase is nothing but $\exp(2\pi i\eta(\gamma,\gamma'))$. Indeed, we can excise a small disk on $U_\gamma$ at the expense of inserting a line with magnetic charge $\gamma$ and electric charge $\eta(\gamma,\cdot)$. This line winds around $U_{\gamma'}$ and thus its insertion amount to a phase factor $\exp(2\pi i\eta(\gamma,\gamma'))$.

Assuming for simplicity that the homology of $X$ has no torsion, we can choose a basis of 2-cycles $\ell^a$ with intersection form $I_{ab}$. Then a 2-form gauge field $B\in H^2(X,\Gamma)$ can be represented by its periods $B^a\in \Gamma$ , and the above phase factor can be written as
\eqn\oneformSPTT{
\exp \pi i \left[ \sum_{a,b}I_{ab} \eta(B^a, B^b) \right]
}
Here the summation is over all $a,b$, including $a=b$, since a 2-cycle $\ell^a$ can have self-intersections. It is clear now that the partition function may have a sign ambiguity, precisely because the self-intersection numbers may be odd. If the intersection form $I_{ab}$ is even (which happens for spin manifolds), this does not happen, and the partition function is well-defined. In general, we need to specify the numbers $\eta(B^a,B^a)$ modulo even integers rather than modulo integers. This is equivalent to choosing a quadratic refinement of $\eta$: we simply set $\sigma(\gamma)={1\over 2} \eta(\gamma,\gamma)$. It is easy to see that $\sigma$ is a quadratic refinement of $\eta$.

In this discussion we have neglected torsion in (co)homology; this restriction can be removed as follows. In general, a $\Gamma$-valued 2-form gauge field on $X$ is a 2-cocycle  $B\in C^2(X,\Gamma)$. We can use either \v{C}ech or simplicial cocycles. Gauge symmetry can shift this cocycle by an arbitrary exact cocycle, so gauge-invariant information is contained in a cohomology class $[B]\in H^2(X,\Gamma)$ . The phase factor must be an integral of some element $H^4(X,\R/\Z)$, which is quadratic in $B$ and changes by an exact cocycle under gauge transformations. Such quadratic expressions can be classified and have the form
\eqn\oneformSPTPont{
\int_X{\frak P}_\sigma(B),
}
where ${\frak P}_\sigma: H^2(X,\Gamma)\ra H^4(X,\R/\Z)$ is a certain quadratic cohomology operation on $H^2(X,\Gamma)$ depending on $\sigma$ called the Pontryagin square. It is completely characterized by the above properties plus the requirement
\eqn\oneformSPTPontq{
{\frak P}_\sigma(B+B')-{\frak P}_\sigma(B)-{\frak P}_\sigma(B')=\eta(B,\cup B').
}
In the absence of torsion there is a simple formula for $\frP_\sigma$. In this case we have $H^2(X,\Gamma)=H^2(X,\Z)\otimes\Gamma$, and thus we can represent $B$ by its periods $B^a\in\Gamma$ along a basis of integral 2-cycles $\ell^a$. Then we have
\eqn\Pontryaginsq{
\frP_\sigma(B)=\sum_a \sigma(B^a) \hat\ell_a\cup\hat\ell_a+\sum_{a<b} \eta(B^a,B^b) \hat\ell_a\cup \hat\ell_b,
}
where $\hat\ell_a$ is a basis in $H^2(X,\Z)$ dual to the basis $\ell^a$ in $H_2(X,\Z)$.

\appendix{F}{Non-commuting fluxes}

In this appendix we make a connection to the works \refs{\FreedYA, \FreedYC}, which discussed the quantization of the free $U(1)$ gauge theory in four dimensions.  These authors placed the theory on $\CM_3 \times \R$ and viewed $\R$ as time.  Then they constructed operators that measure discrete electric and magnetic fluxes and found that when $\CM_3$ contains torsion cycles, these operators do not always commute.  We would like to understand this in the language of $U_g$ defects.

Recall that this theory has $U(1)$ electric and magnetic $1$-form $U(1)$ symmetries. From the discussion in section $3$, we can construct unitary operators $U^E(\alpha)$ and $U^M(\alpha)$ implementing these symmetries for each $\alpha \in H^1(\CM_3,U(1))$.  The latter group is Pontryagin-dual to $H^2(\CM_3,\Z)$, and so the eigenvalues of these operators take values in this group.  These eigenvalues are precisely the discrete electric and magnetic fluxes of \refs{\FreedYA, \FreedYC}.  For example, the magnetic flux is given by the first Chern class of the $U(1)$ bundle in which a given wavefunction is supported and takes values in $H^2(\CM_3,\Z)$.

An important observation of \refs{\FreedYA, \FreedYC} was that there is not, in general, a simultaneous grading of the Hilbert space by these fluxes.  In other words, the operators $U^E(\alpha)$ and $U^M(\beta)$ need not commute.  Rather, one has (eq. (1.21) in \FreedYA):
\eqn\commfail{
U^E(\alpha) U^M(\beta)= \langle\phi(\alpha),\beta\rangle U^M(\beta) U^E(\alpha)
}
Here $\phi:H^1(\CM_3,U(1))\ra H^2(\CM_3,\Z)$ is the Bockstein homomorphism, which is defined as follows. Let us choose a lift of $\alpha$ to a cochain in $C^1(\CM_3, \R)$. The coboundary of this cochain must project to zero in $C^2(\CM_3, \R/\Z)$ and so is valued in $C^2(\CM_3, \Z)$.  It is clearly closed, and one can check that this procedure defines uniquely a cohomology class $\phi(\alpha)$ in $H^2(\CM_3, \Z)$. In fact the image of $\phi$ can be checked to be contained in the torsion subgroup of $H^2(\CM_3, \Z)$.

To give a concrete example, suppose $\CM_3$ contains a single torsion cycle $\gamma$ of order $r$, e.g.\ $\CM_3 = \S^3/\Z_r$.  Then $H_1(\CM_3,\Z)\simeq H^2( \CM_3,\Z)\simeq \Z_r$.  The cycle $\gamma$ generating $H_1(\CM_3,\Z)$ is not a boundary, but $r \gamma$ is, so that there is a surface $\Sigma$ whose boundary wraps $r$ times around $\gamma$. If $\alpha(\gamma)=g\in U(1)$, we must have $g^r=1$. That is, if we identify $U(1)$ with $\R/\Z$, we must have $\alpha(\gamma)=k/r$, where $k$ is an integer defined modulo $r$. The integral 2-cocycle $\phi(\alpha)$ assigns $k$ to the 2-chain $\Sigma$. Although $k$ is defined modulo $r$, the corresponding class in $H^2(\CM_3,\Z)$ is well-defined, thanks to the relation $\partial\Sigma=r\gamma$. The Poincare-dual 1-cycle is homologous to $k\gamma$ in this case.

In the language of topological defects, an operator $U^E(\alpha)$ is represented by a network of codimension-1 defects in $\CM_3$, with one-dimensional junctions. In the above example, the network consists of a single surface $\Sigma$, while the junction is along $\gamma$. Objects charged under electric and magnetic one-form symmetries are electric and magnetic loops in $\CM_3$. The commutation relation \commfail\ can be interpreted as the statement that junctions in the electric network are charged under magnetic one-form symmetry, and vice versa. In the above example, an electric defect $U^E(\Sigma,k)$ labeled by $k\in\Z/r\Z$ has magnetic charge $k[\gamma]\in H_1(\CM_3,\Z)$. According to the discussion in section 3, this signals that $U(1)\times U(1)$ global one-form symmetry has a mixed 't Hooft anomaly.

In order to show that this phenomenon is not specific to free field theories, we demonstrate it in a non-Abelian gauge theory.  Specifically, we study an $su(4)$ gauge theory and consider several distinct theories: $SU(4)$ with the genuine lines generated by $W$, $(SU(4)/\Z_4)_{p=0,1,2,3}$ with the genuine lines generated by $HW^p$, $(SU(4)/\Z_2)_{-}$ with the genuine lines generated by $H^2W$ and $(SU(4)/\Z_2)_{+}$ with the genuine lines generated by $W^2$ and $H^2$ \AharonyHDA.  The global one-form symmetries of the $SU(4)$, $(SU(4)/\Z_4)_{p}$ and $(SU(4)/\Z_2)_{-}$ theories is $\Z_4$ generated by $U^E, U^M$ and $U^E$ respectively.  And the one-form symmetry of the $(SU(4)/\Z_2)_{+}$ theory is $\Z_2\times \Z_2$ generated by $U^E$ and $U^M$.

Next, we place these theories on the spatial manifold $\CM_3 = \S^3/\Z_r$ and construct the Hilbert space.  For $r=4$ we can construct generators of the global symmetries out of an open version of the generators $U^E$ or $U^M$ on the open surface $\Sigma$ whose boundary $\partial \Sigma=4\gamma$ with $\gamma$ the nontrivial one cycle of $\S^3/\Z_4$.  The charge operator is constructed by appropriate gluing along $\gamma$.  These generators generate the $\Z_4$ or $\Z_2\times \Z_2$ symmetry.  In all these theories we find four sectors in the Hilbert space \RazamatOPA\ transforming under the different representations of their corresponding global symmetries.

The situation with $r=2 $ is different.  In the $SU(4)$, $(SU(4)/\Z_4)_{p}$ and $(SU(4)/\Z_2)_{-}$ theories we cannot construct generators using $U^E$ or $U^M$, because in order to make them topological we need to attach to them $H^2(\gamma)$ or $W^2(\gamma)$, but these are not genuine lines in these theories.  Instead, we can consider only operators constructed out of $(U^E)^2$ and $(U^M)^2$ and therefore the global symmetry is only $\Z_2$ and correspondingly, there are only two sectors in the Hilbert space \RazamatOPA.  In the $(SU(4)/\Z_2)_{+}$ theory we can construct symmetry generators using $\CO^E=U^E(\Sigma)H^2(\gamma)$ and $\CO^M=U^M(\Sigma)W^2(\gamma)$ because $W^2$ and $H^2$ are genuine lines in this theory.  As in \refs{\FreedYA, \FreedYC}, these operators satisfy a central extension of $\Z_2\times \Z_2$; i.e.\ the discrete electric and magnetic fluxes do not commute.  Indeed, the analysis of \RazamatOPA\ shows that if we diagonalize the magnetic flux the Hilbert space of the theory contains only two sectors.

\listrefs
\end